\documentclass[a4paper,12pt,useAMS]{emulateapj}
\usepackage{apjfonts,graphicx,graphics}

\usepackage{natbib}
 
\newcommand{\simgt}{\lower.5ex\hbox{$\; \buildrel > \over \sim \;$}}
\newcommand{\simlt}{\lower.5ex\hbox{$\; \buildrel < \over \sim \;$}}

\def\btheta{\mbox{\boldmath $\theta$}}

\def\bx{\mbox{\boldmath $x$}}

\def\bC{\mbox{\boldmath $C$}}
\def\bL{\mbox{\boldmath $L$}}

\def\singlebond{\@makechembond\@ne}
\def\doublebond{\@makechembond\tw@}
\def\triplebond{\@makechembond\thr@@}

\lefthead{Umetsu et al.}

\righthead{Mass Structure of Cl0024+1654 from a Full Lensing Analysis}

\begin{document}

\title{
The Mass Structure of the 
Galaxy Cluster Cl0024+1654
from a Full Lensing Analysis of Joint Subaru and ACS/NIC3 Observations
\altaffilmark{1}}

\author{
Keiichi Umetsu\altaffilmark{2}, 
Elinor Medezinski\altaffilmark{3},
Tom Broadhurst\altaffilmark{3},
Adi Zitrin\altaffilmark{3},
Nobuhiro Okabe\altaffilmark{2},
Bau-Ching Hsieh\altaffilmark{2},
Sandor M. Molnar\altaffilmark{2}
} 

\altaffiltext{1}
 {Based in part on data collected at the Subaru Telescope,
  which is operated by the National Astronomical Society of Japan.}
\altaffiltext{2}
 {Institute of Astronomy and Astrophysics, Academia Sinica,
  P.~O. Box 23-141, Taipei 10617, Taiwan.}
\altaffiltext{3}
  {School of Physics and Astronomy, Tel Aviv University, Tel Aviv 69978, Israel.
   }



\begin{abstract}
We derive an accurate mass distribution of the rich galaxy cluster 
Cl0024+1654 ($z=0.395$) based on deep Subaru $BR_{\rm c}z'$ imaging and
our recent comprehensive strong lensing analysis of HST/ACS/NIC3
observations.  
We obtain the weak lensing distortion and magnification of
undiluted samples of red and blue background
galaxies by carefully combining all color and positional information.
Unlike previous work, the weak and strong lensing are
in excellent agreement where the data overlap. The joint mass
profile continuously steepens out to the virial
radius with only a minor contribution $\sim 10\%$ in the mass from
 known subcluster at a projected distance of 
 $\simeq 700\,$kpc$\,h^{-1}$.   
The cluster light profile closely resembles the mass profile,
and our model-independent $M/L_R$
profile shows an overall flat behavior with a mean of $\langle
M/L_R\rangle\simeq 230h (M/L_R)_\odot$, but exhibits a mild declining
trend with increasing radius at cluster outskirts,
$r\simgt 0.6r_{\rm vir}$.
The projected mass distribution for the entire cluster is well
fitted with a single Navarro-Frenk-White model with 
a virial mass, $M_{\rm vir}=
 (1.2\pm 0.2)\times 10^{15}M_{\odot}\,h^{-1}$,
and
a concentration,  $c_{\rm vir}=9.2^{+1.4}_{-1.2}$.
This model fit
is fully consistent with the depletion of the red background counts,
providing independent confirmation.
Careful examination and interpretation 
of X-ray and dynamical data, based on recent
high-resolution cluster collision simulations,
strongly suggest that 
this cluster system is in a post collision state, 
which we show is consistent with our well-defined mass profile
for a major merger occurring along the line of sight, viewed
 approximately $2-3$\,Gyr after impact 
when the gravitational potential has had time to relax in the center,
before the gas has recovered and before the outskirts are fully
virialized.
Finally, our full lensing analysis provides a model-independent
 constraint of $M_{\rm 2D}(<r_{\rm vir})=(1.4\pm 0.3)\times
 10^{15}M_\odot \,h^{-1}$ for the projected mass of the whole 
 system, including any currently unbound material beyond the virial
 radius,
which can constrain the sum of the two pre-merger cluster masses when
designing simulations to explore this system.
\end{abstract}
 
\keywords{cosmology: observations --- galaxies: clusters: individual
(Cl0024+1654) ---  gravitational lensing} 


\section{Introduction} 
\label{sec:intro}

 Cl0024+1654 ($z=0.395$) is the most distant
 cluster of galaxies discovered by \citet{Zwicky1959},
 and is the focus of some of the
 most thorough studies of cluster properties, including the internal
 dynamical \citep{Czoske+2002,Diaferio+2005},
 X-ray emission
 \citep{Soucail+2000_CL0024,2004ApJ...601..120O,Zhang+2005_CL0024}, 
 and both weak \citep{2003ApJ...598..804K,Hoekstra2007,Jee+2007_CL0024}
 and strong
\citep{Colley+1996_CL0024,Tyson+1998_CL0024,Broadhurst+2000_CL0024,2006ApJ...642...39C,Zitrin+2009_CL0024} 
 lensing work.  

 Despite the round and concentrated appearance of this cluster, several
 independent lines of evidence point to recent merging of a substantial
 substructure. 
 The internal dynamics of about $300$ spectroscopically-measured
 galaxies 
 has been modeled by a high-speed, line-of-sight collision of 
 two systems with a mass ratio of the order of 2:1,
 leading to a compressed distribution of
 velocities along the line of sight \citep{Czoske+2001,Czoske+2002}. 
 A direct line-of-sight merger is
 also used to account for the ``ring'' of dark matter claimed by 
\citet{Jee+2007_CL0024}, based on the central mass distribution derived
 from deep Hubble Space Telescope (HST) Advanced Camera for
 Surveys (ACS) images.
 On a larger scale, the mass distribution derived from
 a mosaic of HST/WFPC2 pointings \citep{2003ApJ...598..804K}
 reveals an additional substantial subcluster in the mass 
 at a projected radius of $3\arcmin$ 
 coincident with a noticeable concentration of galaxies.
 This substructure is however not along the line of sight, but 
 is associated with the main cluster component in redshift space
 \citep{Czoske+2002}, and
 lies  $R\sim 700\,{\rm kpc}\,h^{-1}$ 
 northwest (NW) in projection from the center of the main cluster.

No clear evidence of excess X-ray emission is found at the location of
the NW galaxy clump, but interaction may be implied instead by the
anomalously low level of X-ray emission relative to the standard
X-ray luminosity--mass relation. 
The measured gas temperature is also unusually
low, only $T_X \simeq 4.5$\,keV \citep{2004ApJ...601..120O}, 
over the full range of radius ($R\simlt 300\,{\rm kpc}\,h^{-1}$) 
probed by deep Chandra data, whereas recent weak and strong lensing
observations indicate that the cluster is a high-mass system with a
total projected mass of $M_{\rm 2D}\simgt 10^{15}M_{\odot}$
\citep{Hoekstra2007,Jee+2007_CL0024,Mahdavi+2008_CL0024,Zitrin+2009_CL0024}.
The careful hydrodynamical simulations of
\citet{2001ApJ...561..621R}
predict that in a period of 1--3\,Gyr
(or a timescale of the sound crossing time, $\tau_{sc}$) 
after a substantial merger 
the hot gas associated with the whole system is
extensively distributed so that the emissivity is actually markedly
reduced by virtue of the lowered gas density, once the shock
associated with the collision has dissipated. Furthermore, substantial
proportion of the gas may escape from the system together with high
velocity galaxies, so that the velocity dispersion of the remainder is
significantly reduced. These lack of any observed hot shocked gas component
implies very clearly that no merger has happened very recently, within
a couple of Gyr, unlike the bullet cluster \citep{Clowe+2006_Bullet}
and other clusters caught in
the first collisional encounter \citep{Okabe+Umetsu2008}.

Cl0024+1654 displays one of the finest examples of gravitational lensing
forming a symmetric 5-image system of a well-resolved galaxy, which
was first noted by \citet{Koo1988_CL0024}
and later resolved into a close triplet
of arcs by \citet{Kassiola+1992_CL0024},
with two additional
images found by
\citet{Smail+1996_arcs}  
and by \citet{Colley+1996_CL0024}
using HST WFPC1 and WFPC2 data, respectively. These arcs
have been used by \citet{Colley+1996_CL0024}
to construct an image
of the source, whose redshift $z=1.675$ 
\citep{Broadhurst+2000_CL0024}
permits an accurate and model-independent enclosed mass for the central 
$R<100\,{\rm kpc}\,h^{-1}$ area of 
$M(<R) = (1.11\pm0.03)\times 10^{14}M_{\odot}\,h^{-1}$, with a
central mass-to-light ratio of
$(M/L)_B=(320\pm 30) h (M_\odot/L_\odot)_B$ 
\citep{Broadhurst+2000_CL0024}.

 More recently the central mass profile has been constrained by
 lensing with the identification of many new multiply lensed images
 \citep{Zitrin+2009_CL0024}
 in very deep multi-color imaging with HST/ACS as
 part of the ACS/GTO program \citep{Holland+2003_ACS}. 
 Here a joint fit was
 made to 33 lensed images and their photometric redshifts with a
 relatively simple 6-parameter description of the deflection
 field. 
 This modeling method has been recently applied to two unique, X-ray
 luminous  high-$z$ clusters, MACS J1149.5+2223 and MACS J0717.5+3745,
 uncovering  
 many sets of  multiply-lensed images
 \citep{Zitrin+Broadhurst2009,Zitrin+2009_0717}. 
 The high resolution and accuracy of colors allow a
 secure identification of counter images by delensing the pixels of
 candidate lensed images to form a source which is then relensed to
 predict the detailed appearance of counter images, a technique
 developed for similar high quality ACS/GTO data of A1689 
\citep{2005ApJ...621...53B}.
 This model has been used to estimate the magnification of
 high-redshift candidate galaxies 
 with photometric redshifts $z_{\rm phot}\sim 6-7$
 identified in combining this data
 with deep near-infrared images \citep{Zheng+2009_CL0024}.
 Here we add to this new strong-lensing information with 
 new weak-lensing data from deep, multi-color  imaging with the 
 Subaru telescope
 to examine the mass distribution in detail over the full
 profile of the cluster. These high-quality Subaru images 
 span the full optical range in the $B$, $R_{\rm c}$, and $z'$ 
 passbands, allowing us to
 define a background sample of red and blue galaxies free from cluster
 member and foreground galaxies. We have learned from our earlier work
 that without adequate color information, the weak-lensing signal can be
 heavily diluted particularly towards the cluster center by the
 presence of unlensed cluster members
 \citep{BTU+05,Medezinski+07,UB2008}, 
 so that weak lensing underpredicts the Einstein radius
 derived from 
 strong-lensing studies and the gradient of the inner mass profiles based on
 weak lensing is also underestimated. Unfortunately, many examples of
 this problem are present in the literature, and here we carefully
 explore the weak-lensing signal in color-color space and by
 comparison with the deep photometric redshift survey in
 the COSMOS field \citep{Ilbert+2009_COSMOS}.

A major motivation for pursuing improved lensing measurements is the
increased precision of model predictions for the mass density profiles of
cluster-size massive dark-matter halos based on $N$-body simulations in
the standard $\Lambda$ cold dark-matter (hereafter $\Lambda$CDM)
model 
\citep{2007ApJ...654..714H,2007MNRAS.381.1450N,Duffy+2008}.
Clusters of galaxies provide a definitive test of the
standard structure-formation model because their mass density profiles,
unlike galaxies, are not 
expected to be significantly affected by cooling of baryons 
\citep[e.g.,][]{Blumenthal+1986,Broadhurst+Barkana2008}.
This is because the high temperature and low density of 
the intra-cluster gas (hereafter ICG) prevents efficient cooling and
hence 
the majority of baryons simply trace the gravitational potential of
the dominant dark matter. Massive clusters are of particular interest
in the context of this model, because they are predicted to have a
distinctively shallow mass profile (or low concentration) described by
the form proposed by \citet{1997ApJ...490..493N}
and this question
has been the focus of our preceding work 
\citep{BTU+05,UTB08,BUM+08,UB2008,2009ApJ...694.1643U}.

The paper is organized as follows.  
We briefly summarize in \S 2 the basis of cluster weak gravitational
lensing.
In \S 3 we describe the observations, the photometry procedure,
the sample selection,
and the weak-lensing shape analysis.
In \S 4 we present our weak lensing
methods, and derive the cluster lensing distortion and convergence profiles 
from Subaru weak lensing data.
In \S 5 we examine in detail the cluster 
mass and light profiles 
based on the joint weak and strong lensing analysis. 
In \S 6 we compare our results with previous studies of Cl0024+1654 to
examine the long-standing puzzle on large mass discrepancies between
lensing and X-ray/dynamical methods, and investigate
the implications of observed discrepancies and anomalies.
In \S 7 we explore and discuss a possible interpretation of the observed 
X-ray features and mass discrepancies
Finally, a summary is given in \S 8.

Throughout this paper, 
we use the AB magnitude system,
and  adopt a
concordance $\Lambda$CDM cosmology with 
$\Omega_{m}=0.3$, $\Omega_{\Lambda}=0.7$, and
$h\equiv H_0/(100\, {\rm km\, s^{-1}\, Mpc^{-1}})=0.7$.
In this cosmology, $1\arcmin$ corresponds to $224$\,kpc\,$h^{-1}$
(and $1\arcsec$ to $3.73$\,kpc\,$h^{-1}$)
at the cluster redshift.
All quoted errors are 68.3\% confidence limits unless otherwise stated.
The reference sky position is
the center of the central bright elliptical galaxy 
(the galaxy 374 in the spectroscopic catalog of \cite{Czoske+2002}):  
${\rm R.A.} =$ 00:26:35.69, 
${\rm Decl.} =$ +17:09:43.12 (J2000.0).
We refer to this position as the optical cluster center, hereafter.
The cluster center of mass for our radial profile analysis is chosen to
be the {\it dark-matter} center 
at
$\Delta {\rm R.A.} = -2.32\arcsec$,
$\Delta {\rm Decl.} = -1.44\arcsec$
of \cite{Zitrin+2009_CL0024}.

\section{Basis of Cluster Weak Lensing}
\label{sec:basis}

Weak gravitational lensing is responsible for the weak shape distortion
and magnification of the images of background sources due to the
gravitational field of intervening foreground clusters of galaxies
and large scale structures in the universe
\citep[e.g.,][]{1999PThPS.133...53U,2001PhR...340..291B}. 
The deformation of the image can be described by the $2\times 2$
Jacobian matrix $\cal{A}_{\alpha\beta}$
($\alpha,\beta=1,2$) of the lens mapping.\footnote{Throughout the paper 
we assume in our weak lensing analysis that the angular size of
background galaxy images is sufficiently small 
compared to the scale over which the underlying lensing fields vary, so
that the higher-order weak lensing effects, such as {\it flexion}, can
be safely neglected; see, e.g.,
\cite{2005ApJ...619..741G,HOLICs1,HOLICs2}.} 
The Jacobian ${\cal A}_{\alpha\beta}$ is real and symmetric, so that
it can be decomposed as
\begin{eqnarray}
\label{eq:jacob}
{\cal A}_{\alpha\beta} &=& (1-\kappa)\delta_{\alpha\beta}
 -\Gamma_{\alpha\beta},\\
\Gamma_{\alpha\beta}&=&
\left( 
\begin{array}{cc} 
+{\gamma}_1   & {\gamma}_2 \\
 {\gamma}_2  & -{\gamma}_1 
\end{array} 
\right),
\end{eqnarray}
where 
$\delta_{\alpha\beta}$ is Kronecker's delta,
$\Gamma_{\alpha\beta}$ is the trace-free, symmetric shear matrix
with $\gamma_{\alpha}$ being the components of 
spin-2
complex gravitational
shear $\gamma:=\gamma_1+i\gamma_2$,
describing the anisotropic shape distortion,
and $\kappa$ is the 
lensing convergence responsible for the 
trace-part of the Jacobian matrix, describing the isotropic area
distortion.
In the weak lensing limit where $\kappa,|\gamma|\ll 1$, 
$\Gamma_{\alpha\beta}$ induces a quadrupole anisotropy of the 
background image, which can be observed from ellipticities 
of background galaxy images.
The flux  magnification due to gravitational lensing
is given by the inverse Jacobian determinant,
\begin{equation}
\label{eq:mu}
\mu = 
\frac{1}{{\rm det}{\cal A}}
=
\frac{1}{(1-\kappa)^2-|\gamma|^2},
\end{equation}
where we assume subcritical lensing, i.e., 
${\rm det}{\cal A}(\btheta)>0$.

The lensing convergence is expressed as a line-of-sight projection
of the matter density contrast $\delta_m=(\rho_m-\bar{\rho})/\bar{\rho}$ 
out to the source plane ($s$)
weighted by certain combination of co-moving angular diameter
distances $r$
\citep[e.g.,][]{2000ApJ...530..547J},
\begin{eqnarray}
\label{eq:kappa}
&&\kappa =
\frac{3H_0^2\Omega_m}{2c^2}
\int_0^{\chi_s}\!d\chi\, 
{\cal G}(\chi,\chi_s)
\frac{\delta_m}{a} 
\equiv \int\!d\Sigma_m\,\Sigma_{\rm crit}^{-1},\\
\label{eq:gdistance}
&&{\cal G}(\chi,\chi_s)=
\frac{r(\chi)r(\chi_s-\chi)}{r(\chi_s)},
\end{eqnarray}
where $a$ is the cosmic scale factor,
$\chi$ is the co-moving distance,
$\Sigma_m$ is the surface mass density of matter, $\Sigma_m
=\int\!d\chi \, a(\rho_m-\bar{\rho})$,
with respect to the cosmic mean density $\bar{\rho}$, and
$\Sigma_{\rm crit}$ 
is the critical surface mass density for gravitational lensing,
\begin{equation} 
\label{eq:sigmacrit}
\Sigma_{\rm crit} = \frac{c^2}{4\pi G}\frac{D_{s}}{D_d D_{ds}}
\end{equation}
with $D_s$, $D_d$, and $D_{ds}$ being the (proper) angular diameter distances
from the observer to the source, from the observer to the deflecting
lens, and from the lens to the source, respectively.
For a fixed background cosmology and a lens redshift $z_d$,
$\Sigma_{\rm crit}$ is a function of background source redshift
$z_s$.
For a given mass distribution $\Sigma_m(\btheta)$, the lensing signal is
proportional to the angular diameter distance ratio,
\begin{equation}
\label{eq:dratio}
\beta(z_s) =  \max\left[0,
\frac{D_{ds}(z_s)}{D_s(z_s)}
\right],
\end{equation}
where $\beta(z_s)$ is zero for unlensed objects with $z_s\le z_d$.

In the present weak lensing study we aim to reconstruct
the dimensionless surface mass density $\kappa(\btheta)$ 
from weak lensing distortion data.
For a two-dimensional mass reconstruction,
we utilize the relation between the gradients of 
$\kappa$ and $\gamma$
\citep{1995ApJ...439L...1K,2002ApJ...568...20C}, 
\begin{equation}
\label{eq:local}
\triangle \kappa (\btheta)
= \partial^{\alpha}\partial^{\beta}\Gamma_{\alpha\beta}(\btheta)
= 2\hat{\cal D}^*\gamma(\btheta)
\end{equation}
where 
$\hat{\cal D}$ is the complex differential operator
$\hat{\cal D}=(\partial_1^2-\partial_2^2)/2+i\partial_1\partial_2$.
The Green's function for the two-dimensional Poisson equation is
$\triangle^{-1}(\btheta,\btheta')=\ln|\btheta-\btheta'|/(2\pi)$,
so that equation (\ref{eq:local}) can be solved to yield the following
non-local relation between $\kappa$ and $\gamma$ 
\citep{1993ApJ...404..441K}:
\begin{equation}
\label{eq:gamma2kappa}
\kappa(\btheta) = 
\frac{1}{\pi}\int\!d^2\theta'\,D^*(\btheta-\btheta')\gamma(\btheta')
\end{equation}
where $D(\btheta)$ is the complex kernel defined as 
$D(\btheta)=(\theta_2^2-\theta_1^2-2i\theta_1\theta_2)/|\btheta|^4$.
In general, the observable quantity is not the 
gravitational shear $\gamma$ but the complex {\it reduced} shear,
\begin{equation}
\label{eq:redshear}
g=\frac{\gamma}{1-\kappa}
\end{equation}
in the subcritical regime where ${\rm det}{\cal A}>0$
(or $1/g^*$ in the negative parity region with ${\rm det}{\cal A}<0$). 
We see that the reduced shear $g$ is invariant under the following
global transformation:
\begin{equation}
\label{eq:invtrans}
\kappa(\btheta) \to \lambda \kappa(\btheta) + 1-\lambda, \ \ \ 
\gamma(\btheta) \to \lambda \gamma(\btheta)
\end{equation}
with an arbitrary scalar constant $\lambda\ne 0$ 
\citep{1995A&A...294..411S}.
This transformation is equivalent to scaling 
the Jacobian matrix ${\cal A}(\btheta)$ with $\lambda$, 
$\cal {A}(\btheta) \to \lambda {\cal
A}(\btheta)$. This mass-sheet degeneracy can be unambiguously broken
by measuring the magnification effects, because the magnification $\mu$
transforms under the invariance transformation (\ref{eq:invtrans}) as
\begin{equation}
\mu(\btheta) \to \lambda^2 \mu(\btheta).
\end{equation}

\section{Subaru Data and Analysis}
\label{sec:subaru}

In this section we present a technical description of our weak
lensing analysis of Cl0024+1654
based on deep Subaru $BR_{\rm c}z'$ images.
The reader only interested in the main result may skip directly to
\S \ref{sec:wl+sl}.

\subsection{Subaru Data and Photometry}
\label{subsec:data}

For our weak-lensing analysis of Cl0024+1654 we retrieved from the
Subaru archive, SMOKA, \footnote{http://smoka.nao.ac.jp.}
imaging data in $B$, $R_{\rm c}$, and $z'$ taken with the wide-field
camera Suprime-Cam 
\citep[$34^\prime\times 27^\prime$;][]{2002PASJ...54..833M}
at the prime-focus of the 8.3m Subaru telescope.
The cluster was observed in the course of the PISCES program
\citep{2005PASJ...57..309K,2005PASJ...57..877U,Tanaka+2005_PISCES}. 
The FWHM in the co-added mosaic image is
$1.27\arcsec$ in $B$, 
$0.80\arcsec$ in $R_{\rm c}$,
and 
$0.82\arcsec$ in $z'$ 
with $0.202\arcsec$ pixel$^{-1}$, covering a field of $\simeq  34\arcmin
\times 26\arcmin$. 
The observation details of Cl0024+1654 are listed in
Table~\ref{tab:subaru}. We use the $R_{\rm c}$-band data for our weak
lensing shape measurements (described in \S\ref{subsec:shear}) 
for which the instrumental response, sky background and seeing conspire
to provide the best-quality images. 
The standard pipeline reduction software for Suprime-Cam 
\citep[SDFRED, see][]{Yagi+2002_SDFRED,Ouchi+2004_SDFRED}
is used for flat-fielding, instrumental distortion correction,
differential refraction, sky subtraction, and stacking. Photometric
catalogs are constructed from stacked and matched images using
SExtractor \citep{1996A&AS..117..393B}.
A composite $BR_{\rm c}z'$ color image of the central $8\arcmin\times
8\arcmin$ region of the cluster is shown in Figure \ref{fig:BRZ}.
Since our lensing work relies much on the colors of galaxies, special
care has to be paid to the measurement of colors from $BR_{\rm c}z'$
images with different seeing conditions.  For an accurate measurement
of colors the Colorpro \citep{colorpro}
routine is used;
this allows us to use 
SExtractor's 
AUTO magnitudes for total magnitudes of galaxies
and
its isophotal magnitudes for estimation of colors, 
and applies a further
correction for the different seeing between different bands. From the
colors and magnitudes of galaxies we can safely select background
galaxies (see \S\ref{subsec:color}) for the weak-lensing analysis.

Astrometric correction is done with the SCAMP tool
\citep{Bertin+2006_SCAMP}
using reference objects in the NOMAD catalog \citep{NOMAD}.
Photometric zero-points were calculated from associated standard star
observations taken on the same night. Since for $B$ and $R_{\rm c}$
only one standard field was taken, and for $z'$-band only a few
spectrophotometric standards are available, we further calculated
zero-points by fitting galaxy templates to bright elliptical galaxies
in this cluster, using BPZ \citep{BPZ}
in the Subaru three-band photometry combined with HST/ACS
images taken in the $F435W, F625W$, and $F850LP$ bands, for
which accurate zero-points are well known. Stars were also compared in
both Subaru and the HST/ACS images.
Finally, we find consistency of our magnitude zero-points to within $\pm
0.05$\,mag.

\subsection{Weak Lensing Distortion Analysis}
\label{subsec:shear}

We use the IMCAT package developed by
N. Kaiser\footnote{http://www.ifa.hawaii.edu/\~{}kaiser/imcat.} 
to perform object detection, photometry and shape measurements,
following the formalism outlined in  \citet[][KSB]{1995ApJ...449..460K}.
Our analysis pipeline is implemented based on the procedures described
in \citet{2001A&A...366..717E}
and on verification tests with STEP1 and STEP2 data of mock ground-based
observations  
\citep{2006MNRAS.368.1323H,2007MNRAS.376...13M}.
For details of our implementation of the KSB+ method,
see also \citet{UB2008,2009ApJ...694.1643U}.

\subsubsection{Object Detection}
\label{subsubsec:detection}

Objects are first detected as local peaks in the 
co-added mosaic image in $R_{\rm c}$
by using the IMCAT hierarchical peak-finding
algorithm {\it hfindpeaks}
which for each object yields object parameters such as a peak position,
$\bx$, 
an estimate of the object size, $r_g$
(where $r_g$ is the Gaussian scale length of the object as given
by {\it hfindpeaks}; see KSB for details),
the significance of
the peak detection, $\nu$.  
The local sky level and its gradient are 
measured around each object 
using the IMCAT {\it getsky} routine.
Total fluxes and half-light radii, $r_h$, are then
measured 
on sky-subtracted images using the IMCAT {\it apphot} routine.
For details of our IMCAT-based photometry, see \S 4.2.1 of 
\cite{2009ApJ...694.1643U}.
We removed from our detection catalog 
extremely small objects with $r_g<1$\,pixel,
objects with low detection significance, $\nu<10$,
objects with large raw ellipticities, $|e|>0.5$ 
\citep[see][]{2000ApJ...539..540C,2006MNRAS.368.1323H,2007MNRAS.376...13M}
where $|e|=(a^2-b^2)/(a^2+b^2)$ with axis-ratio $0<b/a\le 1$ for
images with elliptical isophotes,
noisy detections with unphysical negative fluxes, and
objects containing more than $10$ bad pixels, ${\tt nbad}>10$.
This selection procedure yields an object catalog with
$N=82431$ ($93.2$ arcmin$^-2$).

\subsubsection{Weak Lensing Distortion Measurements}
\label{subsubsec:shape}

In the KSB algorithm we measure 
the complex image ellipticity 
$e_{\alpha} = \left\{Q_{11}-Q_{22}, Q_{12} \right\}/(Q_{11}+Q_{22})$ 
from the weighted quadrupole moments of the surface brightness of
individual galaxies, 
\begin{equation}
\label{eq:Qij}
Q_{\alpha\beta} = \int\!d^2\theta\,
 W({\theta})\theta_{\alpha}\theta_{\beta} 
I({\btheta})
\ \ \ (\alpha,\beta=1,2)
\end{equation} 
where $I(\btheta)$ is the surface brightness distribution of an object,
and $W(\theta)$ is a Gaussian window function matched to the size of the
object.
In equation (\ref{eq:Qij})
the object centroid is chosen as the coordinate
origin, and the maximum radius of integration is chosen to be
$\theta_{\rm max}=4r_g$ from the centroid.
In our practical implementation of equation (\ref{eq:Qij}),
we iteratively refine the Gaussian-weighted centroid of each object
to accurately measure the object shapes.
An initial guess for the centroid is provided with the IMCAT {\it
hfindpeaks} routine (see \ref{subsubsec:detection}).
During the process objects with
offsets in each iteration larger than 3 pixels are removed
\citep{Oguri+2009_Subaru}.

The next step is to correct observed image ellipticities for the 
point-spread function (PSF)
anisotropy using a sample of stars as references:
\begin{equation}
e'_{\alpha} = e_{\alpha} - P_{sm}^{\alpha \beta} q^*_{\beta} 
\label{eq:qstar}
\end{equation}
where $P_{sm}$ is the {\it smear polarizability} tensor  
which is close to diagonal, and
$q^*_{\alpha} = (P_{sm}^*)^{-1}_{\alpha \beta}e_*^{\beta}$ 
is the stellar anisotropy kernel.
We select bright
($20\simlt R_{\rm c}\simlt 22$), 
unsaturated 
stellar objects 
identified in a branch
of the $r_h$ versus $R_{\rm c}$ diagram to measure $q^*_{\alpha}$.
In order to obtain a smooth map of $q^*_{\alpha}$ which is used in
equation (\ref{eq:qstar}), we divided the co-added mosaic image 
of $\sim 10{\rm K} \times 7.7{\rm K}$ pixels
into $4\times 3$ rectangular blocks, where
the block length is based on the coherent
scale of PSF anisotropy patterns
\citep[see, e.g.,][]{UB2008,2009ApJ...694.1643U}. 
In this way the PSF anisotropy in
individual blocks can be well described by fairly low-order
polynomials.
We then fitted the $q^*$ in each block independently with
second-order bi-polynomials, $q_*^{\alpha}(\btheta)$, in
conjunction with iterative outlier rejection on each
component of the residual:
$\delta e^*_\alpha =
e^*_{\alpha}-(P_{sm}^*)^{\alpha\beta}q^*_{\beta}(\btheta)$.   
The final stellar sample contains 672 stars (i.e., $N_*\sim 60$ stars
per block), 
or the mean surface number density of $n_*\simeq
0.75$\,arcmin$^{-2}$,
with $\nu\simgt 800$.
We note that the mean stellar ellipticity before correction is
$\bar{e_1}^* \simeq -3.5\times 10^{-3}$
and
$\bar{e_2}^* \simeq -5.0\times 10^{-2}$
over the data field, while the residual
$e^*_{\alpha}$ after correction
is reduced to
$ {\bar{e}^{*{\rm res}}_1} = (-0.07\pm 1.18)\times 10^{-4}$
and
$ {\bar{e}^{*{\rm res}}_2} = (+2.10\pm 1.58)\times 10^{-4}$.
The mean offset from the null expectation is reduced down to
$|\bar{e}^{* \rm res}| = (2.1\pm 1.6) \times 10^{-4}$, which is almost
two-orders of magnitude smaller than the weak lensing signal in cluster
outskirts.
We show in Figure \ref{fig:anisopsf1}
the quadrupole PSF anisotropy field
as measured from stellar ellipticities before and after the anisotropic
PSF correction.
Figure \ref{fig:anisopsf2} shows
the distributions of stellar ellipticity
components before and after the PSF anisotropy correction.
From the rest of the object
catalog, we select objects with 
$r_h > \overline{r_h^*} + \nu_h \sigma(r_h^*)$
as a $R_{\rm c}$-selected weak lensing galaxy sample, 
where $\overline{r_h^*}$
is the median value of stellar half-light radii $r_h^*$,
corresponding to half the median width of the circularized PSF
over the data field, $\sigma(r_h^*)$ is the rms dispersion of
$r_h^*$, and $\nu_h$ is conservatively taken to be $1.5$ in the present
weak lensing analysis.

We then need to correct image ellipticities for
the isotropic smearing effect 
caused by atmospheric seeing and the window function used for the shape
measurements. The pre-seeing reduced shear
$g_\alpha$ ($\alpha=1,2$) can be
estimated from 
\begin{equation}
\label{eq:raw_g}
g_{\alpha} =(P_g^{-1})_{\alpha\beta} e'_{\beta}
\end{equation}
with the {\it pre-seeing shear polarizability} tensor
$P^g_{\alpha\beta}$ defined as
\cite{1998ApJ...504..636H},
\begin{equation}
\label{eq:Pgamma}
P^g_{\alpha\beta} = P^{sh}_{\alpha\beta}- 
\left[
P^{sm} (P^{sm*})^{-1} P^{sh*}
\right]_{\alpha\beta}
\approx 
 P^{sh}_{\alpha\beta}-  P^{sm}_{\alpha\beta} 
\frac{{\rm tr}[P^{sh*}]}{{\rm tr}[P^{sm*}]}
\end{equation}
with $P^{sh}$ being the {\it shear polarizability} tensor.
In the second equality we have used a trace approximation to the stellar
shape tensors, $P^{sh*}$ and $P^{sm*}$.
To apply equation (\ref{eq:raw_g}) the quantity ${\rm tr}[P^{sh*}]/{\rm
tr}[P^{sm*}]$ must be known for each of the galaxies with different
size scales. 
Following
\citet{1998ApJ...504..636H}, we recompute 
the stellar shapes $P^{sh*}$ and $P^{sm*}$ in a range of filter scales
$r_g$ spanning that of the galaxy sizes.
At each filter scale $r_g$, 
the median 
 $\langle {\rm tr}[P^{sh*}]/{\rm tr}[P^{sm*}]\rangle$ 
over the stellar sample is calculated, and used in equation 
(\ref{eq:Pgamma}) as an estimate of  ${\rm tr}[P^{sh*}]/{\rm
tr}[P^{sm*}]$.
Further, we adopt a scalar correction scheme, namely
\begin{equation}
\label{eq:Pg}
(P_g)_{\alpha\beta}
=\frac{1}{2}{\rm tr}[P_g]\delta_{\alpha\beta}\equiv
P^{\rm s}_{g}\delta_{\alpha\beta}
\end{equation}
\citep{2001A&A...366..717E,1998ApJ...504..636H,UB2008,2009ApJ...694.1643U}. 

Following the prescription in \citet{UB2008}
and \citet{2009ApJ...694.1643U}, we compute for each object the variance
$\sigma_g^2$ of $g=g_1+ig_2$ from $N$ neighbors identified in the 
object size ($r_g$) and magnitude ($R_{\rm c}$) plane. We take $N=30$.
The dispersion $\sigma_g$ is used as an
rms error of the shear estimate for individual galaxies.
With the dispersion $\sigma_g$ we define for each object
the statistical weight $u_g$ by
\begin{equation}
\label{eq:ug}
u_g \equiv \frac{1}{\sigma_g^2+\alpha_g^2},
\end{equation}
where $\alpha_g^2$ is the softening constant variance 
\citep[e.g.,][]{2003ApJ...597...98H}. 
We choose $\alpha_g=0.4$, which is a typical value of 
the mean rms $\overline{\sigma}_g$ over the background
sample \citep[see, e.g.,][]{UB2008,2009ApJ...694.1643U}. 
The case with $\alpha_g=0$ corresponds to an inverse-variance weighting.
On the other hand,
the limit  $\alpha_g\gg \sigma_{g}$ yields a uniform weighting.
Figure \ref{fig:weight} shows the mean statistical weight
$\langle u_g\rangle$
as a function of object size $r_g$ ({\it left}) and of object magnitude
$R_{\rm c}$ ({\it right})
for the $R_{\rm c}$-selected weak-lensing galaxy sample, where
in each panel the mean weight is normalized to unity in the first bin. 

\subsubsection{Shear Calibration}
\label{subsubsec:calib}

In practical observations, the measurement of $P_g^{\rm s}$ 
is quite noisy for an individual faint galaxy.
Further, $P_g^{\rm s}$ depends nonlinearly on galaxy
shape moments (see, e.g., \citet{2001PhR...340..291B} 
for an explicit expression for the shape tensors), so that this
nonlinear error propagation may lead to a systematic bias in weak
lensing distortion measurements (T. Hamana, in private communication).
Indeed, it has been found that smoothing noisy $P_g^{\rm s}$ estimates 
does not necessarily improve the shear estimate but can lead to a
systematic underestimate of the distortion amplitude
by $\sim 10\%$--$15\%$
\citep[e.g.,][]{2001A&A...366..717E,2006MNRAS.368.1323H,2007MNRAS.376...13M}.  
In order to improve the precision in shear recovery, 
we have adopted the following shear calibration strategy:
First, we select as a sample of {\it shear calibrators}
those galaxies with a detection significance $\nu$
greater than a certain limit $\nu_{\rm c}$
and with a positive raw $P_g^{\rm s}$ value. 
Note that the shear calibrator sample
is a subset of the target galaxy sample (see \S
\ref{subsubsec:detection} and \S \ref{subsubsec:shape}).
Second, 
we divide the calibrator size ($r_g$) and magnitude ($R_{\rm
c}$) plane into a grid of $10\times 3$
cells\footnote{We note that $P_g$ is a strong function of the object
size $r_g$ (see, e.g., Figure 3 of Umetsu \& Broadhurst (2008)), and
only weakly depends on the object magnitude through $P^{sh}$.}
each containing approximately equal numbers of calibrators, and  compute
a median value of $P_g^{\rm s}$ at each cell in order to improve the
signal-to-noise ratio. 
Finally, each object in the target sample is matched to the nearest
point on the ($r_g,R_{\rm c}$) calibration grid to obtain the filtered
$P_g^{\rm s}$ measurement.
In this study, we take $\nu_{\rm c} = 30$ as the significance threshold
for the calibrator sample.\footnote{The significance threshold for the
target sample is 10 (see \S \ref{subsubsec:detection})}
Finally, we use the estimator 
$g_{\alpha} = e'_{\alpha}/\left< P_g^{\rm s}\right>$
for the reduced shear.
The final $R_{\rm c}$-selected galaxy sample contains
$N=38758$ objects, or the
mean surface number density of $\bar{n}_g=43.7$ galaxies arcmin$^{-2}$.
The mean variance over the 
galaxy sample is obtained as $\overline{\sigma_g^2}\simeq 0.20$, or
$\overline{\sigma}_g\equiv (\overline{\sigma_g^2})^{1/2}\simeq 0.46$.

We have tested our weak lensing analysis pipeline including the galaxy
selection procedure
using simulated  Subaru Suprime-Cam images of the STEP2 project
\citep{2007MNRAS.376...13M}. We find that we can recover
the weak lensing signal with good precision: typically,
$m\sim -5\%$ of the shear calibration bias, 
and $c\sim 10^{-3}$ of the residual
shear offset which is about one-order
of magnitude smaller than the weak-lensing signal in cluster outskirts
($|g|\sim 10^{-2}$).
This level of calibration bias is subdominant compared to the
statistical uncertainty ($\sim 15\%$) 
due to the intrinsic scatter in galaxy shapes, and the degree of bias
depends on the quality of the PSF in a complex manner. Therefore, we do
not apply shear calibration corrections to our data.
Rather we emphasize that
the most critical source of systematic uncertainty in cluster weak
lensing is dilution of the distortion signal due to
the inclusion of unlensed foreground and cluster member galaxies,
which can lead to an underestimation of the true signal for $R\simlt
400\,$kpc$\,h^{-1}$ by a factor of $2-5$ \citep[see Figure 1
of][]{BTU+05}.


\subsection{Galaxy Sample Selection from the Color-Color Diagram}
\label{subsec:color}


It is crucial in the weak lensing analysis
to make a secure selection of background galaxies in order
to minimize contamination by {\it unlensed}
cluster and foreground galaxies;
otherwise dilution of the distortion signal arises from
the inclusion of unlensed galaxies, particularly at small radius
where the cluster is relatively dense
\citep{BTU+05,Medezinski+07,UB2008}.
This dilution effect is simply to reduce the strength of
the lensing signal when averaged over a local ensemble of 
galaxies, in proportion to the fraction of unlensed galaxies
whose orientations are randomly distributed, thus diluting the lensing
signal relative to 
the true background level, derived from the uncontaminated background
level we will establish below.

To separate unlensed galaxies from the background
and hence minimize the weak lensing dilution,
we follow the background selection method 
recently developed by \cite{Medezinski+2009}, which relies on 
empirical correlations in color-color (CC) space 
derived from the deep Subaru photometry,
by reference to the deep photometric redshift survey in the COSMOS field
(see \S \ref{subsec:depth}).
For Cl0024+1654, we have a wide wavelength coverage ($BR_{\rm c}z'$) of
Subaru Suprime-Cam. 
When defining color samples, we require that objects are detected in all
three Subaru bands.
Further, we limit the data to $z' \simlt 25.5$ AB mag 
in the reddest band available for
the cluster.
Beyond this limit incompleteness creeps into the
bluer bands, complicating color measurements, in particular 
of red galaxies.
Our CC-selection criteria yielded a total of $N=8676, 1655$, and 5004
galaxies for the red, green, and blue samples, respectively, 
usable for our weak lensing distortion analysis; these correspond to
mean surface number densities of $\bar{n}_g=10.9, 2.1$, and 6.3 galaxies 
arcmin$^{-2}$, respectively.
In Table \ref{tab:color}, we list the magnitude limits, the
total number of galaxies ($N$), 
the mean surface number density ($\bar{n}_g$),
and the mean rms error for the galaxy shear estimate
($\overline{\sigma}_g$), for our color samples.
The resulting color boundaries for respective galaxy samples are shown 
in Figure \ref{fig:cc}.
We emphasize that this model-independent empirical method
allows us to clearly distinguish these distinct blue and red populations
(see Figures \ref{fig:cc} and \ref{fig:ccr}),
by reference to the well-calibrated COSMOS photometry,
as well as to separate unlensed foreground/cluster galaxies from the
background (see also \S \ref{subsec:depth} and Figure \ref{fig:nz}).

\subsubsection{Cluster Galaxies}
\label{subsubsec:green}

Following the prescription by \cite{Medezinski+2009}
we construct a CC diagram and
first identify in this space where the cluster lies by virtue of
the concentration of cluster members. 
Figure \ref{fig:ccr} (left panel) shows in CC space the distribution of 
mean projected distances ($\theta$) from the cluster center for all
galaxies in the Cl0024+1654 field. As demonstrated by 
\cite{Medezinski+2009}, this diagram clearly reveals the presence of strong
clustering of galaxies with lower mean radius, confined in a distinct
and relatively well-defined region of CC space.
This small region corresponds to an overdensity of galaxies
in CC space comprising the red sequence of the cluster and
a blue trail of later type cluster members, as clearly demonstrated in
the right panel of Figure \ref{fig:ccr}.
In Figures \ref{fig:cc} (green points)
and \ref{fig:ccr} (left panel)
we mark this overdense region in CC space to safely encompass
the region significantly dominated by the cluster.
We term the above sample,
which embraces all cluster member galaxies, the {\it green} sample,
as distinct from well separated redder and bluer galaxies identified in
this CC space (see the right panel of Figure \ref{fig:ccr}).
Naturally, a certain fraction of background galaxies must be also
expected in this region of CC space, where 
the proportion of these galaxies can be estimated by comparing the
strength of their weak lensing signal with that of the reference
background samples \citep{Medezinski+07,Medezinski+2009}. 
Figure \ref{fig:rgb} (crosses)
demonstrates that the level of the tangential
distortion for the green sample is consistent with zero to the outskirts
of the cluster, indicating that the proportion of background galaxies in
this sample is small compared with cluster members.

\subsubsection{Red Background Galaxies}
\label{subsubsec:red}

To define the foreground and background populations,
we utilize the combination of the strength of the 
weak lensing signal and 
the number density distribution
of galaxies in CC space.
With the $BR_{\rm c}z'$ photometry we can improve upon the simple
color-magnitude selection previously performed in our weak lensing
analyses of $z\sim 0.2$ clusters
\citep{Medezinski+07,UB2008,2009ApJ...694.1643U},
where we had defined a ``red'' population, which comprises
mainly objects lying redward of the E/S0 color-magnitude sequence.
This can be done by properly identifying and selecting in CC space
the reddest population dominated by an obvious overdensity and a red
trail (see the right panel of Figure \ref{fig:ccr}).
For this red sample we define a conservative diagonal
boundary relative to the green sample 
(see Figure \ref{fig:cc}, red points), 
to safely avoid contamination by cluster members (\S
\ref{subsubsec:green})
and also foreground galaxies (see \S~\ref{subsubsec:blue}).
To do this,  
we measure the average distortion strength
as a function of the distance from the cluster sequence in CC space,
and take the limit to where no evidence of dilution of the weak lensing
signal is visible. We also define a color limit 
at $R_{\rm c}-z'\simeq 0.35$ that separates what
appears to be a distinct density maximum of very blue objects.
The boundaries of the red sample as defined above are
marked on Figure \ref{fig:cc} (red points), 
and can be seen to lie well away from
the green cluster sample. For this red
sample we show below a clearly rising weak-lensing signal all the
way to the smallest radius accessible (Figure \ref{fig:rgb}, 
triangles), with no sign of a central turnover which would indicate the 
presence of unlensed cluster members.

\subsubsection{Blue Background and Foreground Galaxies}
\label{subsubsec:blue}

Special care must be taken in defining blue background galaxies,
as objects lying bluer than the E/S0 sequence of a cluster can comprise
blue cluster members, foreground objects, and background blue galaxies. 
This is of particular concern where only one
color (i.e., 2 bands) is available. We have shown in our earlier
work \citep{BTU+05,Medezinski+07}
that this can lead to a dilution of the weak lensing signal
relative to the red background galaxies due to unlensed foreground and
cluster galaxies, the relative proportion of which will depend on the
cluster redshift.
Encouragingly, it has been demonstrated by \cite{Medezinski+2009} that
the foreground unlensed population is well defined in CC space as a
clear overdensity (Figure \ref{fig:cc}, magenta points; Figure
\ref{fig:ccr}, right panel) and we can therefore simply exclude these
objects in this region of CC space from our analysis by setting the
appropriate boundaries relative to this overdensity, found to be where
the weak lensing signal starts showing dilution by these foreground
galaxies. 
The bluer galaxy overdensity in CC space, seen in Figures \ref{fig:cc}
and \ref{fig:ccr}, is also unclustered (Figure \ref{fig:cc}, blue
points) and mainly concentrated in one obvious cloud in CC space.
This blue cloud has a continuously rising weak lensing signal (Figure
\ref{fig:rgb}, circles), towards the center of the cluster, with an
amplitude which is consistent with the red background population defined
and hence we can safely conclude that these objects lie in the
background with negligible cluster or foreground contamination, which
would otherwise drag down the central weak lensing signal. 
The boundaries
of this blue background sample are plotted in Figure \ref{fig:cc} 
(blue points) which we extend to include object lying
outside the main blue cloud but well away from the foreground and
cluster populations defined above.

\subsection{Depth Estimation}
\label{subsec:depth}

An estimate of the background depth is required when converting the
observed lensing signal into physical mass units, 
because the lensing signal depends on the source redshifts 
through the distance ratio $\beta(z_s)=D_{ds}/D_s$.

Since we cannot derive complete samples of reliable photometric
redshifts from our limited three-band ($BR_{\rm c}z'$) images of
Cl0024+1654,
we instead make use of deep field photometry covering
a wider range of passbands, sufficient for photometric redshift
estimation of faint field redshift distributions, appropriate for
samples with the same color-color/magnitude limits as our red
and blue populations.
The 30-band COSMOS photometry 
\citep{Ilbert+2009_COSMOS} is very suited for
our purposes, consisting of deep optical and near-infrared photometry
over a wide field, producing reliable
photometric redshifts for the majority of field galaxies to faint
limiting magnitudes: $m<25$ AB mag in the Subaru $i'$ band. 
The public 30-band COSMOS photometric catalog contains
about 380000 objects 
over 2\,deg$^2$ covering Subaru $BVg'r'i'z'$ photometry. 
The photometric zero-point offsets given in Table 13 of
\cite{Capak+07_COSMOS} were applied to the COSMOS catalog.
From this we select $\sim 3\times 10^5$ galaxies with
reliable photometric redshifts as a COSMOS galaxy sample.  
Since the COSMOS photometry does not cover the Subaru $R_{\rm
c}$ band, we need to estimate $R_{\rm c}$-band magnitudes for this
COSMOS galaxy sample.
We use a new version of the HyperZ template fitting code
\citep[New-HyperZ ver.11, Roser Pell\'{o}, private
communication;][]{hyperz}
to obtain for each galaxy 
the best-fitting spectral template, from which the $R_{\rm c}$ magnitude
is derived with the transmission curve of the Subaru $R_{\rm c}$-band
filter.  
Note, since Subaru $BVg'r'i'z'$ magnitudes are available for all these
galaxies, 
this $R_{\rm c}$ estimation can be regarded as an interpolation. 
Therefore, the $R_{\rm c}$ magnitudes obtained with this method will be
sufficiently accurate for our purpose,
even if photometric redshifts derived by HyperZ 
(which will not be used for our analysis)
suffer from
catastrophic errors.

To assess the effective redshift depth for our blue and red background
populations, we apply for each sample our color-color/magnitude selection 
to the COSMOS multiband photometry, and obtain the redshift distribution
$N(z)$ of the background population with the same color and magnitude
cuts.
The resulting photometric-redshift
distributions $N(z)$ of the CC/magnitude-selected 
red, green, and blue samples in the COSMOS field 
are displayed in Figure \ref{fig:nz}
\citep[see also][]{Medezinski+2009}.
We then calculate moments of the redshift distribution of 
the distance ratio $\beta(z_s)=D_{ds}/D_s$ for each
background population as 
\begin{equation}
\langle \beta^n \rangle =
  \frac{\int\!dz\,N(z)\beta^n(z)}{\int\!dz\,N(z)}.
\end{equation}
The first moment $\langle \beta\rangle$ represents the mean lensing
depth in the weak lensing limit ($\kappa, |\gamma|\ll 1$, $g\approx
\gamma$), where the relationship between the surface mass density and
the lensing observables is linear (see \S \ref{sec:basis}). In this
case, one can safely assume that all background galaxies lie in a single
source plane at redshift $\overline{z}_{s,\beta}$ corresponding to the
mean depth $\langle\beta\rangle$, defined as
$\langle\beta\rangle=\beta(\overline{z}_{s,\beta})$. 
Table \ref{tab:color} lists for respective color samples 
the mean source redshift $\overline{z}_s$, the effective single-plane
redshift $\overline{z}_{s,\beta}$, 
and the first and second moments
of the distance ratio. 
From this, we find the blue sample to be deeper than the
red by a factor of $\langle\beta{({\rm
Blue})}\rangle/\langle\beta{({\rm Red})}\rangle=1.21\pm 0.08$, which is
consistent with the corresponding ratio,
$\langle g_+{({\rm
Blue})}\rangle/\langle g_+{({\rm Red})}\rangle=1.17\pm 0.20$ (see Figure
\ref{fig:rgb}),
of the mean tangential distortion averaged over the full radial extent
of the cluster.

In general,
a wide spread of the redshift distribution of background
galaxies, in conjunction with the single plane approximation, 
may lead to an overestimate of the gravitational shear in the
nonlinear regime
\citep{2000ApJ...532...88H}. To the first order of
$\kappa$, this bias in the observed reduced shear is written as
\citep{2000ApJ...532...88H,1997A&A...318..687S} 
\begin{equation}
\frac{\Delta g}{\tilde{g}}\approx 
\left(\frac{\langle
\beta^2\rangle}{\langle\beta\rangle^2}-1\right)\kappa,
\end{equation}
where $\tilde{g}$ is the reduced shear from the single source plane
assumption, namely, $\tilde{g}\equiv g(\overline{z}_{s,\beta})
=\gamma(\overline{z}_{s,\beta})/[1-\kappa(\overline{z}_{s,\beta})]$, and
$\kappa=\kappa(\overline{z}_{s,\beta})$. 
With the 30-band COSMOS photometry,
the level of bias is estimated as
$\Delta g/\tilde{g}\approx 0.020\kappa$,
$0.016\kappa$, and $0.021\kappa$ for the red, blue, and 
blue+red background samples, respectively. 

By virtue of the great depth of the Subaru imaging and of the moderately
low redshift of the cluster,
for Cl0024+1654,
this effect turns out to be 
quite negligible at all radii in the subcritical regime ($\kappa<1$).
Finally, 
taking into account the photometric zero-point errors 
in our Subaru photometry (\S \ref{subsec:data}) and 
the photometric redshift errors of individual COSMOS galaxies,
we estimate the uncertainty in the mean depth
$\langle\beta\rangle$ 
to be $\sim 4\%$ for the red galaxies,
and $\sim 6\%$ for the blue galaxies;
for the composite
blue+red background sample with a blue-to-red ratio of 
0.58 (see Table \ref{tab:color}), it is about $5\%$.

\section{Cluster Weak Lensing Analysis}
\label{sec:wl}

\subsection{Two-Dimensional Mass Map}
\label{subsec:2dmap}

Weak lensing measurements of the gravitational shear field can be used
to reconstruct the underlying projected mass density field,
$\Sigma_m(\btheta)$. 
In the present study, we use the dilution-free, $BR_{\rm
c}z'$-selected  blue+red background sample (\S\ref{subsec:color}) both
for the 2D mass reconstruction and the lens profile measurements.
We follow the prescription described in \citet[][see \S
4.4]{2009ApJ...694.1643U}
to derive the projected mass distribution of the cluster
from Subaru distortion data.

Figure \ref{fig:gmap} displays the spin-2 reduced-shear field,
$g(\btheta)$,
obtained from the blue+red background sample, 
where for visualization purposes $g(\btheta)$
is smoothed with a Gaussian with $1.41\arcmin$ FWHM.
In the left panel of Figure \ref{fig:knl} 
we show the reconstructed two-dimensional
map of lensing convergence
$\kappa(\btheta)=\Sigma_m(\btheta)/\Sigma_{\rm crit}$
in the central $26\arcmin\times 22\arcmin$ region.
A prominent mass peak is visible in the cluster center,
around which the lensing distortion pattern is clearly tangential
(Figure \ref{fig:gmap}).
This first maximum
in the $\kappa$ map is detected at a significance level 
of $16\sigma$, 
and coincides well with the optical cluster center
within the statistical
uncertainty: 
$\Delta{\rm R.A.}=4.5\arcsec \pm 4.6\arcsec$,
$\Delta{\rm Decl.}=0.0\arcsec \pm 5.6\arcsec$,
where $\Delta{\rm R.A.}$ and $\Delta{\rm Decl.}$ are right-ascension and
declination offsets, respectively, from the center 
of the central bright elliptical galaxy, or
the galaxy 374 in the spectroscopic catalog of \cite{Czoske+2002}.
Our mass peak is also in spatial agreement 
with the X-ray emission peak 
revealed by Chandra ACIS-S observations 
of \citet[][G1 peak]{2004ApJ...601..120O},
${\rm R.A.}\simeq$ 00:26:36.0, 
${\rm Decl.}\simeq$ +17:09:45.9 (J2000.0),
which is offset to the northeast by $\sim 5\arcsec$
from the optical center,
and is close to the galaxy 380 in \citet{Czoske+2002}.

Also compared in Figure \ref{fig:knl} are
member galaxy distributions in Cl0024+1654,
Gaussian smoothed to the same
resolution of ${\rm FWHM}=1.41\arcmin$.
The middle and right panels display the
$R_{\rm c}-$band number and luminosity density maps, respectively,
of green cluster galaxies (see Table \ref{tab:color}). 
Overall, mass and light are similarly distributed in the cluster:
The cluster is fairly centrally concentrated in projection,
and contains an extended substructure located about $3\arcmin$ 
($\sim 670$\,kpc\,$h^{-1}$) northwest
of the cluster center, as previously found by
CFHT/WHT spectroscopic observations of \citet{Czoske+2002}
and 
HST/WFPC2 and STIS observations of \cite{2003ApJ...598..804K}.
This NW clump of galaxies is associated with the primary
cluster  in redshift space (peak $A$ at $\overline{z}=0.395$ in 
\cite{Czoske+2002}, and the density peak of the NW galaxy clump is located at
$\Delta{\rm R.A.} \simeq -2.29\arcmin$,
$\Delta{\rm Decl.} \simeq 2.48\arcmin$
in our galaxy number density map.

\subsection{Cluster Center Position}
\label{subsec:center}

In order to obtain meaningful radial profiles
one must carefully define the center of the cluster.
For this purpose we rely on the improved strong-lens model
of \cite{Zitrin+2009_CL0024}, constructed using deep 
HST/ACS and NIC3 images.
Their mass model accurately reproduces the well-known,
spectroscopically-confirmed 5-image system of a source galaxy at 
$z=1.675$ \citep{Broadhurst+2000_CL0024}, 
and confirms the tentative 2-image system 
identified 
by \citet{Broadhurst+2000_CL0024},
finding an additional third image associated
with this source. 
In addition, 9 new multiple-image systems were identified by their
improved mass model, bringing
the total known for this cluster to 33 multiply-lensed images 
spread fairly evenly over the central region, $\theta \simlt
48\arcsec$. 
The mass model of \cite{Zitrin+2009_CL0024}
reveals a fairly round-shaped
radial critical curve with radius $\sim 10\arcsec$ (at $z=1.675$), 
providing a reasonably well defined center
($\Delta{\rm R.A.}\sim -1.5\arcsec$, $\Delta{\rm Decl.}\sim -2.5\arcsec$),
which is slightly offset, but located fairly close to, the optical
center, and was used as the center of mass in the radial mass-profile
analysis of \cite{Zitrin+2009_CL0024}. In this study, we will define the
center of mass in a more quantitative manner as the peak position of
the smooth dark-matter component of the Zitrin et al. mass model.
The resulting center of mass ({\it dark-matter} center, hereafter)
is at offset position
$\Delta{\rm R.A.}\simeq -2.32\arcsec$,
$\Delta{\rm Decl.}\simeq -1.44\arcsec$, consistent with the geometric
center of the inner critical curve within $1.3\arcsec$.
We note that the mass peak in our Subaru $\kappa$ map is in
spatial agreement with this dark-matter center.

The central mass distribution of Cl0024+1654 has been examined
by other authors using high-resolution HST observations. 
\cite{2003ApJ...598..804K} determined the center of mass 
to be at
$\Delta{\rm R.A.}\simeq -2.3\arcsec$,
$\Delta{\rm Decl.}\simeq -4.2\arcsec$,
from their WFPC2/STIS weak
lensing measurements and strong lensing constraints by the 
5-image system ($z=1.675$), and their center of mass is close to our
dark-matter center ($\Delta\theta\simeq 2.7\arcsec$).
\citet{Jee+2007_CL0024} derived a high-resolution mass map from their
lensing analysis of deep 6-band ACS images,
incorporating as strong-lensing constraints 
the 5-image system and
two additional multiple-system candidates (Objects B1-B2 and
C1-C2, in their notation).\footnote{
Regarding 
the validity of the lensing hypothesis of the 
C1-C2 system, see discussions in \cite{Zitrin+2009_CL0024}.} 
Their mass peak is in good agreement with central bright elliptical
galaxies, and close to the 
galaxy 374 in \cite{Czoske+2002}. 
\citet{Jee+2007_CL0024} chose the geometric center of the {\it
ringlike} dark-matter structure (at $\theta\sim 75\arcsec$), 
revealed by their non-parametric mass reconstruction,
as the cluster center for their radial mass-profile analysis.
Their ring center is located at
$\Delta{\rm R.A.}\simeq 3.3\arcsec$,
$\Delta{\rm Decl.}\simeq -7.6\arcsec$, and is about $6.5\arcsec$ offset
from our dark-matter center.
This offset is about $9\%$ of the dark-matter ring radius ($\theta\sim
75\arcsec$); this level of discrepancy can be reconciled by noting
that the ringlike structure revealed by \cite{Jee+2007_CL0024} is
diffuse and not perfectly round in shape.

\subsection{Lens Distortion Profile}
\label{subsec:gt}

The spin-2 shape distortion of an object 
due to gravitational lensing
is described by
the complex reduced shear, $g=g_1+i g_2$ (see equation [\ref{eq:redshear}]),
 which is coordinate dependent.
For a given reference point on the sky, one can instead 
form coordinate-independent
quantities, 
the tangential distortion $g_+$ and the $45^\circ$ rotated component,
from linear combinations of the distortion coefficients
$g_1$ and $g_2$ as
\begin{equation}
g_+ = -(g_1 \cos 2\phi + g_2\sin 2\phi), \ \ 
g_{\times} = -(g_2 \cos 2\phi - g_1\sin 2\phi),
\end{equation}
where $\phi$ is the position angle of an object with respect to
the reference position, and the uncertainty in the $g_+$ and
$g_{\times}$ 
measurement 
is $\sigma_+ = \sigma_{\times } = \sigma_{g}/\sqrt{2}\equiv \sigma$ 
in terms of the rms error $\sigma_{g}$ for the complex shear
measurement.
Following the ACS strong lensing analysis of \cite{Zitrin+2009_CL0024},
we take their dark-matter center
as the cluster center of mass for our radial profile analysis (see \S
\ref{subsec:center}).  
To improve the statistical significance of the distortion measurement,
we calculate the weighted average of 
$g_+$ and $g_\times$ as
\begin{eqnarray}
\label{eq:gt}
\langle g_+(\theta_m)\rangle &=& \frac{\sum_i u_{g,i}\, g_{+,i}}
{\sum_i u_{g,i}},\\
\langle g_\times(\theta_m)\rangle &=& \frac{\sum_i u_{g,i}\, g_{\times,i}}
{\sum_i u_{g,i}},
\end{eqnarray}
where the index $i$ runs over all of the objects located within 
the $m$th annulus,
and $u_{g,i}$ is the statistical weight (see equation [\ref{eq:ug}])
for the $i$th object,
and $\theta_m$ is the weighted center of the $m$th radial bin,
\begin{equation}
\label{eq:rm}
\theta_m =
\displaystyle\sum_{i\in {\rm bin}\,m} u_{g,i}|\btheta_i|
\Big/ 
\displaystyle\sum_{i\in {\rm bin}\,m} u_{g,i}
\end{equation}
with $\btheta_i$ being the offset vector of the $i$th galaxy position
from the cluster center (\S \ref{subsec:center}). 
In our practical analysis, we use the
continuous limit of equation (\ref{eq:rm}): See Appendix A of
\citet{UB2008}.

In Figure \ref{fig:rgb} we compare
azimuthally-averaged
radial profiles of the tangential distortion
$g_+$ ($E$ mode)
and the $45^\circ$ rotated ($\times$) component $g_{\times}$
($B$ mode)
as measured for our red,  blue,  and green 
galaxy samples (Table \ref{tab:color}).
The error bars represent $68.3\%$ confidence limits, $\sigma_+$,
estimated by bootstrap resampling of the original data set.
The red and blue populations show a very similar form of the radial
distortion profile
which declines smoothly from the cluster center, remaining positive to the
limit of our data, $\theta_{\rm max}=16\arcmin$.
The mean distortion amplitude of the blue population is consistent with,
but slightly higher than, that of the red population,
which is 
related to the greater depth of the blue population relative to the 
red 
(\S \ref{subsec:color} and \ref{subsec:depth}). 
This smooth overall trend
suggests that the NW substructure identified in
Figure \ref{fig:knl} has only a minor effect on the overall profile,
as found in the WFPC2 weak-lensing analysis of \citet{2003ApJ...598..804K}.
On the other hand, the tangential distortion profile for the
green galaxies is consistent with a null signal at all radii, while 
this population is strongly clustered at small radius (Figures
{\ref{fig:ccr} and \ref{fig:knl}), 
indicating 
that the green galaxies mostly consists of cluster member
galaxies.  This convincingly demonstrates the power of our color
selection method.

Now we assess the tangential distortion profile from 
the blue+red background sample
(\S \ref{subsec:color}) 
in order to examine the form of the underlying cluster mass profile and to  
characterize cluster mass properties.   
In the weak lensing limit ($\kappa,|\gamma| \ll 1$), the azimuthally
averaged tangential distortion profile $\langle g_+(\theta)\rangle$
(equation [\ref{eq:gt}]) is related to the projected mass density profile
\citep[e.g.,][]{2001PhR...340..291B} as
\begin{equation} 
\label{eq:gt2kappa}
\langle g_+(\theta)\rangle \approx 
\langle \gamma_+ (\theta)\rangle
=\bar{\kappa}(<\theta)-\langle\kappa(\theta) \rangle,
\end{equation}
where $\langle \cdots \rangle$ denotes the azimuthal average,
and $\bar{\kappa}(<\theta)$ is the mean convergence within a circular
aperture of radius $\theta$ defined as
$\bar{\kappa}(<\theta)=(\pi\theta^2)^{-1}\int_{|\btheta'|\le
\theta}\!d^2\theta'\,\kappa(\btheta')$. Note that the second equality 
in equation (\ref{eq:gt2kappa}) holds for an arbitrary mass distribution.
With the assumption of quasi-circular symmetry in the
projected mass distribution, one can express the tangential distortion
as 
\begin{equation}
\label{eq:gt2kappa_nlin}
\langle g_+(\theta)\rangle \approx 
\frac{\bar{\kappa}(<\theta)-\langle\kappa(\theta)\rangle}
{1-\langle\kappa(\theta)\rangle}
\end{equation}
in the non-linear but sub-critical
(${\rm det}{\cal A}(\btheta)>0$) regime.
Figure \ref{fig:gt} shows 
the tangential and $45^\circ$-rotated distortion profiles
for the blue+red background sample.
Here the presence of $B$ modes can be used to check for systematic
errors. 
The observed $E$-mode signal is significant with
a total detection
signal-to-noise ratio (S/N) of $\simeq 14$,
remaining positive to the
 limit of our data, $\theta_{\rm max}=16\arcmin$, or a projected distance
 of 3.6\,Mpc\,$h^{-1}$.
The $\times$-component 
is consistent with a null signal at all radii, indicating
the reliability of our weak lensing analysis.
The significance level of the $B$-mode detection is
about $2.8\sigma$, which is about a factor of $5$ smaller than $E$-mode. 


To quantify and characterize the mass distribution of Cl0024+1654,
we compare the measured $g_+$ profile with 
the physically and observationally motivated NFW model
\citep[e.g.,][]{BTU+05,UB2008,BUM+08,2009ApJ...694.1643U,2009arXiv0903.1103O}.  
The NFW universal density profile has a two-parameter
functional form as 
\begin{eqnarray}
\label{eq:NFW}
 \rho_{\rm NFW}(r)= \frac{\rho_s}{(r/r_s)(1+r/r_s)^2}, 
\end{eqnarray}
where $\rho_s$ is a characteristic inner density, and $r_s$ is a
characteristic inner radius.
The logarithmic gradient $n\equiv d\ln\rho(r)/d\ln r$
 of the NFW density profile flattens
continuously towards the center of mass, with a flatter central slope
$n=-1$ and a steeper outer slope ($n\to -3$ when $r\to \infty$)
than a purely isothermal structure ($n=-2$).
A useful index, the concentration, compares
the virial radius,
$r_{\rm vir}$, to $r_s$ of the NFW profile, $c_{\rm vir}=r_{\rm
vir}/r_s$. 
We specify the NFW model with the halo virial mass $M_{\rm vir}$ and the
concentration $c_{\rm vir}$ instead of $\rho_s$ and $r_s$.
We employ the radial dependence of the NFW lensing profiles,
$\kappa_{\rm NFW}(\theta)$ and $\gamma_{+,{\rm NFW}}(\theta)$,
given by
\citet{1996A&A...313..697B} and \citet{2000ApJ...534...34W}.\footnote{
Note that the Bartelmann's formula for the NFW lensing profiles are
obtained assuming that the projection integral to extend to
infinity. Alternatively, a truncated NFW profile can be used to model
the lensing profiles\cite{2003MNRAS.340..580T}.
We have confirmed that the best-fit NFW parameters obtained using
the above two different formulae agree to within $0.5\%$ for the case of
Cl0024+1654 lensing; for detailed discussions, see 
\cite{2007arXiv0705.0682B,2007ApJ...654..714H}.
}
The NFW density profile can be further generalized to
describe a profile with an arbitrary power-law shaped central cusp,
$\rho(r)\propto r^{-\alpha}$, and an asymptotic outer slope of $n=-3$
\citep{Zhao1996,Jing+Suto2000},
\begin{equation}
\label{eq:gnfw}
\rho_{\rm gNFW}(r)=\frac{\rho_s}{(r/r_s)^\alpha(1+r/r_s)^{3-\alpha}},
\end{equation} 
which reduces to the NFW model for $\alpha=1$. We refer to the profile
given by equation (\ref{eq:gnfw}) as the generalized NFW (gNFW)
profile. 
It is useful to introduce the radius $r_{-2}$ at which the logarithmic
slope $n$ of the density is isothermal, i.e., $n=-2$. For the gNFW
profile, $r_{-2}=(2-\alpha)r_s$, and thus the corresponding
concentration parameter reduces to $c_{-2}\equiv r_{\rm
vir}/r_{-2}=c_{\rm vir}/(2-\alpha)$.
We specify the gNFW model with the central cusp slope, $\alpha$, 
the halo virial mass, $M_{\rm vir}$, and the concentration,
$c_{-2}=c_{\rm vir}/(2-\alpha)$.

Table \ref{tab:nfw} summarizes the results of fitting with the NFW
(gNFW)
model, listing the lower and upper radial limits of
the data used for fitting,
the best-fit parameters 
$(M_{\rm vir}, c_{\rm vir})$ and their errors (68.3\% CL),
the minimized $\chi^2$ value ($\chi^2_{\rm min}$)
with respect to the degrees
of freedom (dof), and the predicted Einstein radius 
$\theta_{\rm E}$\footnote{For a given source redshift,
the Einstein radius $\theta_{\rm E}$ is defined
as $1=\bar{\kappa}(<\theta_{\rm E}, z_s)$. For an NFW model, this
equation for $\theta_{\rm E}$ can be solved numerically, for example, by
the Newton-Raphson method. See Appendix B of \cite{UB2008}.}
for a background source at $z_s=1.675$, 
corresponding to the 5-image system.
The quoted errors in Table \ref{tab:nfw} include the 
uncertainty in the mean depth $\langle\beta\rangle$ of the background
galaxies (see Table \ref{tab:color}). 
The best-fit NFW model is given by $M_{\rm
vir}=1.14^{+0.22}_{-0.19}\times 10^{15}M_\odot\,h^{-1}$ and $c_{\rm
vir}=10.6^{+2.9}_{-2.0}$,
with $\chi^2_{\rm min}/{\rm dof}=2.2/8$.
This model yields an Einstein radius of $\theta_{\rm
E}=37\arcsec^{+11}_{-10}$ 
at $z_s=1.675$, consistent within the errors with the observed
location of the Einstein radius $\theta_{\rm E}=30\arcsec\pm 3\arcsec$ 
\citep[$z_s=1.675$,][]{Zitrin+2009_CL0024}.  

Assuming a singular isothermal sphere, this Einstein
radius constraint is readily translated into the equivalent 
one-dimensional velocity dispersion of $\sigma_v=1245\pm 62\,$km\,s$^{-1}$.
From a fit of 
the truncated, nonsingular isothermal sphere profile 
to the strong-lensing mass model of 
\citet{Tyson+1998_CL0024},
\citet{Shapiro+Iliev2000_CL0024} obtained an average velocity
dispersion of $\sigma_v\simeq 1200$\,km\,s$^{-1}$
within a sphere of $r=600\,$kpc$\,h^{-1}$, 
in close agreement with the value of $\sigma_v= 1150\pm 100$\,km\,s$^{-1}$
measured by \cite{1999ApJS..122...51D}
from 107 galaxy redshifts.
A similar value of $\sigma_v=1050\pm 75\,$km\,s$^{-1}$
was derived by \citet{Czoske+2001,Czoske+2002}
from 193 galaxy redshifts within $5\arcmin$ from the cluster center.
\citet{Diaferio+2005} used their caustic method 
to study the internal velocity structure of Cl0024+1654. This method
allows one to identify in a radius-redshift phase space diagram 
the cluster boundaries, which serve as a
measure of the local escape velocity
\citep[see][]{Diaferio1999,2008MNRAS.386.1092L}.  
With the spectroscopic galaxy
catalog of \citet{Czoske+2001,Czoske+2002} they found an average
one-dimensional velocity dispersion of $\sigma_v\simeq 937$\,km\,s$^{-1}$
for this cluster, which is lower but consistent with the previous
results.  
A comparison between their caustic and our lensing mass profiles will be
given in \S \ref{subsec:m3d}.

\subsection{Lensing Convergence Profile}
\label{subsec:kappa}

Here we examine the lensing convergence ($\kappa$) profile
using the one-dimensional, non-parametric 
reconstruction method developed by
\citet[][Appendix A]{UB2008} based on the nonlinear extension of
aperture mass densitometry which measures the projected mass interior to
a given radius from distortion data. See also Appendix of
\cite{2009ApJ...694.1643U} for details of this reconstruction method. 

We use a variant of the aperture mass
densitometry, or the so-called $\zeta$-statistic
\citep{1994ApJ...437...56F,2000ApJ...539..540C}, 
of the form:
\begin{eqnarray}
\label{eq:zeta}
\zeta_{\rm c}(\theta)
&\equiv &
2\int_{\theta}^{\theta_{b1}}\!d\ln\theta' 
\langle\gamma_+(\theta')\rangle\nonumber\\
&&+ \frac{2}{1-(\theta_{b1}/\theta_{b2})^2}
\int_{\theta_{b1}}^{\theta_{b2}}\! d\ln\theta'  
\langle\gamma_+(\theta')\rangle
\nonumber\\
&=& 
\bar{\kappa}(<\theta) - \bar{\kappa}_b,
\end{eqnarray}
where 
$\bar{\kappa}(<\theta)$ is the average convergence interior to 
radius $\theta$,
and ($\theta_{b1}, \theta_{b2}$) are the inner and
outer radii, respectively, of the annular background region 
in which the mean background contribution, 
$\bar{\kappa}_b\equiv \bar{\kappa}(\theta_{b1}<\vartheta <\theta_{b2})$,
is defined. 
The substructure
contribution to $\kappa$ is local, 
whereas the inversion
from the observable distortion to $\kappa$ involves a non-local process
(\S \ref{sec:basis}).
Consequently the one-dimensional inversion method requires a boundary
condition specified in terms of the mean background convergence
$\bar{\kappa}_b$.
The inner and outer radii of the annular background region are set to
$\theta_{b1}=14\arcmin$ (3.1\,Mpc\,$h^{-1}$) 
and $\theta_{b2}=16\arcmin$ (3.6\,Mpc\,$h^{-1}$), respectively,
sufficiently larger than the cluster virial radius of massive clusters
($\simlt 2\,$Mpc\,$h^{-1}$), so that the weak-lensing approximation 
$\langle\gamma_+(\theta)\rangle \approx \langle
g_+(\theta)\rangle$ is valid in the background region. 
In the nonlinear but subcritical regime 
(i.e., outside the critical curves),
$\langle\gamma_+(\theta)\rangle$ can be expressed in terms of 
the averaged tangential reduced shear as
$\langle \gamma_+(\theta) \rangle \approx
\langle g_+(\theta)\rangle [1-\langle\kappa(\theta)\rangle]$
assuming a quasi-circular symmetry in the projected mass distribution.
Hence, for a given boundary condition $\bar{\kappa}_b$,
the non-linear equation (\ref{eq:zeta}) 
for $\zeta_{\rm c}(\theta)$ can be solved by an iterative procedure:
\begin{eqnarray}
\label{eq:1drec}
\zeta_{\rm c}^{(k+1)}(\theta)
&\approx&
2\int_{\theta}^{\theta_{b1}}\!d\ln\theta' 
\langle g_+(\theta)\rangle [1-\langle\kappa^{(k)}(\theta)\rangle] \nonumber\\
&+&
 \frac{2}{1-(\theta_{b1}/\theta_{b2})^2}\int_{\theta_{b1}}^{\theta_{b2}}\!
 d\ln\theta'   
\langle g_+(\theta')\rangle,\\
\langle\kappa^{(k)}(\theta)\rangle&=&
\hat{\cal L}_\theta \zeta_{\rm c}^{(k)}(\theta)
+\bar{\kappa}_b
\end{eqnarray}  
where we have introduced a differential operator
defined as $\hat{\cal L}_\theta\equiv
\frac{1}{2\theta^2}\frac{d}{d\ln\theta}\theta^2$ 
that satisfies $\hat{\cal L}_\theta 1=1$,
and $\zeta_{\rm c}^{(k)}$ 
represents the 
aperture densitometry 
 in the $k$th  step of the iteration
$(k=0,1,2,...)$.
This iteration is preformed by
starting with $\langle\kappa^{(0)}(\theta)\rangle=0$ 
for all radial bins, and repeated 
until convergence is reached at all radial bins. 
We compute the bin-to-bin error covariance matrix
$C_{ij}=\langle\delta \kappa(\theta_i) \delta
\kappa(\theta_j)\rangle$ 
for $\langle\kappa_i\rangle\equiv \langle\kappa(\theta_i)\rangle$
with the linear approximation by propagating the rms error
$\sigma_+(\theta)$ for the
averaged tangential shear measurement $\langle g_+(\theta)\rangle$.
Finally, we determine the best value of $\bar{\kappa}_b$ iteratively in
the following way: The iterations start with $\bar{\kappa}_b=0$. At each
iteration, we update the value of $\bar{\kappa}_b$ using the best-fit NFW
model (i.e., assuming $\rho(r)\propto r^{-3}$ at large radii)
for the reconstructed $\kappa$ profile. 
This iteration is repeated until convergence is obtained.

In Figure \ref{fig:kappa} we show
the resulting surface-mass density profile 
reconstructed from the lens-distortion measurements of 
our blue+red background galaxies registered in deep Subaru $BR_{\rm
c}z'$ images.
The error bars are correlated, with neighboring bins having 
$\simlt 10\%$ cross-correlation coefficients at inner radii,
and a $\sim 40\%$ cross-correlation coefficient at the outermost radii. 
The best-fit NFW model to the $\kappa$ profile is obtained as $M_{\rm
vir}=1.08^{+0.21}_{-0.19}\times 10^{15}M_\odot\,h^{-1}$
and 
$c_{\rm vir}=9.7^{+14.6}_{-4.6}$ ($\chi^2_{\rm min}/{\rm dof}=3.6/8$, 
see Table \ref{tab:nfw}),
with the resulting $\bar{\kappa}_b$ value of $2.74\times 10^{-3}$, fully
consistent with the results from the $g_+$ profile (\S
\ref{subsec:gt}), 
ensuring 
the validity of the boundary condition for a shear-based mass reconstruction 
\citep{UB2008}.
We note that simply assuming $\bar{\kappa}_b=0$ 
yields very similar results, 
$M_{\rm vir}=1.06^{+0.20}_{-0.18}\times 10^{15}M_\odot\,h^{-1}$
and
$c_{\rm vir}=10.9^{+19.0}_{-5.3}$ with $\chi^2_{\rm min}/{\rm dof}=3.4/8$,
in agreement well within the $1\sigma$ uncertainty.
Our reconstructed $\kappa$ profile is fairly smooth and
well approximated by a single NFW profile, but exhibits a slight excess
at $\theta\sim 3\arcmin$,
although with large scatters, with respect to the best-fit NFW
profile. This excess feature roughly coincides with the projected
distance of the NW clump identified in Figure \ref{fig:knl}.
We will come back to this in \S \ref{subsec:2dshear}.

From the NFW fit to the $\kappa$ profile 
the statistical uncertainty on 
$M_{\rm vir}$ is about 20\%, comparable
to that from the $g_+$ profile, while the constraint on
$c_{\rm vir}$ is rather weak, allowing a wide range of the concentration
parameter: 
$5\simlt c_{\rm vir}\simlt 25$ ($68.3\%$ CL). 
This reflects the fact that
the reconstruction error tends to increase towards the central region as
a result of inward error propagation.
Consequently, the Einstein radius is poorly constrained from the
one-dimensional mass reconstruction: 
$\theta_{\rm E}=33\arcsec^{+25}_{-20}$ at $z_s=1.675$.
Nevertheless, this one-dimensional, non-parametric inversion method
allows us to derive a $\kappa$ profile with a full covariance matrix
from weak-lensing distortion data alone, which can be readily compared and
combined with inner strong lensing data to provide a full mass profile
for the entire cluster (see \S \ref{sec:wl+sl}).
Furthermore, such non-parametric mass profiles are useful when
comparing the total matter distribution with cluster properties obtained
from other wavelengths
\citep{2008MNRAS.386.1092L,Lemze+2009,Lapi+Cavaliere2009}. 

\subsection{Lensing Depletion Profile}
\label{subsec:magbias}

Lensing magnification, $\mu(\btheta)$, influences the observed
surface density of background sources, 
expanding the area of
sky, and enhancing the observed flux of background sources
\citep{1995ApJ...438...49B,UB2008}.
The former effect reduces the effective observing area in the source plane,
decreasing the number of background sources per solid angle; on the other
hand, the latter effect amplifies the flux of background sources,
increasing the number of sources above the limiting flux. 

For the number counts to measure magnification, 
we use our red background sample based on the
SExtractor photometry (\S \ref{subsec:data}).
For these the intrinsic count slope $s\equiv d\log N(<m)/dm$ 
at faint magnitudes
is relatively flat, $s\sim 0.1$,
so that a net count depletion results \citep{BTU+05,UB2008,BUM+08}.
For depletion analysis, we have a total of 18561
red galaxies down to a limiting magnitude of $z'_{\rm lim}=25.5$  AB mag 
(see \S \ref{subsec:color}), where the sample size 
is about twice as large as that for distortion analysis;
the smaller sample for the distortion analysis is due to the fact that
it requires
the galaxies used are well resolved to make reliable shape
measurements \citep{BTU+05,UB2008}.
%

The number counts for a given magnitude cutoff $m_{\rm cut}$, 
approximated locally as a
power-law cut with slope, $s= d\log_{10}N(<m)/dm$,
are modified in the presence of lensing
as $N(<m_{\rm cut})\approx N_0(<m_{\rm cut})\mu^{2.5s-1}$
\citep{1995ApJ...438...49B},
where $N_0(<m_{\rm cut})$ is the unlensed counts.
Thanks to 
the large field of view of Subaru/Suprime-Cam, the
normalization and slope of the unlensed counts 
for our red galaxy sample are reliably estimated as 
$n_0=20.2\pm 0.4$\,arcmin$^{-2}$
and 
$s=0.12\pm 0.06$, respectively, 
from the outer region of the cluster, 
$11\arcmin \simlt \theta \simlt 16\arcmin$.
The slope is less than the lensing invariant
slope, $s=0.4$, so a net deficit of background galaxies is expected.

The count-in-cell statistic is measured from the flux-limited
red background sample on a regular grid
of $70\times 52$ equal-area cells covering a field of 
$35\arcmin\times 26\arcmin$ \citep{UB2008}. 
We then calculate the radial profile of the red
galaxy counts by azimuthally averaging the count-in-cell
distribution, where each cell is weighted by
the fraction of its area lying within the respective annular bins
\citep{UB2008}
and a tail of $>3\sigma$ cells is excluded in each annulus to remove 
inherent small-scale clustering of the background 
\citep{BUM+08}. 
Here the masking effect due to bright cluster galaxies and bright
foreground objects has been properly taken into
account and corrected for, following the prescription of \citet[][see \S
4.2]{UB2008}.   
We conservatively account for the masking of observed sky
by excluding a generous area $\pi a b$ around each masking object, where
$a$ and $b$ are defined as $\nu_{\rm mask}\equiv 3$ times 
the major ({\tt A\_IMAGE}) and minor axes ({\tt B\_IMAGE}) computed
from SExtractor, corresponding roughly
to the isophotal detection limit \citep[see][]{UB2008}.
We calculate the correction factor for this masking effect
as a function of radius from the cluster center, 
and renormalize the number density of each radius accordingly.
The masking area is estimated as 
about $3\%$ at large radii, and found to increase up to
$\sim 25\%$ of the sky close to the center, $R\simlt 170\,$kpc$\,h^{-1}$. 
Note that if we use the masking factor $\nu_{\rm mask}$ of $2$ or $4$,
instead of 3, 
the results shown below remain almost unchanged \citep[see for
details,][\S 5.5.3]{UB2008}.

Figure \ref{fig:magbias} shows the resulting count-depletion profile
derived from the red background sample based on the SExtractor
photometry. The error bars include not only 
the Poisson contribution but also the variance due to
variations of the counts
along the azimuthal direction, i.e., contributions from 
the intrinsic clustering of red galaxies and
departure from circular symmetry \citep[similar to the second term of
equation (42) of][]{UB2008}.
A strong depletion of the red galaxy
counts is shown in the central, high-density region of the cluster,
and clearly detected out to a few arcminutes from the cluster center.
The statistical significance of the detection of the depletion signal is
about $8\sigma$.
The detection significance of the distortion derived from the blue+red
background sample ($14\sigma$, see \S \ref{subsec:gt})
is better than the counts,
so that here we use our depletion measurements only as a consistency check.
The magnification measurements with ({\it squares}) and without ({\it
triangles}) the masking 
correction are roughly consistent with each other. 
To test the consistency between our distortion and depletion
measurements, 
we calculate the depletion of the counts,
$n(\theta)=n_0 \mu^{2.5s-1}$,
expected for the
best-fitting NFW profile derived from the distortion
measurements (Figure \ref{fig:gt}),
normalized to the observed density $n_0$.
A slight dip at $\theta=2\arcmin-3\arcmin$ in the depletion profile
corresponds to the contribution of the NW clump, which is also seen in
the Subaru distortion data.
This comparison shows 
clear consistency between two independent lensing observables
with different systematics,
which strongly supports the reliability and accuracy of our weak lensing
analysis.
The count depletion of red galaxies is
seen in all our earlier work (A1689, A1703, A370, RXJ1347-11),
and the clear result 
($8\sigma$ detection) 
found here strengthens
the use of this information when testing for consistency with weak
distortions
\cite[e.g.,][]{BTU+05,UB2008,BUM+08}.

\section{Combining Strong and Weak Lensing Data}
\label{sec:wl+sl}

The Subaru data allow the weak lensing profiles to be
accurately measured in several independent radial bins in the
subcritical regime ($\theta > \theta_{\rm E}$).
Here we examine the form of the projected mass density profile for the
entire cluster, 
by combining the Subaru weak-lensing measurements with the inner
strong-lensing information from deep, high-resolution HST/ACS/NIC3
observations \citep{Zitrin+2009_CL0024}.

\subsection{One Dimensional Analysis}
\label{subsec:1d}

\subsubsection{HST/ACS/NIC3 Strong Lensing Constraints}
\label{subsubsec:acs}

As demonstrated in \citet{BTU+05} and \citet{UB2008}
it is crucial to have information on the 
central mass distribution in order to 
derive useful constraints on the degree of concentration in the cluster
mass distribution.

To do this, we constrain the two NFW parameters
from $\chi^2$ fitting to the 
combined weak and strong lensing data: 
\begin{equation}
\label{eq:chi2_combined}
\chi^2=\chi^2_{\rm WL} + \chi^2_{\rm SL},
\end{equation}
where the $\chi^2_{\rm WL}$ for weak lensing is defined by
\begin{equation} 
\label{eq:chi2_wl} 
\chi^2_{\rm WL} = \displaystyle\sum_{i,j}
 \left(
\langle\kappa_i\rangle-\hat{\kappa}_i
 \right)
\left(C^{-1}\right)_{ij} 
 \left(
 \langle\kappa_j\rangle-\hat{\kappa}_j\right),
\end{equation}  
with 
$\hat{\kappa}_i\equiv \hat{\kappa}(\theta_i)$ 
being the NFW model prediction for the
lensing convergence at $\theta_i$
and $(C^{-1})_{ij}$ being the inverse of the bin-to-bin error
covariance matrix for the one-dimensional mass reconstruction. 
The Subaru outer $\kappa$ profile is given in 
10 logarithmically-spaced bins in radius
$\theta=[0.66\arcmin,10\arcmin]$.  
For the strong-lensing data,
we utilize the improved strong-lens model of \cite{Zitrin+2009_CL0024}, well
constrained by 33 multiply-lensed images.
With this model,
we calculate the inner $\kappa$ profile
around the dark-matter
center (\S \ref{subsec:center}) in 16 linearly-spaced radial bins
spanning from $\sim 0.13\arcmin$ to $\sim 0.80\arcmin$,
and the amplitude
is scaled to the mean depth $\langle\beta\rangle \simeq 0.612$
of our blue+red background sample (Table \ref{tab:color}).
For a joint fit, 
we exclude the strong-lensing
data points at radii overlapping with the Subaru data,
yielding 12 independent data points as strong-lensing constraints
for out joint fit.
Finally,
the $\chi^2$ function for the strong-lensing constraints is expressed as
\begin{equation}
\chi^2_{\rm SL} = \displaystyle\sum_{i}
\frac{
(\langle\kappa_i\rangle-\hat\kappa_i)^2
}
{\sigma_i^2}
\end{equation} 
where $\hat{\kappa}_i\equiv \hat{\kappa}(\theta_i)$ 
is the model prediction of the 
NFW halo for the $i$th bin,
and $\sigma_i$ is the $1\sigma$ error for $\langle\kappa_i\rangle\equiv
\langle\kappa(\theta_i)\rangle$;  
the bin width of the inner $\kappa$ profile is sufficiently broad to 
ensure that the errors between different bins are independent.
By combining the full lensing constraints from the ACS/NIC3 and Subaru
observations (22 data points),
we can trace the cluster mass distribution over a 
large range in amplitude $\kappa\sim [10^{-2},1]$ and
in projected radius $R\equiv D_d\theta \simeq [40,2300]$\,kpc$\,h^{-1}$.

In Figure \ref{fig:kprof} we show, for the entire cluster, 
the radial profile of the dimensionless surface mass density
 $\kappa(\theta)=\Sigma_m(\theta)/\Sigma_{\rm crit}$
from our combined strong- and weak-lensing observations, where all of the
radial profiles are scaled to a fiducial redshift of $z_s=1$. 
For comparison purposes,
the inner $\kappa$ profile ({\it triangles}) is shown to the maximum
radius $\sim 0.83\arcmin$ probed by the 33 multiply-lensed images,
allowing for a direct comparison of the strong and weak lensing
measurements in the overlapping region.
This comparison convincingly shows that
our strong and weak lensing measurements are fully consistent in the
overlapping region.

The resulting constraints on the NFW model parameters
and the predicted Einstein radius $\theta_{\rm E}$ (at $z_s=1.675$)
are shown in Table \ref{tab:nfw}.
We show in Figure \ref{fig:CM_kappa} 
the $68.3\%$, $95.4\%$, and $99.7\%$ confidence levels in
the $c_{\rm vir}$-$M_{\rm vir}$ plane,
estimated from $\Delta\chi^2=2.3, 6.17$, and $11.8$, respectively,
for each of the 
Subaru ({\it dashed contours}), 
ACS/NIC3 ({\it solid contours}),
and joint Subaru and ACS/NIC3 ({\it filled gray areas}) $\kappa$ data sets.
Apparently the constraints are strongly degenerate when only the inner or
outer $\kappa$ profile is included in the fits.
The virial mass $M_{\rm vir}$ is well constrained
by the Subaru data alone, while the Subaru constraint on 
$c_{\rm vir}$ is rather weak, allowing a wide range of 
the concentration parameter, $c_{\rm vir}$ (see \S \ref{subsec:kappa}). 
On the other hand, the inner $\kappa$ profile from ACS/NIC3 observations
probes up to $\theta\sim 0.8\arcmin$,
or the projected radius of $R\sim 0.18$\,Mpc$\,h^{-1}$ at the cluster
redshift, 
which however is only about one-tenth of the cluster virial radius inferred from
our full lensing analysis of Subaru and ACS/NIC3 data,
$r_{\rm vir}\simeq 1.8$\,Mpc\,$h^{-1}$,
resulting in a rather weak constraint on $M_{\rm vir}$.
Combining complementary strong- and weak-lensing information
significantly narrows down the statistical uncertainties on the NFW
parameters, placing stringent constraints on the entire mass profile: 
$M_{\rm vir}=1.15^{+0.18}_{-0.15}\times 10^{15}M_\odot\,h^{-1}$
and 
$c_{\rm vir}=9.2^{+1.4}_{-1.2}$ ($\chi^2_{\rm min}/{\rm dof}=4.1/20$). 
All the sets of NFW models considered here
are consistent with each other within the statistical uncertainty, and
properly reproduce the observed location of the Einstein radius (see
Table \ref{tab:nfw}).

Our high-quality lensing data, covering the entire cluster,
allow us to place useful constraints on the gNFW structure parameters,
namely the central cusp slope $\alpha$ as well as the NFW virial mass
and concentration parameters. Using our full lensing constraints, we
obtain  the best-fit gNFW model with the following parameters (see also
Table \ref{tab:nfw}):
$\alpha=0.04^{+0.93}_{-0.04}$,
$M_{\rm vir}=1.06^{+0.19}_{-0.16}\times 10^{15}M_\odot\,h^{-1}$,
and
$c_{-2}=9.8\pm 1.2$. 
The resulting best-fit $\kappa$ profile from our full lensing analysis
is shown in Figure \ref{fig:kprof}
as a thin solid ({\it gray}) curve, along with the best-fit NFW profile
($\alpha=1$).  
Our combined weak and strong lensing data 
of CL0024+1654
favor a shallower inner density slope 
with $\alpha\le 0.97$ (68.3\% CL).
The two-dimensional marginalized constraints ($68.3\%$, $95.4\%$, and
$99.7\%$ CL) on $(M_{\rm vir},\alpha)$ and
$(c_{-2},\alpha)$ are shown in Figure \ref{fig:CM_gNFW}.
We note that the deviation in the inner density profile between the
best-fit gNFW and NFW models becomes significant only below the
innermost radius of our lensing data, corresponding to the size of the
radial critical curve. This results in a rather poor constraint on the
central cusp slope  \citep[cf.][]{Newman+2009_A611}.

\subsubsection{Einstein-Radius Constraint}
\label{subsubsec:rein}

As a model-independent constraint,
we can utilize the observed location of tangential critical curves
traced by giant arcs and multiply-lensed images of background galaxies,
defining an approximate Einstein radius, $\theta_{\rm E}$
\citep{2005ApJ...621...53B,Oguri+Blandford2009,Zitrin+2009_CL0024,Richard+2009_A1703}.  
For an axially-symmetric lens, the Einstein-radius constraint is written
as $\bar{\kappa}(<\theta_{\rm E})=1$, or
$g_+(\theta_{\rm E})=1$ (see equation [\ref{eq:gt2kappa}]),
corresponding to the maximum distortion,
and provides an integrated constraint
on the inner mass profile interior to $\theta_{\rm E}$.
More generally, an
effective Einstein radius can be defined by axially averaging
the projected surface-mass density, which itself is well determined
when a large number of constraints are available
\citep{Broadhurst+Barkana2008,Richard+2009_A1703}.
The Einstein-radius constraint for Cl0024+1654 is shown in Figure
\ref{fig:gt} as the innermost data point (see also Figure
\ref{fig:kappa}).
Here we follow the method described in 
\citet[][\S 5.4.2]{UB2008} for incorporating the inner Einstein-radius
information into lensing constraints: See also \citet{Oguri+2009_Subaru}.

We constrain the NFW model parameters ($c_{\rm vir}, M_{\rm vir}$) 
by combining the Subaru $g_+$ profile and 
the Einstein-radius information. 
We define the $\chi^2$ function for combined 
lens-distortion and Einstein-radius constraints by\footnote{Strictly
speaking, the rms dispersion $\sigma_i$ for the distortion measurement
$g_i=g(\btheta_i)$ is given as 
$\sigma_i \approx \sigma_{g,i}(1-|\hat{g}(\btheta_i)|^2)$ in the
subcritical, nonlinear regime \citep{2000A&A...353...41S}. 
We however neglect this nonlinear correction for the shear dispersion,
and adopt a weighting scheme as described in \S \ref{subsubsec:shape}. }   
\begin{equation}
\label{eq:chi2_g_rein} 
\chi^2 = \displaystyle\sum_i
\frac{\left(\langle g_{+,i}\rangle - \hat{g}_{+,i}
      \right)^2} 
  {\sigma_{+,i}^2} + 
\frac{
\left(1-\hat{g}_{+,E}\right)^2
}  
 {\sigma_{+,{\rm E}}^2},
\end{equation}
where the first term is the $\chi^2$-function for the Subaru
tangential shear measurements 
$\langle g_{+,i}\rangle\equiv \langle g_+(\theta_i)\rangle$
and the second term 
is the $\chi^2$ function for the Einstein radius constraint;
$\hat{g}_{+,i}\equiv\hat{g}_{+}(\theta_i)$ is the NFW model prediction
for the reduced tangential shear at $\theta=\theta_i$ calculated for the
blue+red background sample (see \S  \ref{subsec:color}),
$\hat{g}_{+,E}\equiv\hat{g}_{+}(\theta_{\rm E}, z_{\rm E})$ is the  
model prediction for the reduced tangential shear at $\theta=\theta_{\rm
E}$,
evaluated at the arc redshift, $z_s=z_{\rm E}$.
Following \cite{Zitrin+2009_CL0024}, 
we take $\theta_{\rm E}=30\arcsec$ with an rms error of
$\sigma_{\theta_{\rm E}}=3\arcsec$,
corresponding to the observed 5-image system at $z_{\rm E}=1.675$.
We then propagate this error to $g_+$ as
$\sigma_{+,{\rm
E}} =  \sigma_{\theta_{\rm E}}|\partial g_+/\partial\theta_{\rm
E}|_{\theta=\theta_{\rm E}} \simeq 0.2$
\citep[see][]{UB2008}.
Similarly,
one can combine a $\kappa$ profile reconstructed from
lens distortion measurements with 
inner Einstein-radius information 
\cite[see \S 5.4.2 of][]{UB2008}.

We show in Figure \ref{fig:CM_gt} the resulting constraints
in the  $c_{\rm vir}$-$M_{\rm vir}$ plane
obtained from the Subaru $g_+$ profile
with ({\it filled gray areas}) and without ({\it dashed contours})
the inner Einstein-radius information,
together with the NFW $c_{\rm vir}$-$M_{\rm vir}$ relation for the
observed Einstein-radius constraint ($\theta_{\rm E}=30\arcsec$ at
$z_{\rm E}=1.675$).
The constraints from strong and weak lensing 
are fairly consistent with each other, showing similar degeneracy
directions in the $c_{\rm vir}$-$M_{\rm vir}$ plane.
Our joint fit to the Subaru $g_+$ profile and the inner Einstein-radius 
constraint tightly constrains the NFW model parameters,
$M_{\rm vir}=1.19^{+0.23}_{-0.20}\times 10^{15}M_\odot\,h^{-1}$ and
$c_{\rm vir}=8.6^{+1.8}_{-1.4}$ $(\chi^2_{\rm min}/{\rm dof}=4.9/9)$,
in good agreement with those from 
a joint fit to the ACS/NIC3 and Subaru mass profiles
(\S \ref{subsubsec:acs}).

\subsection{Two-Dimensional Analysis: Two-Component Lens Model}
\label{subsec:2dshear}

Our one-dimensional treatment thus far does not take into account the
effect of the NW clump located at a projected distance of $\theta\simeq
3\arcmin$ (\S \ref{subsec:2dmap}). 
Here we aim to account for this
effect by two-dimensional lens modeling with
a multi-component shear fitting technique
\citep{2003ApJ...598..804K},
in conjunction with the inner strong-lensing constraints on $\kappa(\theta)$.
This method utilizes unbinned distortion measurements for individual
galaxies, and hence does not require any binning of distortion data (cf. \S
\ref{subsec:1d}).

Assuming particular mass profiles for lens components,  
the model reduced-shear field
$\hat{g}(\btheta)=\hat{g}_1(\btheta)+i\hat{g}_2(\btheta)$ is simply
calculated as 
\begin{equation}
 \hat{g}(\btheta)=
\frac{\sum_{l=1}^{N_l} \hat\gamma_l(\btheta)}
{1-\sum_{l=1}^{N_l}
\hat\kappa_l(\btheta)},
\end{equation}
where $N_l$ is the number of lens halo components.
\citet{2003ApJ...598..804K} analyzed 
a sparse-sampled mosaic of 2-band (F450W, F814W) WFPC2 observations in
the Cl0024+1654 field,
and found
that two lens-components are necessary to match
their WFPC2 lens-distortion data (i.e., $N_l=2$), 
responsible for the central and NW clumps in projection space.

Following \citet{2003ApJ...598..804K}, 
we use a circularly-symmetric NFW profile to 
describe the projected lensing fields of the central
component, which has been shown to be a good approximation from our
one-dimensional full lensing analysis (\S
\ref{subsec:1d}).  
Further, we assume that the central component is
responsible for the central strong-lensing constraints on $\kappa(\theta)$
derived from the ACS/NIC3 observations.
For the NW component, we use a truncated form of 
the NFW profile \citep[tNFW, see][]{2003MNRAS.340..580T},
which approximates a structure of a stripped halo by a sharp truncation
at the halo virial radius.
The location of the central component is fixed at the dark-matter center
of mass (\S \ref{subsec:center}), around which the inner $\kappa$
profile (at $0.17\arcmin \simlt \theta \simlt 0.80\arcmin$ with 16 bins) 
is defined (\S \ref{subsubsec:acs}).
Further, the location of the NW component is fixed at 
the observed density peak position 
of the NW clump of green cluster galaxies (\S \ref{subsec:2dmap}).

We constrain two sets of NFW model parameters ($M_{\rm vir},c_{\rm vir}$)
for the central and NW lens components by minimizing the total
$\chi^2$ function for our 
combined two-dimensional weak-lensing distortion and central
strong-lensing constraints.
The total  $\chi^2$ function 
is given as
equation (\ref{eq:chi2_combined}), but with the following $\chi^2$
function for Subaru weak lensing:
\begin{equation}
\chi^2_{\rm WL, 2D}=
\displaystyle\sum_{i=1}^{N_g}
\displaystyle\sum_{\alpha=1,2}  
\frac{
\left[
g_{\alpha}(\btheta_i)-\hat{g}_{\alpha}(\btheta_i)
\right]^2
}
{\sigma_{g,\alpha}^2(\btheta_i)},
\end{equation}
where the index $i$ runs over all objects in our blue+red sample,
but excluding those at radii overlapping with the inner strong-lensing
data ($\theta \simlt 0.80\arcmin$), and $\sigma_{g,\alpha}(\btheta_i)$
($\alpha=1,2$) is the rms error for the real/imaginary component of the
$i$th reduced-shear measurement $g_{\alpha}(\btheta_i)$, which we take as
$\sigma_{g,1}(\btheta_i)=\sigma_{g,2}(\btheta_i)=
\sigma_{g}(\btheta_i)/\sqrt{2}$. 
We have 11647 such usable objects in the blue+red sample, i.e., a
total of $2\times 11647$ independent measurements for the spin-2
distortion field. Thus, we have a total of 23310 joint constraints from
strong and weak lensing, and 23306 dof.

Table \ref{tab:2dshear} lists the resulting best-fitting parameters 
of our two-component lens mass model. 
The two-component lens mass model provides an acceptable fit 
with the minimized $\chi^2$ value of
$\chi^2_{\rm min}/{\rm dof}=1.577$,
and with the best-fit NFW parameters for the central component,
$M_{\rm vir}=(1.11\pm 0.18)\times 10^{15}M_\odot \, h^{-1}$
and
$c_{\rm vir}=8.1\pm 1.2$,
fairly consistent with those from the
corresponding one-dimensional analysis (\S \ref{subsec:1d}).
For the NW halo component,
we find a best-fit set of NFW parameters,
$M_{\rm vir}=(1.28\pm 0.51)\times 10^{14}M_\odot \, h^{-1}$ and
$c_{\rm vir}=5.0\pm 3.5$, 
with the virial radius (and hence the truncation radius), 
$r_{\rm vir}= (0.83\pm 0.11)\,$Mpc$\,h^{-1}$, 
corresponding to the angular radius of  
$\theta_{\rm vir}=3.7\arcmin\pm 0.5\arcmin$.
It is found that the best-fitting NFW parameters 
obtained with and without the truncation at the virial radius
agree to within $\sim 2$\% for the NW clump.

Our results can be directly compared with those from 
\citet{2003ApJ...598..804K}, who obtained
$M_{200}=1.7^{+0.9}_{-0.8}\times 10^{14}M_\odot\,h^{-1}$
and 
$c_{200}=4^{+2}_{-1}$
for the NW clump, where the quantities here are
determined at $r_{200}$ corresponding
to a mean interior overdensity of 200, relative to the critical
density $\rho_{\rm crit}(z_d)$ of the universe at the cluster redshift
$z_d=0.395$. These parameters can be translated into the corresponding 
virial parameters,
$M_{\rm vir}=2.0^{+0.9}_{-1.0}\times 10^{14}M_\odot\,h^{-1}$
and
$c_{\rm vir}=5^{+2}_{-1}$, consistent with our results within the
statistical uncertainty.
For the central component, on the other hand,
\citet{2003ApJ...598..804K} obtained an NFW model with very high
concentration,
$M_{200}=4.0^{+0.8}_{-0.7} \times 10^{14}M_\odot\,h^{-1}$
and
$c_{200}=22^{+9}_{-5}$,
or in terms of the virial parameters,
$M_{\rm vir}=4.3^{+0.9}_{-0.8} \times 10^{14}M_\odot\,h^{-1}$
and
$c_{\rm vir}=26^{+11}_{-6}$.
This high concentration
may be due to the inclusion of the Einstein radius constraint in their fit
to their outer weak lensing data which would otherwise be underestimated
by weak lensing data alone \citep[][\S 5.5.2]{UB2008}. 
It could also be partly explained by the irregular, non-axisymmetric
mass distribution in the cluster core
\citep{2006ApJ...642...39C,2007MNRAS.379..190C}. 
Simply adding the two virial masses of \citet{2003ApJ...598..804K}
yields $M_{\rm tot}=6.3^{+1.3}_{-1.2}\times 10^{14}M_\odot\,h^{-1}$,
representing a large discrepancy with respect to our results.

The radial mass profile of Cl0024+1654 has also been examined by
\cite{Hoekstra2007} using ground-based weak-lensing shape
measurements from CFHT/CFH12K $R$-band data.
By fitting an NFW profile to the tangential distortion signal at
$R=[0.25,1.5]\,$Mpc$\,h^{-1}$,
\citet{Hoekstra2007} obtained a virial mass of $M_{\rm
vir}=1.76^{+0.62}_{-0.53}\times 10^{15}M_\odot \,h^{-1}$ for this
cluster, which is higher than, but consistent with, our results.

\subsection{Mass-to-Light Ratio}
\label{subsec:ML}

Having obtained the radial mass density profile,
we now turn to examine the cluster mass-to-light ratio ($M/L$)
in a model independent approach.
For this we utilize the {\it weak lensing dilution} method developed by 
\cite{Medezinski+07} \citep[see also][]{Medezinski+2009}, and derive the
cluster luminosity density profile $\Sigma_l(\theta)$
to large radius, with the advantage
that no subtraction of far-field background counts
is required. 
We weight each ``green'' galaxy flux $F_i$ by its tangential distortion,
$g_{+,i}$, and subtract this ``$g_+$-weighted'' luminosity contribution
of each galaxy, which when averaged over the distribution will have zero
contribution from the unlensed cluster members.
This will account for any difference in the
brightness distributions of the cluster members to that of the
background galaxies, in particular the skewness of the cluster
sequence to brighter magnitudes. The total flux of the cluster in the
$n$th radial bin is then given as
\begin{equation}
\label{eq:flux_dilution}
F_{\rm tot}(\theta_n)=\displaystyle\sum_{i\in {\rm bin}\,n}{F_i}
-
\frac{ \langle \beta^{({\rm B})}\rangle/\langle\beta^{({\rm G})}\rangle }
{\langle g_{+,n}^{({\rm B})}\rangle}
\displaystyle\sum_{i\in{\rm bin}\,n}{g_{+,i}^{({\rm G})}\,F_i},
\end{equation}
where $\langle g_{+,n}^{({\rm B})}\rangle
\equiv\langle g_+^{({\rm B})}(\theta_n)\rangle$ is the true background
level of the tangential distortion, averaged over the $n$th radial bin,
and $\langle\beta^{({\rm G})}\rangle$ and $\langle\beta^{({\rm B})}\rangle$ 
are the source-averaged distance ratios (see equation [\ref{eq:dratio}] 
and Table \ref{tab:color}) for
green and reference background samples, 
\footnote{In general, {\it background} samples may contain foreground
field galaxies with $z_s<z_d$ and $\beta(z_s)=0$.}
respectively (see Appendices A and B of \cite{Medezinski+07} for a 
derivation of this equation). 
Here we take the blue+red sample as our reference background. 
The flux is then translated to luminosity.
 First we calculate the absolute magnitude,
\begin{equation}
M=m-5\log{d_L(z)}-K(z)+5,
\end{equation}
where $d_L(z)$ is the luminosity distance to the cluster, and
the $K$-correction, $K(z)$,
is evaluated for each radial bin according to
its colour. 
We use the $R_{\rm c}$-band data to calculate the cluster
light profile.

The results of the $K$-corrected cluster luminosity density measurements 
are shown in Figure \ref{fig:lprof}, along with our lensing constraints
on the surface mass density
profile, $\Sigma_m(\theta)$ ({\it shaded gray area}), 
converted into a luminosity density profile
assuming a constant $M/L_R$ of $230h (M/L_R)_\odot$,
corresponding to the mean cluster mass-to-light ratio
interior to $\theta=3\arcmin$.
There are two sets of $\Sigma_l$ profiles shown in 
Figure \ref{fig:lprof}, namely with ({\it
squares}) and without ({\it circles}) the contribution from the
NW galaxy clump located at a projected distance of $\theta\sim
3\arcmin$ (\S \ref{subsec:2dmap}). 
The observed light profile, including the NW clump,
closely resembles the mass profile 
derived from our joint weak and 
strong lensing analysis, and shows a shoulder feature at 
$\theta\sim 3\arcmin$. This feature disappears in the NW-clump corrected
light profile, and hence is caused by the excess luminosity due to the
NW galaxy clump. The total luminosity of the NW clump is estimated as 
$L(<1\arcmin)=(3.84\pm 0.20)\times 10^{11}L_\odot\,h^{-2}$
and
$L(<1.5\arcmin)=(5.47\pm 0.45)\times 10^{11}L_\odot\,h^{-2}$
for apertures of radius $1\arcmin$ and $1.5\arcmin$, respectively. 
Assuming the same mean 
mass-to-ratio of $\langle M/L_R\rangle=230h (M/L_R)_\odot$ 
for the NW galaxy clump, we have projected mass estimates of
$M_{\rm 2D}(<1\arcmin)=(0.86\pm 0.05)\times 10^{14}M_\odot\,h^{-1}$
and
$M_{\rm 2D}(<1.5\arcmin)=(1.23\pm 0.10)\times 10^{14}M_\odot\,h^{-1}$;
these values 
are consistent, within the errors,
with the predictions of our NFW model constrained by the
combined weak and strong lensing data.

We show in Figure \ref{fig:ml} our model-independent radial
profile of the differential mass-to-light ratio,
$\delta M(\theta)/\delta L_R(\theta)=\Sigma_m(\theta)/\Sigma_l(\theta)$,
obtained by dividing the non-parametric $\Sigma_m$ profile 
from our Subaru distortion data by the $R_{\rm c}$-band
$\Sigma_l$ profile.
Our direct, model-independent approach yields larger errors
than those from conventional parametric procedures assuming a particular
form (e.g., an NFW profile) for the underlying mass density profile. 
The resulting $\delta M/\delta L_R$ profile is noisy and
consistent with a constant
mass-to-light ratio, but shows a local peak around $R\sim 1\,$Mpc$\,h^{-1}$
with $M/L_R=400^{+300}_{-180} h (M/L_R)_\odot$, and
then falls off to larger radius. 
A similar behavior has been convincingly shown by a recent lensing-based
analysis \citep{Medezinski+2009}
of several high-mass clusters (A1689, A1703, A370, RXJ1347-11)
with $M_{\rm vir}\simgt 10^{15}M_\odot \,h^{-1}$
using deep multi-color Subaru images. 
\footnote{\cite{Medezinski+2009} used their best-fit
NFW $\Sigma_m$ profile to derive cluster mass-to-light ratio
profiles.}  
For these relaxed, high-mass clusters,
it has been shown that the $\Sigma_l$ profiles all decline smoothly as
$\sim \theta^{-1}$ in projection, whereas the $\Sigma_m$ profiles are well
described by the continuously varying NFW profile, so that
the $\delta M/\delta L$
profile peaks around 20\% of the virial radius in the range
$(300-500)M_\odot/L_\odot$, and then steadily falls to a mean field value
of $\sim 100 M_\odot/L_\odot$, consistent with careful dynamical work by
\cite{Rines+2000}.  
In contrast, the $M/L_R$ peak for
Cl0024+1654 is found to be at $\sim 0.6r_{\rm vir}$, much larger than
those found for other massive, relaxed clusters with centrally
peaked $M/L$ profiles. 
This different radial trend of Cl0024+1654
may reflect a different
dynamical state from relaxed systems.

\section{Mass Profile and Comparison}
\label{sec:comparison}

\subsection{Comparisons with Other Lensing Studies}
\label{subsec:comparison}

First we compare our results with those from previous lensing studies
of Cl0024+1654
to assess the consistency over different regimes of gravitational
lensing (weak and strong regimes)
and to highlight any potential problems or discrepancies arising from
systematic effects, such as weak lensing dilution due to 
contamination by unlensed foreground and cluster member galaxies,
shear calibration biases, and spurious boundary effects.

\subsubsection{Tangential Distortion and Lensing Convergence Profiles}
\label{subsubsec:gtcomp}

In Figure \ref{fig:gtcomp} we compare our Subaru 
tangential-distortion
profile  ({\it squares})
with the results from 
different authors and different observations/methods:
WFPC2 weak-lensing analysis of \cite{2003ApJ...598..804K},
ACS weak-lensing analysis of \cite{Jee+2007_CL0024},
and
ACS/NIC3 strong-lensing analysis of \cite{Zitrin+2009_CL0024}.
For this comparison, we have converted the azimuthally-averaged 
convergence profile $\kappa(\theta)$
of \cite{Zitrin+2009_CL0024}
into a distortion profile ({\it shaded region}) for a fiducial source
at $z_s=1.3$, roughly matching 
the mean depth of our Subaru blue+red background sample
(Table \ref{tab:color}).

Our Subaru and ACS/NIC3 results 
are in good agreement in the overlapping region ($0.6\arcmin
\simlt \theta \simlt 1\arcmin$).
Furthermore, a simple extrapolation of 
the best-fitting NFW profile for the inner ACS/NIC3 observations
({\it solid}; see \S \ref{subsubsec:acs}) fits well with
the outer Subaru distortion information 
over a wide range of radius, but
 somewhat overpredicts the distortion profile at $3\arcmin \simlt
 \theta \simlt 5\arcmin$, as our Subaru data prefer a slightly steeper
 profile.
The triangles show the $g_+$ profile as obtained
from the ACS weak-lensing analysis of \cite{Jee+2007_CL0024},
where the data are limited to the positive-parity region ($\theta\simgt
 40\arcsec$) in this comparison. Their overall $g_+$ profile is steeper
 than our combined ACS/NIC3 and Subaru results, but is roughly
 consistent out to 
$\theta\sim 1.5\arcmin$, beyond which it drops
more rapidly than our Subaru profile.
The open circles show the $g_+$ profile at $0.9\arcmin \simlt \theta
 \simlt 11\arcmin$ from the weak-lensing analysis by 
\cite{2003ApJ...598..804K}. 
The mean depth of their background sample is 
$\overline{z}_s=1.15\pm 0.3$ from photometric redshift
 measurements in the Hubble Deep Field-South
 \citep{2003ApJ...598..804K}.
The flattened slope of Kneib et al. profile at $\theta\simlt 2\arcmin$
 is likely due to contamination of the weak lensing signal by unlensed
 cluster member galaxies. Indeed, \cite{2003ApJ...598..804K}
 made an empirical correction for this dilution effect in their
 two-dimensional lens  modeling.  Nevertheless, taking into account the
 difference in 
 redshift depths, our Subaru and Kneib et al. results are in
 agreement at $\theta\simgt 3\arcmin$, where the weak-lensing
 approximation is valid  ($|g|\simlt 0.1$) and
 the dilution effect due to contamination by 
 unlensed cluster galaxies is less significant.
 At the sensitivity and resolution of ground-based weak-lensing
 measurements, 
 our Subaru distortion profile shows no evidence for the {\it dip} at
 $\theta\simeq 1.25\arcmin$, corresponding to the ringlike structure
 revealed by the ACS lensing analysis of \cite{Jee+2007_CL0024}.

Similarly, we compare in Figure \ref{fig:kcomp}
the radial profiles of lensing convergence
$\kappa(\theta)=\Sigma_m(\theta)/\Sigma_{\rm crit}$, 
or the dimensionless surface mass density, derived from the
lensing studies shown in Figure \ref{fig:gtcomp}.
Here all of the $\kappa$ profiles are scaled to the same fiducial
source redshift of $z_s=1$ for direct comparison. As we have already
shown in Figure \ref{fig:kprof}, our weak and strong
lensing results are fully consistent with each other in the overlapping
region, and the combined $\kappa$ profile of the full lensing
constraints, for the entire cluster,
is well fitted with a single NFW profile (see \S
\ref{subsubsec:acs}).  At larger radii of $\theta\simgt 2\arcmin$,
our Subaru-derived $\kappa$ profile agrees well with 
the two-component NFW lens model of \cite{2003ApJ...598..804K}, 
constrained from their weak-lensing
measurements combined with the inner strong-lensing information on the
Einstein radius,
whereas their $\kappa$ values at small radii,
$0.3\arcmin \simlt \theta \simlt 2\arcmin$,
 are systematically lower than our Subaru results.
On the other hand, the $\kappa$ profile reconstructed from the ACS
lensing analysis of \cite{Jee+2007_CL0024} is in agreement with our
Subaru and ACS/NIC3 results at radii $0.3\arcmin \simlt \theta\simlt
0.8\arcmin$, beyond which, however,
the surface-mass density of \cite{Jee+2007_CL0024}
stays almost constant at $\kappa\sim 0.5$ out to the maximum radius
of the ACS observations ($R_{\rm max}\sim 370\,{\rm kpc}\,h^{-1}$
at $z=0.395$), 
and largely disagrees with both our and Kneib et al.'s results.
The overall shallow profile obtained by \cite{Jee+2007_CL0024} could be
due to the mass-sheet degeneracy (\S \ref{sec:basis}), as demonstrated
by \cite{Liesenborgs+2008_CL0024}.

\subsubsection{Cumulative Projected Mass Profiles}
\label{subsubsec:m2dcomp}

Now we derive a non-parametric projected mass profile
$M_{\rm 2D}(<\theta) =
\pi(D_d\theta)^2\Sigma_{\rm crit}\bar{\kappa}(<\theta)$
of Cl0024+1654 from our combined $\kappa$ profile of the 
ACS/NIC3 and Subaru lensing observations 
described in \S \ref{subsubsec:acs}; 
there are a total of 12 and 10 data points 
from the ACS/NIC3 and Subaru observations, respectively,
over the radial range $R\simeq [40,2300]\,{\rm kpc}\,h^{-1}$.
We interpolate $\kappa(\theta)$ between 
measured points with a cubic spline function.
We use Monte-Carlo techniques 
to estimate the confidence limits on $M_{\rm 2D}(<\theta)$, 
by properly propagating the measurement errors encoded in the
joint 
covariance matrix $C_{ij}=\langle \delta\kappa_i\delta\kappa_j\rangle$,
containing both diagonal and off-diagonal elements,
constructed from the combined ACS/NIC3 and Subaru $\kappa$ data.
To do this,
we Cholesky-decompose this covariance matrix as
$\bC=\bL\bL^{T}$, where $\bL$ is a lower triangular matrix, and assign
a random noise fluctuation $\delta\kappa(\theta_i)$ in each radial bin
by $\delta\kappa_i=L_{ij}\xi_j$ with $\xi_j$ being drawn from the Gaussian
distribution with zero mean and unit variance
\cite[e.g., see][]{Park+2003}.  
We then generate 1000 random realizations of the $\kappa$ profile, 
and perform an identical analysis
on the Monte-Carlo data sets.

Figure \ref{fig:m2d} shows the resulting 
projected mass profile $M_{\rm 2D}$
of Cl0024+1654, along with the results from 
\cite{2003ApJ...598..804K} and \cite{Jee+2007_CL0024}.
The resulting 68.3\% confidence limits 
from our combined ACS/NIC3 and Subaru observations 
are shown as the gray-shaded area, in good agreement with the results
from 
\citet[][Figure 12]{Jee+2007_CL0024}
at radii up to $\theta
\simlt 1\arcmin$ ($R\simlt 220\,{\rm kpc}\,h^{-1}$). 
Beyond this radius, however, their mass profile
shows a much steeper increase  out to the boundaries
of the ACS observations (see also Figure \ref{fig:kcomp}), 
owing to the flat behavior of their $\kappa$ profile at $\theta\simgt
0.8\arcmin$ (i.e., $M_{\rm 2D}\propto \theta^2$), 
and thus disagrees with our mass profile.
Beyond $\theta_{\rm E}\simeq 0.5\arcmin$ for the 5-image system
($z_s=1.675$), 
the $M_{\rm 2D}$ profile from the best-fit two-component NFW model of  
\cite{2003ApJ...598..804K}
is systematically lower than our and Jee et al.'s mass profiles.
Finally, all of these results are in close agreement at radii around the
Einstein radius $\theta_{\rm E}\simeq 30\arcsec$ at $z_s=1.675$,
tightly constrained by the 5-image system.
From our non-parametric $M_{\rm 2D}$ profile, we find a projected
mass of
$M_{\rm 2D}(<30\arcsec)=(1.32\pm 0.12) \times 10^{14}\,h^{-1}$,
in excellent agreement with $M_{\rm 2D}(<30\arcsec)=(1.27\pm0.09)\times 
10^{14}M_\odot \,h^{-1}$ obtained by \cite{Jee+2007_CL0024}.

\subsection{Mass Discrepancy}
\label{subsec:m3d}

A comparison of cluster mass estimates from X-ray, dynamical, and
strong/weak lensing observations is of particular importance to 
assess the validity of fundamental assumptions made in various methods 
of mass determination.
As compared with gravitational lensing,
which is purely geometrical and sensitive to 
the projected mass along the line of sight,
X-ray and dynamical methods infer the three-dimensional
structure of clusters under the 
assumptions of symmetry and dynamical equilibrium.
In particular, the X-ray method can be affected strongly during
mergers where the assumption of
hydrostatic equilibrium is no longer valid \citep{Okabe+Umetsu2008}.
The redshift-space caustic method of \cite{Diaferio1999}, 
in contrast to the traditional Jeans approach,
does not rely on the assumption of dynamical equilibrium, and can
be used to estimate the cluster mass even in non-equilibrium regions.

A detailed dynamical analysis by
\cite{Czoske+2001,Czoske+2002} revealed a fairly complex,
bimodal velocity structure in Cl0024+1654, with two distinct redshift
components peaked at $z=0.395$ (component $A$) and $z=0.381$ (component
$B$). The former is the primary cluster, and the latter
is a foreground, loose aggregate of galaxies physically separated from 
the primary component.
They showed that the redshift distribution of cluster members in the
central region is strongly skewed towards negative velocities.
\citet{Czoske+2002} suggested that
the diffuse foreground component could be the result of a high-speed, head-on
encounter of two similar-mass clusters with a merger axis very nearly
parallel to the line of sight. 
In this context, the relative velocity of the two components
implied by the redshift difference is $v\sim
3000\,$km\,s$^{-1}$.  
With $N$-body simulations of two colliding CDM halos
with a mass ratio of about 2:1,
they also demonstrated that such a scenario can 
explain the observed peculiar redshift distribution in Cl0024+1654.
In their simulations they adopted 
$M_{A,{\rm vir}}=9.5\times 10^{14}M_\odot \,h^{-1}$
and
$M_{B,{\rm vir}}=5.0\times 10^{14}M_\odot \,h^{-1}$
as initial virial masses for the components $A$ and $B$, respectively.
It is interesting to note that the sum of the two virial masses,
$M_{\rm vir}=1.45\times
10^{15}M_\odot\,h^{-1}$, is close to our $1\sigma$ upper limit on 
the total virial mass $M_{\rm
vir}\simeq 1.4\times 10^{15}M_\odot\,h^{-1}$,
and is consistent with the total projected mass estimate
$M_{\rm 2D}(<r_{\rm vir})=1.36^{+0.27}_{-0.33}$
for the entire cluster system, including any currently unbound material
beyond the virial radius.

The collision scenario of \citet{Czoske+2002}
has been also supported by the presence of a 
ringlike structure in the
projected dark matter distribution revealed by the 
ACS lensing analysis of 
\cite{Jee+2007_CL0024}.
They speculated from the dissipation of the shock in the
Chandra and XMM-Newton X-ray 
observations and the size of their observed ringlike structure 
($R\sim 280\,$kpc$\,h^{-1}$) in their mass map
that the two clusters in Cl0024+1654 would have undergone the first
core passage about $1-2$\,Gyr ago,
and that the dense gas cores of the two clusters would have
survived the collision with a distinct separation.
If the cluster is not considered to represent a single relaxed system, 
but is a superposition of two separated clusters, 
then the single-component assumption should lead to a substantial
underestimation of the X-ray and dynamical mass,
whereas the lensing mass should be the sum of the two cluster components.

To highlight this problem, we now compare our results with
integrated three-dimensional mass profiles $M_{\rm 3D}(<r)$ derived
from X-ray and dynamical, as well as lensing, observations
with the {\it single cluster assumption}.
Here we deproject the two-dimensional mass profile
and obtain a non-parametric $M_{\rm 3D}$ profile
simply assuming spherical symmetry, following the method introduced by 
\cite{Broadhurst+Barkana2008}.
This method is based on the fact that
the surface-mass density $\Sigma_m(R)$ is related to the
three-dimensional mass density $\rho(r)$ by an Abel integral transform;
or equivalently, one finds that
the three-dimensional mass $M_{\rm 3D}(<r)$
out to spherical radius $r$ is written in terms of $\Sigma_m(R)$ as
\begin{equation}
\label{eq:m3d}
M_{\rm 3D}(<r) = 2\pi\int_0^r\! dRR\Sigma_m(R)
-4\int_r^\infty\!dRR f
\left(
\frac{R}{r}
\right)
\Sigma_m(R),
\end{equation}
where 
$f(x)=(x^2-1)^{-1/2}-\tan^{-1}(x^2-1)^{-1/2}$
\citep{Broadhurst+Barkana2008}.
\footnote{
This integral transformation has 
an integrable singularity at the lower limit of the
second integrand ($R=r$), which can be readily avoided by a suitable
coordinate transformation. 
}
To perform the integrals in equation
(\ref{eq:m3d}), we assume smoothness of 
$\Sigma_m(R)$ and use the interpolation method described in \S
\ref{subsubsec:m2dcomp}.
We then extrapolate inward assuming $\Sigma_m(R)$ = const. 
from the innermost point and outward assuming $\Sigma_m(R)\propto
R^{-2}$  from the outermost point, 
where these exponents are motivated by the NFW profile,
which provides an excellent description of our data (\S
\ref{subsubsec:acs}). 
We find that even varying these exponents by $\pm
1$ would change the mass profile at all relevant radii
by only $\simlt 0.3\%$, which is
negligible compared to the effect of the measurement errors.
Finally we propagate errors on $\Sigma_m(R)$ using the same
Monte-Carlo method as in \S \ref{subsubsec:m2dcomp}.
This deprojection method allows us to derive in a non-parametric way
three-dimensional virial quantities ($r_{\rm vir}, M_{\rm vir}$)
and values of $M_{\Delta}=M_{\rm 3D}(<r_\Delta)$ 
within a sphere of a fixed mean interior overdensity $\Delta$
with 
respect to the critical density of the universe at the cluster
redshift. Table \ref{tab:mdelta} gives a summary of the
cluster mass estimates $M_{\rm 3D}(<r_\Delta)$
from our non-parametric deprojection analysis.

In Figure \ref{fig:m3d}, the resulting deprojected mass profile $M_{\rm
3D}(<r)$  and  68.3\% confidence level region
are shown as the solid line and gray-shaded area, respectively,
and compared with results from other studies.
Also indicated in Figure \ref{fig:m3d}
is the location of the virial radius 
$r_{\rm vir}=1.75^{+0.08}_{-0.11}$\, Mpc$\,h^{-1}$, 
derived from the non-parametric
deprojected mass profile, corresponding to the virial mass of 
$M_{\rm vir}=1.21^{+0.18}_{-0.22}\times 10^{15}M_{\odot}\,h^{-1}$,
which is in excellent agreement with those from our best-fit NFW models, 
$M_{\rm vir}=(0.9-1.3)\times 10^{15}M_\odot\,h^{-1}$ ($1\sigma$).
Our non-parametric estimate of $M_{500}$ is
$M_{500}=7.2^{+1.7}_{-2.0}\times 10^{14}M_{\odot}\,h^{-1}$
within the spherical radius
$r_{500}=0.93^{+0.07}_{-0.10}\,$Mpc$\,h^{-1}$,
in good agreement with the model-independent mass estimate 
$M_{500}=(7.4\pm 2.4)\times 10^{14}M_\odot\,h^{-1}$ by \citet{Hoekstra2007}.

Clearly, the comparisons in Figure \ref{fig:m3d}
exhibit a large discrepancy between our lensing and
the X-ray/dynamical results based on the single cluster assumption,
indicating a substantial contribution from additional mass components.
The dashed curve shown is
the spherical NFW mass profile from the best-fit lens model of 
\cite{2003ApJ...598..804K},
where only the central halo component has been included.
The X-ray derived mass profile ({\it dotted curve})  
of \cite{2004ApJ...601..120O}
is taken from Figure 1 of 
\cite{Diaferio+2005}.
Overall,
the caustic mass estimates ({\it triangles}) of
\cite{Diaferio+2005} lie between the results of 
\cite{2003ApJ...598..804K} and \cite{2004ApJ...601..120O}.

Our $M_{\rm 3D}$ profile is only marginally consistent with the caustic
mass estimates of 
\cite{Diaferio+2005} at $r\simlt 0.7\,$Mpc$\,h^{-1}$, beyond which
the mass discrepancy increases with increasing radius.
The outermost radius probed by \cite{Diaferio+2005} is close to our
estimated virial radius (Table \ref{tab:mdelta}), at which
the lensing to caustic mass ratio is about $3.0\pm 1.5$.
This mass discrepancy could be largely reconciled within the collision
scenario of \cite{Czoske+2002}, in which Cl0024+1654 is the
result of a high-speed collision of two massive clusters along the line
of sight. 
Simply assuming that the total virial mass of the whole system is
conserved before and after the collision,
with a fiducial mass ratio of 2:1
\citep{Czoske+2002,Jee+2007_CL0024,Zuhone+2009_CL0024_X-ray,Zuhone+2009_CL0024},   
the total mass of the primacy cluster ($A$), excluding
the NW substructure, is then $M_{A,{\rm vir}}=(7.3\pm 1.5)\times
10^{14}M_\odot\,h^{-1}$, marginally consistent with the caustic mass
estimate within the uncertainty. 
Further, such a collision may  disrupt the dark matter from the
cluster cores, and eject a substantial amount of mass into cluster
outskirts, while the velocity dispersion at large radii can be
significantly reduced \citep[by $10\%-15\%$,][]{Czoske+2002}; a detailed
discussion will be presented in \S \ref{subsec:collision}.
This may lead to a systematic mass discrepancy between lensing and
caustic methods at cluster outskirts.

Under the hypothesis of hydrostatic equilibrium,
\cite{2004ApJ...601..120O} derived a spherical hydrostatic mass of 
$M_{200}=3.2^{+1.0}_{-0.6}\times 10^{14}M_\odot\,h^{-1}$
within a sphere of $r_{200}\simeq 1.0\,$Mpc$\,h^{-1}$
($\theta_{200}=4.3\arcmin$). 
At this radius, our derpojected mass
estimate is $M_{\rm 3D}=7.4^{+1.1}_{-1.4}\times
10^{14}M_\odot\,h^{-1}$. 
This comparison yields a lensing to X-ray mass ratio of
$2.3\pm 0.7$ at $r\sim 1$\,Mpc\,$h^{-1}$.
Interestingly,
the Chandra and XMM-Newton observations of this cluster
revealed that the X-ray surface brightness distribution is better fitted
by a superposition of two ICG models rather than one
\citep{2004ApJ...601..120O,Zhang+2005_CL0024}.
\citet{Jee+2007_CL0024} demonstrated that assuming a superposition of
two isothermal models for the X-ray brightness distribution
can indeed
 bring the Chandra data into closer agreement with the lensing mass
estimates, $M_{\rm 2D}^{X}(<30\arcsec)\sim 1.1\times
10^{14}M_\odot\,h^{-1}$ (see \S \ref{subsubsec:m2dcomp}).

\section{Discussion}
\label{sec:discussion}

\subsection{Post-Collision Scenario}
\label{subsec:collision}

Our full lensing analysis of joint Subaru and ACS/NIC3 observations 
has shown that the radial structure of the projected mass in 
Cl0024+1654 is consistent
with a continuously steepening density profile, 
with only a minor contribution from the NW substructure,
and well described by the general NFW profile,
as previously found for 
a number of relaxed, massive clusters
\citep{BTU+05,UB2008,2009ApJ...694.1643U,2009arXiv0903.1103O},
expected for collisionless CDM halos.  
Mass and light in the cluster are similarly distributed 
and exhibit a prominent peak at the cluster center in projection space. 
All these observed features 
might lead one to conclude that Cl0024+1654 is a fairly relaxed, massive
cluster with a total virial mass of $M_{\rm vir}=(1.2\pm 0.2)\times
10^{15}M_\odot\,h^{-1}$,  but with noticeable substructure of a
fractional mass $(11\pm 4)\%$, 
located at a projected radius $R\simeq 700\,$kpc$\,h^{-1}$.

Several independent lines of 
evidence, however, suggest that the cluster Cl0024+1654
is the result of a high-velocity, near head-on collision
between two similar-mass clusters 
occurring along the line of sight,
viewed approximately $2-3$\,Gyr after impact:\\


\noindent
(i) Dynamical data\\

\noindent
The first piece of evidence is the existence of a bimodal redshift
distribution in the system \citep{Czoske+2002}.
In addition, the central velocity distribution of cluster members 
is strongly skewed negative with a long tail towards the foreground,
secondary component, suggesting strong recent gravitational interaction
\citep{Czoske+2002}.\\

\noindent
(ii) X-ray emission features\\

\noindent
Next, anomalously low levels of X-ray emission and temperature
\citep{2004ApJ...601..120O,Zhang+2005_CL0024},
relative to the standard X-ray observable and mass relations,
may also point to interaction, and an incomplete merger.
In a period of $1-3$\,Gyr ($\simgt \tau_{sc}$) after the first passage
of a close encounter between two massive clusters,
the shock-heated gas associated with this whole
system could be in a much expanded phase with correspondingly lower
X-ray luminosity and temperature \citep{2001ApJ...561..621R}.
In this scenario, the dense gas cores of the two clusters
are thought to have survived the collision with a distinct separation, 
and have not settled down to a single system 
\citep{Jee+2007_CL0024}. 
This is supported by the high-resolution Chandra/XMM-Newton X-ray data
which favor a superposition of two gas components over a
single one
\citep{2004ApJ...601..120O,Jee+2007_CL0024}.
Recently,
such X-ray brightness features observed in the cluster have been remarkably
reproduced by a high-resolution simulation of a high-velocity 
head-on collision of 2:1 mass-ratio clusters 
\citep{Zuhone+2009_CL0024_X-ray}, motivated by the scenario of
\cite{Czoske+2002}. 
In particular, 
it has been shown that a hydrodynamic mass estimate 
in a post-collision state, assuming one cluster component,
underestimates the actual mass by a factor of $\sim 2-3$
\citep{Zuhone+2009_CL0024_X-ray}, where the X-ray bias is found to be
due mainly to the projection of denser, colder gas along the line of sight of
hotter gas in the main cluster.\\

\noindent
(iii) Mass discrepancy between lensing and caustic methods\\

\noindent
Similarly,
a mass discrepancy between lensing and caustic methods found in
cluster outskirts (\S \ref{subsec:m3d}) may indicate the existence of an
extended, diffuse mass component with lowered velocity dispersions
due to evaporation of faster moving dark-matter particles and galaxies
\citep{2001ApJ...561..621R,Czoske+2002}.\\

%

The collision scenario of \cite{Czoske+2002} appears to be
consistent, 
at least qualitatively, with all these observations, 
and provides a feasible explanation for the long-standing puzzle on the
mass discrepancies and anomalies found in this cluster (\S
\ref{subsec:m3d}). 
Nevertheless, a detailed and quantitative understanding of the observed
phenomena 
and the physical conditions of the hot gas and dark matter has yet to be
obtained. 
Recent $N$-body/hydrodynamic simulations of
\cite{Zuhone+2009_CL0024_X-ray}  
have been successful at
reproducing qualitatively 
the observed features of X-ray brightness and temperature structure in 
Cl0024+1654. Their simulated X-ray observations show that 
the observed X-ray spectral temperatures can be lower than the actual
temperature of the hotter gas by a factor of $\sim 2$ due to the
superposition of denser, colder gas along the line of sight.
However, the resulting X-ray temperature from their mock observations is
only $T\sim 2.6-2.8\,$keV, which is  $\sim 60\%$ of the observed
spectral temperature of $T_X=4.47^{+0.83}_{-0.54}$ by
\cite{2004ApJ...601..120O}. 
Without resorting to further ad hoc explanation, this quantitative
disagreement, in the context of the collision scenario of
\cite{Czoske+2002}, indicates that the initial conditions for their
simulation 
are different from the pre-merger conditions in the real
cluster \citep{Zuhone+2009_CL0024_X-ray}. 
In particular, 
the sum of the initial masses of two clusters
adopted by \cite{Zuhone+2009_CL0024_X-ray} is 
$M_{200}=9\times 10^{14}M_\odot$ ($h=0.5,\Omega_m=1$ in their simulations)
with a mass ratio of
$M_{A,200}/M_{B,200}=2$, 
and is $\sim 60\%$ of our lensing
mass estimate (Table \ref{tab:mdelta}).
 
Gravitational lensing, on the other hand, can provide
essential information to constrain the total mass of the whole
system in the pre-merger state. 
Our comprehensive lensing analysis provides a model-independent
constraint of 
$M_{\rm 2D}(<r_{\rm vir})=(1.36^{+0.27}_{-0.33})\times 10^{15}M_\odot
\,h^{-1}$ 
for the projected mass of the whole  
system, including not only the primary and secondary cluster components,
but also any currently unbound material beyond the virial radius.
If the collision is occurring along our line of sight, then the
total projected mass associated with this system 
is essentially conserved before
and after the collision. Thus, a constraint on 
$M_{\rm 2D}(<r_{\rm vir})$
can be regarded as a crude upper limit to 
the sum of the two pre-merger cluster masses
when designing 
simulations to explore this system.

The contrast between the largely monolithic lensing based mass profile
and the bimodal post-merger interpretation of the dynamical and X-ray
data may be largely reconciled when considering the radial
dependence of the relaxation timescale. If we take the density at a radius
of $r=1$\,Mpc where the mass density we estimate is
$\rho \sim 4\times 10^{-27}$\,g\,cm$^{-3}$,
corresponds to a dynamical time of $t_{dyn}\sim 1$\,Gyr,
shorter than or comparable to the 
estimated time since the merger occurred --- which must be larger than 
$\sim 1-2$\,Gyr
so that the hot gas shock is no longer present and less than about 3\,Gry
so that the gas properties are still noticeably anomalous:
This means that the central region, interior to 1\,Mpc,
can be expected to be well relaxed at this stage of the merger with a
symmetric inner potential, 
which is born out on inspection of numerical simulations,
in particular \cite{2001ApJ...561..621R}.
The outskirts, however, will not be
relaxed for example at 2\,Mpc 
where the mass density is $\rho \sim 7\times
10^{-28}$g\,cm$^{-3}$,  
corresponding to
a dynamical time of $t_{dyn}\sim 4$\,Gyr,
so that we can expect here to see large scale
velocity  bimodality remaining. 
However, the sum of two
roughly aligned, unrelaxed diffuse outer halos 
of the two merging clusters will not be easily distinguished by lensing
in projection.

Recent results of a weak lensing analysis of
\cite{Jee+2007_CL0024} reveal a ringlike structure in the central
projected mass distribution.  
They interpreted
this is the result of a high-speed collision between two dark-matter
halos along the line of sight.
They demonstrated that such ring features could be reproduced
in a simulation of a high-speed collision of two pure dark-matter halos.
More recently, however, detailed $N$-body simulations
show that such features require significant fine-tuning of initial
conditions, such that the initial particle velocity distribution is
purely circular \citep{Zuhone+2009_CL0024}, which however is unlikely in
the collisionless CDM model. 
Our joint weak and strong lensing analysis 
shows no evidence for the presence of ringlike features at the
sensitivity and resolution of ground-based weak lensing observations.
Except for the presence of ringlike features,
our joint distortion profile is in overall agreement
with the results from a weak lensing analysis of \cite{Jee+2007_CL0024}. 
On the other hand, our self-consistent joint $\kappa$ profile, with a
continuously declining radial trend, is in disagreement with the results of
\cite{Jee+2007_CL0024}. 
These discrepancies between our results and those reported by
\cite{Jee+2007_CL0024} may be explained by a combination of the
mass-sheet degeneracy (\S \ref{subsubsec:gtcomp}) and the monopole 
degeneracy, as suggested by
\cite{Liesenborgs+2008_ring,Liesenborgs+2008_CL0024}.

\subsection{Implications for the $Y_X-M_{\rm tot}$ Relation}
 
Recent cosmological hydrodynamic/$N$-body simulations
\citep{Kravtsov+2006_Yx} suggest that a simple product of the X-ray
spectroscopic temperature and the X-ray derived hot-gas mass,
namely, $Y_X= T_{500}\times M_{{\rm gas},500}$,
is a robust X-ray indicator of the total cluster mass $M_{\rm tot}$,
being insensitive to the cluster dynamical state.
The primary reason for the tight $Y_X-M_{\rm tot}$ relation is that
this X-ray mass proxy is directly related to the
total thermal energy of the ICG, and thus to the depth of the cluster
potential well. As a result, $Y_X$
is not strongly disturbed even by cluster mergers,
unlike  other X-ray mass proxies, $T_X$ or X-ray luminosity
\citep{2001ApJ...561..621R}.  
In addition, the stability of $Y_X$ can be explained by that
fractional deviations of the average gas temperature $T_{500}$ and 
X-ray derived $M_{{\rm gas},500}$
tend to be anticorrelated \citep[see Figure 5 of][]{Kravtsov+2006_Yx},
and hence partially cancelled out in the product $Y_X$. 

Here we use the Chandra X-ray data presented by
\cite{2004ApJ...601..120O} to derive the value of $Y_X=M_{{\rm
gas},500}T_{500}$ for Cl0024+1654, assuming a single cluster component
that represents the primary cluster $A$.
The outer radius for integration of
the Chandra X-ray spectrum is only $r_{\rm out}\simeq
0.5r_{500}$, where $r_{500}$ is based on the virial mass $M_{A,{\rm vir}}$
of the primary cluster component, $M_{A,{\rm vir}}\sim 7.3\times
10^{14}M_\odot \,h^{-1}$, obtained assuming a fiducial 2:1 mass-ratio
collision scenario (\S \ref{subsec:m3d}).
We estimate the value of $T_{500}$ from the measured value $T_X=T(\simlt
0.5r_{500})$ by 
using an empirical $T_{500}/T_X-T_X$ relation found by
Vikhlinin et al. (2009, see equation [5] and Figure 6).
The resulting $Y_X$ from the Chandra X-ray observations is
$Y_X({\rm Chandra}) = (3.8\pm 2.3)\times 10^{13} M_\odot \,{\rm
 keV}\,h^{1/2}$. 
This Chandra measurement of $Y_X$ is compared with that expected from the 
self-similar scaling relation $Y_X=Y_X(M_{500},z)$
calibrated by detailed Chandra observations
 of a sample of nearby, relaxed clusters
\citep{Vikhlinin+2009_CCC2}. 
Using this $Y_X-M_{500}$ relation, 
we obtain the value of $Y_X$, 
predicted for the primary cluster component $A$, as
$Y_X(M_{A,{500}}) = (1.9\pm 0.8)\times 10^{14} M_\odot \,{\rm
 keV}\,h^{1/2}$. 
Hence, the X-ray $Y_X$ measurement only accounts for 
$\simlt 35\%$ ($1\sigma$) of the self-similar prediction based on the
lensing mass estimate: 
$Y_X({\rm Chandra})/Y_X(M_{A,500})=0.20\pm 0.15$.
This discrepancy could be partially reconciled by the superposition
effects of X-ray emission found by \cite{Zuhone+2009_CL0024_X-ray}.

Another plausible mechanism to account for this apparent discrepancy is
an adiabatic cooling in an adiabatic expansion process.
If the ICG is undergoing an adiabatic process,
whose timescale is of the order of
$\tau_{sc}$ and much shorter than the cooling timescale over the
cluster radius (e.g., $r_{500}$),
then the adiabatic condition, $Tn^{1-\gamma_{ad}}={\rm const.}$
with $\gamma_{ad}$ being the adiabatic index of the ICG,
relates the  
fluctuations in gas temperature and density as
$\delta T/T\sim (2/3)\delta n/n$ with $\gamma_{ad}=5/3$ (for
non-relativistic, ideal gas).
In X-ray observations, 
we measure the gas mass $M_{\rm gas}$ at a certain fixed
radius (for example, provided by lensing), $r_{500}$,
so that $\delta n/n \sim \delta M_{{\rm gas},500}/M_{{\rm gas},500}$.
Since the adiabatic
fluctuations $\delta T$ and $\delta n$ are positively correlated, the
deviation in their product, $Y_X$, is enhanced in an adiabatic phase:
\begin{equation}
\label{eq:Ybias}
\frac{\delta Y_X}{Y_X} =  \frac{\delta M_{{\rm gas},500}}{M_{{\rm
 gas},500}}+\frac{\delta T_{500}}{T_{500}} \sim 
 \frac{5}{2}\frac{\delta T_{500}}{T_{500}}.
\end{equation}
During the post-shock adiabatic expansion phase ($t\simgt \tau_{sc}$),
the fractional decrease in the average gas temperature could reach
$-\delta T/T \sim  20\%$ at maximum for an head-on collision of two
similar-mass clusters 
\citep{2001ApJ...561..621R}, suggesting that the maximum fractional
decrease in the $Y_X$ is of the order of $-\delta Y_X/Y_X\sim  50\%$,
provided that the cluster is caught in such an adiabatic expansion
phase.

\subsection{Numerical Simulations}

We evaluate this merger scenario based on numerical simulations of
cluster collisions using FLASH3, an Eurelian three-dimensional 
hydrodynamics/$N$-body simulation code 
\citep{Fryxell+2000_FLASH}.
FLASH uses the Piecewise-Parabolic Method (PPM)
of \cite{Colella+Woodward1984_PPM}
to solve the equations of hydrodynamics,
and a particle-mesh method to solve for the gravitational forces between 
particles in the $N$-body module. 
The gravitational potential is calculated using a multigrid solver
\citep{Ricker2008}.
FLASH uses adaptive mesh refinement (AMR)
with a tree-based data structure allowing recursive grid refinements
on a cell-by-cell basis. With the AMR technique,
our simulations achieve a maximum resolution of 12.7\,kpc.

As pre-collision conditions, we assume two equilibrium
NFW halos 
with a mass ratio of 2:1 following 
\cite{Czoske+2002} and \cite{Jee+2007_CL0024}.
The sum of the virial masses of the two clusters is set to 
$M_{\rm vir}= 1.1 \times 10^{15} M_{\odot} \, h^{-1}
\simeq 1.57\times 10^{15}M_\odot$
based on the results of our full lensing analysis,
so that the initial virial masses for the primary ($A$) and secondary
($B$) clusters are
$M_{A,{\rm vir}}=1.05\times 10^{15}M_\odot$
and 
$M_{B,{\rm vir}}=0.52\times 10^{15}M_\odot$, 
with virial radii of
$r_{A,{\rm vir}}=2.1\,$Mpc
and 
$r_{B,{\rm vir}}=1.7\,$Mpc,
respectively.
The concentration parameters for the two clusters are set to 
$c_{A,{\rm vir}}=4.2$ and $c_{B,{\rm vir}}=4.8$, roughly matching the
median $c_{\rm vir}(M_{\rm vir},z)$ relation for relaxed CDM halos in
the WMAP five-year cosmology \citep{Duffy+2008}.
The clusters are initialized so that their centers are separated by
the sum of their virial radii
with an initial relative velocity of $v=3000$\,km\,s$^{-1}$
\citep{Czoske+2002}. 
We fix the gas mass fraction within $r_{\rm vir}$
to $f_{\rm gas}=0.14$, close to the values obtained from 
recent X-ray and multi-wavelength observations of massive clusters
\citep{Vikhlinin+2009_CCC2,2009ApJ...694.1643U}.
For simplicity,
we ignore the expansion of the universe during the collision. 
For the gas component of each cluster,
we assume a $\beta$ model in hydrostatic equilibrium
with the gravitational potential of the cluster,
with core radius
$r_{c}=0.08 r_{\rm vir}$ and $\beta=1$,
motivated by results of
our analysis of galaxy clusters drawn from high-resolution
cosmological simulations (S. Molnar et al. 2010, submitted to ApJ).
For the primary and secondary cluster,
the maximum values of the three-dimensional temperature
$T_{3\rm D}$
are found to be $8.6$\,keV and $5.8$\,keV, respectively,
in the radial range $0.15r_{500}-r_{500}$.
We use the local Maxwellian approximation for the local velocity
distribution, and determine the dispersion from the Jeans equation
assuming isotropic velocity distribution as a function of radius. 
The gas is simply assumed to have no bulk velocity relative to their 
respective NFW halos; the details of the simulations will be presented 
elsewhere (S. Molnar et al., in preparation).  
In the present study, we consider a head-on collision with zero impact
parameter as done by \cite{Zuhone+2009_CL0024_X-ray}.

We find that during the first core passage,
which occurs at about $t=1$\,Gyr after the beginning of the collision,    
the infalling subcluster gas core pushes the main cluster gas core
out of its equilibrium position. After the first core passage, the hot
gas of the main cluster core, mixed with part of the gas stripped off
the subcluster, falls back into the dark matter potential well of the
main component. 
The infalling material expands adiabatically as it fills up the
available space at the center of the main cluster, resulting in  much
lower gas density and temperature than the original ones.
The subcluster core follows the shock and falls towards the center of
the dark matter potential well of the subcluster. About $1$\,Gyr after
the first core passage ($t=2$\,Gyr) the subcluster core is clearly
separated from the main cluster core, as found by
\cite{Zuhone+2009_CL0024_X-ray}. 

In the present study, we shall focus on global physical properties of
the cluster system, such as X-ray and Sunyaev Zel'dovich effect (SZE)
observables, to be measured during the collision. We derive and compare
values of the X-ray luminosity $L_X$ and the integrated Comptonization
parameter $Y\propto \int_V\!dV\,nT$, or the low-frequency integrated
SZE flux, at times before and after the collision.

We choose the direction of projection such that the centers of the  
two post-collision clusters are aligned along the projection axis, to be
consistent with the high-resolution X-ray observations  
\citep{2004ApJ...601..120O,Zhang+2005_CL0024}.
For the X-ray luminosity we assume
$L_X \propto \epsilon_X\propto n^2 \sqrt{T}$, with $\epsilon_X$ being
the X-ray emissivity. 
The quantities $L_X$ and $Y$ are measured within a fixed aperture radius 
of $r_{A,500}$. 
We find that about $t=2\,(3)$\,Gyr after the collision,
the X-ray luminosity $L_X$ and the integrated Comptonization parameter
$Y$ within $r_{500}$ are reduced to about $11\%\,(19\%)$ and
$33\%\,(47\%)$ of their respective initial values, revealing substantial
reductions in projected X-ray and SZE observables. 

In Figure \ref{fig:tx} we show 
the projected X-ray spectroscopic-like temperature profiles 
\citep{Mazzotta+2004}
for our simulated head-on cluster merger
at three different times, namely 
$t=0, 2$, and $3$\,Gyr.
Here the spectroscopic-like measure 
$T_{sl}= \int_V\!dV\,W_{sl}T/\int_V\!dV\,W_{sl}$
with the weight $W_{sl}=n^2 T^{-3/4}$
provides a good match to the temperature derived
from X-ray spectroscopic data \citep{Mazzotta+2004}.
We find that the projected X-ray temperatures $T_X$ at later times,
$t=2-3$\,Gyr after the beginning of the collision, are dramatically
decreased by factors 
of about 2, showing fairly flat radial profiles and thus resembling
isothermal profiles as found by Chandra and XMM-Newton observations
\citep{2004ApJ...601..120O,Zhang+2005_CL0024}. 
We note that the projected X-ray 
temperature in post-shock states is biased low due to the presence of
low-temperature components in an adiabatic expansion process.
The amplitudes of these post-shock $T_X$ profiles are
in good agreement with the observed average X-ray temperature
$T_X=4.47^{+0.83}_{-0.54}$\,keV ($90\%$ CL) from the Chandra
observations of \cite{2004ApJ...601..120O}, as compared to 
the values $T_X=2.6-2.8$\,keV obtained by
\cite{Zuhone+2009_CL0024_X-ray}.  
Since both the simulations presented
here and in \cite{Zuhone+2009_CL0024_X-ray} adopted the same mass ratio,
impact parameter, and relative velocity as initial conditions, it is
likely that the different post-shock temperatures at $t=2-3$\,Gyr
are due to the different total masses of the two cluster components
adopted. 
We plan to explore in more details the parameter space of our
simulation, using different mass ratios, impact parameters, etc.,
in order to constrain the parameter space of possible solutions.

\section{Summary}
\label{sec:summary}

In this paper, we have presented a joint weak and strong lensing
analysis of the rich, but 
X-ray faint, cluster of galaxies Cl0024+1654 at $z=0.395$ 
based on wide-field Subaru $BR_{\rm c}z'$ imaging combined with 
detailed strong lensing information obtained from deep HST/ACS/NIC3
observations \citep{Zitrin+2009_CL0024}.  

The deep Subaru three-band photometry, in conjunction with our weak
lensing dilution techniques \citep{Medezinski+07,UB2008,Medezinski+2009},
allows for a secure selection of
distant blue and red background populations free from contamination of
unlensed galaxies, providing a greater lensing depth than achievable
in the standard  color-magnitude selection method.

Our non-parametric mass reconstruction from a full lensing analysis of
joint Subaru and ACS/NIC3 observations
reveals a continuously steepening density profile
over a wide radial range from 40 to 2300\,kpc$\,h^{-1}$
with only a minor contribution, $\delta M/M\sim 10\%$ in the mass, from 
known substructure \citep{Czoske+2002,2003ApJ...598..804K}
at a projected distance of $R\simeq 700\,$kpc$\,h^{-1}$.
The cluster light profile closely resembles the mass profile,
and our model-independent $M/L_{R}$
profile shows an overall flat behavior with a mean of $\langle
M/L_R\rangle\simeq 230h (M/L_R)_\odot$, 
in contrast to centrally peaked $M/L$ profiles found for other massive,
relaxed clusters,
and exhibits a mild declining
trend with increasing radius at cluster outskirts,
$r\simgt 0.6r_{\rm vir}$.
We found that the projected  mass distribution for the entire cluster 
can be well fitted with a single NFW profile with virial mass, 
$M_{\rm vir}= (1.2\pm 0.2)\times 10^{15}M_{\odot}\, h^{-1}$,  
in agreement with that obtained from a model independent approach
(\S \ref{subsec:m3d}), and with those from recent lensing observations 
\citep{Hoekstra2007,Zitrin+2009_CL0024},
but in apparent
disagreement with X-ray hydrostatic mass estimates
\citep{2004ApJ...601..120O,Zhang+2005_CL0024}, and with caustic mass 
estimates at cluster outskirts \citep{Diaferio+2005}.

Careful examination and interpretation of X-ray and dynamical data
\citep{Czoske+2002,2004ApJ...601..120O,Jee+2007_CL0024},   
based on detailed high-resolution cluster collision simulations
\citep{Czoske+2002,Zuhone+2009_CL0024,Zuhone+2009_CL0024_X-ray}, 
strongly suggest
that this cluster system is in a post collision state, 
which we have shown is consistent with our well-defined mass profile
for a major merger of two similar-mass clusters
occurring along the line of sight, viewed
approximately $2-3$\,Gyr after impact \citep{Zuhone+2009_CL0024_X-ray}
when the gravitational potential 
has had time to relax in the center,
before the gas has recovered and before the outskirts are fully
virialized. 
Finally, our full lensing analysis provides a model-independent
constraint of $M_{\rm 2D}(<r_{\rm vir})=(1.4\pm 0.3)\times
10^{15}M_\odot \,h^{-1}$ for the projected mass of the whole 
system, including any currently unbound material beyond the virial
radius,
which can constrain the sum of the two pre-merger cluster masses when
designing 
simulations to explore this system.


\acknowledgments
We thank the anonymous referee for a careful reading of the manuscript
and and for providing useful comments.
We are very grateful for discussions with Doron Lemze and Masamune
Oguri, whose comments were very helpful in improving the manuscript.
We thank Nick Kaiser for making the IMCAT package publicly available.
We acknowledge the use of computing resources provided by
the Theoretical Institute for Advanced Research in Astrophysics (TIARA)
in the Academia Sinica Institute of Astronomy \& Astrophysics.
The work is partially supported by the National Science Council of Taiwan
under the grant NSC97-2112-M-001-020-MY3.
The software used in this work was in part developed by
the DOE-supported ASC/Alliance Center for Astrophysical
Thermonuclear Flashes at the University of Chicago.



\begin{thebibliography}{106}
\expandafter\ifx\csname natexlab\endcsname\relax\def\natexlab#1{#1}\fi

\bibitem[{{Baltz} {et~al.}(2007){Baltz}, {Marshall}, \&
  {Oguri}}]{2007arXiv0705.0682B}
{Baltz}, E.~A., {Marshall}, P., \& {Oguri}, M. 2007, ArXiv e-prints, 705,
  0705.0682

\bibitem[{{Bartelmann}(1996)}]{1996A&A...313..697B}
{Bartelmann}, M. 1996, \aap, 313, 697, arXiv:astro-ph/9602053

\bibitem[{{Bartelmann} \& {Schneider}(2001)}]{2001PhR...340..291B}
{Bartelmann}, M., \& {Schneider}, P. 2001, \physrep, 340, 291,
  arXiv:astro-ph/9912508

\bibitem[{{Ben{\'{\i}}tez}(2000)}]{BPZ}
{Ben{\'{\i}}tez}, N. 2000, \apj, 536, 571

\bibitem[{{Bertin}(2006)}]{Bertin+2006_SCAMP}
{Bertin}, E. 2006, in Astronomical Society of the Pacific Conference Series,
  Vol. 351, Astronomical Data Analysis Software and Systems XV, ed.
  C.~{Gabriel}, C.~{Arviset}, D.~{Ponz}, \& S.~{Enrique}, 112--+

\bibitem[{{Bertin} \& {Arnouts}(1996)}]{1996A&AS..117..393B}
{Bertin}, E., \& {Arnouts}, S. 1996, \aaps, 117, 393

\bibitem[{{Blumenthal} {et~al.}(1986){Blumenthal}, {Faber}, {Flores}, \&
  {Primack}}]{Blumenthal+1986}
{Blumenthal}, G.~R., {Faber}, S.~M., {Flores}, R., \& {Primack}, J.~R. 1986,
  \apj, 301, 27

\bibitem[{{Bolzonella} {et~al.}(2000){Bolzonella}, {Miralles}, \&
  {Pell{\'o}}}]{hyperz}
{Bolzonella}, M., {Miralles}, J.-M., \& {Pell{\'o}}, R. 2000, \aap, 363, 476,
  arXiv:astro-ph/0003380

\bibitem[{{Broadhurst} {et~al.}(2005{\natexlab{a}}){Broadhurst},
  {Ben{\'{\i}}tez}, {Coe}, {Sharon}, {Zekser}, {White}, {Ford}, {Bouwens},
  {Blakeslee}, {Clampin}, {Cross}, {Franx}, {Frye}, {Hartig}, {Illingworth},
  {Infante}, {Menanteau}, {Meurer}, {Postman}, {Ardila}, {Bartko}, {Brown},
  {Burrows}, {Cheng}, {Feldman}, {Golimowski}, {Goto}, {Gronwall}, {Herranz},
  {Holden}, {Homeier}, {Krist}, {Lesser}, {Martel}, {Miley}, {Rosati},
  {Sirianni}, {Sparks}, {Steindling}, {Tran}, {Tsvetanov}, \&
  {Zheng}}]{2005ApJ...621...53B}
{Broadhurst}, T. {et~al.} 2005{\natexlab{a}}, \apj, 621, 53,
  arXiv:astro-ph/0409132

\bibitem[{{Broadhurst} {et~al.}(2000){Broadhurst}, {Huang}, {Frye}, \&
  {Ellis}}]{Broadhurst+2000_CL0024}
{Broadhurst}, T., {Huang}, X., {Frye}, B., \& {Ellis}, R. 2000, \apjl, 534,
  L15, arXiv:astro-ph/9902316

\bibitem[{{Broadhurst} {et~al.}(2005{\natexlab{b}}){Broadhurst}, {Takada},
  {Umetsu}, {Kong}, {Arimoto}, {Chiba}, \& {Futamase}}]{BTU+05}
{Broadhurst}, T., {Takada}, M., {Umetsu}, K., {Kong}, X., {Arimoto}, N.,
  {Chiba}, M., \& {Futamase}, T. 2005{\natexlab{b}}, \apjl, 619, L143,
  arXiv:astro-ph/0412192

\bibitem[{{Broadhurst} {et~al.}(2008){Broadhurst}, {Umetsu}, {Medezinski},
  {Oguri}, \& {Rephaeli}}]{BUM+08}
{Broadhurst}, T., {Umetsu}, K., {Medezinski}, E., {Oguri}, M., \& {Rephaeli},
  Y. 2008, \apjl, 685, L9, 0805.2617

\bibitem[{{Broadhurst} \& {Barkana}(2008)}]{Broadhurst+Barkana2008}
{Broadhurst}, T.~J., \& {Barkana}, R. 2008, \mnras, 390, 1647, 0801.1875

\bibitem[{{Broadhurst} {et~al.}(1995){Broadhurst}, {Taylor}, \&
  {Peacock}}]{1995ApJ...438...49B}
{Broadhurst}, T.~J., {Taylor}, A.~N., \& {Peacock}, J.~A. 1995, \apj, 438, 49,
  arXiv:astro-ph/9406052

\bibitem[{{Capak} {et~al.}(2007){Capak}, {Aussel}, {Ajiki}, {McCracken},
  {Mobasher}, {Scoville}, {Shopbell}, {Taniguchi}, {Thompson}, {Tribiano},
  {Sasaki}, {Blain}, {Brusa}, {Carilli}, {Comastri}, {Carollo}, {Cassata},
  {Colbert}, {Ellis}, {Elvis}, {Giavalisco}, {Green}, {Guzzo}, {Hasinger},
  {Ilbert}, {Impey}, {Jahnke}, {Kartaltepe}, {Kneib}, {Koda}, {Koekemoer},
  {Komiyama}, {Leauthaud}, {Lefevre}, {Lilly}, {Liu}, {Massey}, {Miyazaki},
  {Murayama}, {Nagao}, {Peacock}, {Pickles}, {Porciani}, {Renzini}, {Rhodes},
  {Rich}, {Salvato}, {Sanders}, {Scarlata}, {Schiminovich}, {Schinnerer},
  {Scodeggio}, {Sheth}, {Shioya}, {Tasca}, {Taylor}, {Yan}, \&
  {Zamorani}}]{Capak+07_COSMOS}
{Capak}, P. {et~al.} 2007, \apjs, 172, 99, arXiv:0704.2430

\bibitem[{{Clowe} {et~al.}(2006){Clowe}, {Brada{\v c}}, {Gonzalez},
  {Markevitch}, {Randall}, {Jones}, \& {Zaritsky}}]{Clowe+2006_Bullet}
{Clowe}, D., {Brada{\v c}}, M., {Gonzalez}, A.~H., {Markevitch}, M., {Randall},
  S.~W., {Jones}, C., \& {Zaritsky}, D. 2006, \apjl, 648, L109,
  arXiv:astro-ph/0608407

\bibitem[{{Clowe} {et~al.}(2000){Clowe}, {Luppino}, {Kaiser}, \&
  {Gioia}}]{2000ApJ...539..540C}
{Clowe}, D., {Luppino}, G.~A., {Kaiser}, N., \& {Gioia}, I.~M. 2000, \apj, 539,
  540, arXiv:astro-ph/0001356

\bibitem[{{Coe} {et~al.}(2006){Coe}, {Ben{\'{\i}}tez}, {S{\'a}nchez}, {Jee},
  {Bouwens}, \& {Ford}}]{colorpro}
{Coe}, D., {Ben{\'{\i}}tez}, N., {S{\'a}nchez}, S.~F., {Jee}, M., {Bouwens},
  R., \& {Ford}, H. 2006, \aj, 132, 926, arXiv:astro-ph/0605262

\bibitem[{{Colella} \& {Woodward}(1984)}]{Colella+Woodward1984_PPM}
{Colella}, P., \& {Woodward}, P.~R. 1984, Journal of Computational Physics, 54,
  174

\bibitem[{{Colley} {et~al.}(1996){Colley}, {Tyson}, \&
  {Turner}}]{Colley+1996_CL0024}
{Colley}, W.~N., {Tyson}, J.~A., \& {Turner}, E.~L. 1996, \apjl, 461, L83+,
  arXiv:astro-ph/9512128

\bibitem[{{Comerford} {et~al.}(2006){Comerford}, {Meneghetti}, {Bartelmann}, \&
  {Schirmer}}]{2006ApJ...642...39C}
{Comerford}, J.~M., {Meneghetti}, M., {Bartelmann}, M., \& {Schirmer}, M. 2006,
  \apj, 642, 39, arXiv:astro-ph/0511330

\bibitem[{{Comerford} \& {Natarajan}(2007)}]{2007MNRAS.379..190C}
{Comerford}, J.~M., \& {Natarajan}, P. 2007, \mnras, 379, 190,
  arXiv:astro-ph/0703126

\bibitem[{{Crittenden} {et~al.}(2002){Crittenden}, {Natarajan}, {Pen}, \&
  {Theuns}}]{2002ApJ...568...20C}
{Crittenden}, R.~G., {Natarajan}, P., {Pen}, U.-L., \& {Theuns}, T. 2002, \apj,
  568, 20, arXiv:astro-ph/0012336

\bibitem[{{Czoske} {et~al.}(2001){Czoske}, {Kneib}, {Soucail}, {Bridges},
  {Mellier}, \& {Cuillandre}}]{Czoske+2001}
{Czoske}, O., {Kneib}, J.-P., {Soucail}, G., {Bridges}, T.~J., {Mellier}, Y.,
  \& {Cuillandre}, J.-C. 2001, \aap, 372, 391, arXiv:astro-ph/0103123

\bibitem[{{Czoske} {et~al.}(2002){Czoske}, {Moore}, {Kneib}, \&
  {Soucail}}]{Czoske+2002}
{Czoske}, O., {Moore}, B., {Kneib}, J.-P., \& {Soucail}, G. 2002, \aap, 386,
  31, arXiv:astro-ph/0111118

\bibitem[{{Diaferio}(1999)}]{Diaferio1999}
{Diaferio}, A. 1999, \mnras, 309, 610, arXiv:astro-ph/9906331

\bibitem[{{Diaferio} {et~al.}(2005){Diaferio}, {Geller}, \&
  {Rines}}]{Diaferio+2005}
{Diaferio}, A., {Geller}, M.~J., \& {Rines}, K.~J. 2005, \apjl, 628, L97,
  arXiv:astro-ph/0506560

\bibitem[{{Dressler} {et~al.}(1999){Dressler}, {Smail}, {Poggianti}, {Butcher},
  {Couch}, {Ellis}, \& {Oemler}}]{1999ApJS..122...51D}
{Dressler}, A., {Smail}, I., {Poggianti}, B.~M., {Butcher}, H., {Couch}, W.~J.,
  {Ellis}, R.~S., \& {Oemler}, A.~J. 1999, \apjs, 122, 51,
  arXiv:astro-ph/9901263

\bibitem[{{Duffy} {et~al.}(2008){Duffy}, {Schaye}, {Kay}, \& {Dalla
  Vecchia}}]{Duffy+2008}
{Duffy}, A.~R., {Schaye}, J., {Kay}, S.~T., \& {Dalla Vecchia}, C. 2008,
  \mnras, 390, L64, 0804.2486

\bibitem[{{Erben} {et~al.}(2001){Erben}, {Van Waerbeke}, {Bertin}, {Mellier},
  \& {Schneider}}]{2001A&A...366..717E}
{Erben}, T., {Van Waerbeke}, L., {Bertin}, E., {Mellier}, Y., \& {Schneider},
  P. 2001, \aap, 366, 717, arXiv:astro-ph/0007021

\bibitem[{{Fahlman} {et~al.}(1994){Fahlman}, {Kaiser}, {Squires}, \&
  {Woods}}]{1994ApJ...437...56F}
{Fahlman}, G., {Kaiser}, N., {Squires}, G., \& {Woods}, D. 1994, \apj, 437, 56,
  arXiv:astro-ph/9402017

\bibitem[{{Ford} {et~al.}(2003){Ford}, {Clampin}, {Hartig}, {Illingworth},
  {Sirianni}, {Martel}, {Meurer}, {McCann}, {Sullivan}, {Bartko}, {Benitez},
  {Blakeslee}, {Bouwens}, {Broadhurst}, {Brown}, {Burrows}, {Campbell},
  {Cheng}, {Feldman}, {Franx}, {Golimowski}, {Gronwall}, {Kimble}, {Krist},
  {Lesser}, {Magee}, {Miley}, {Postman}, {Rafal}, {Rosati}, {Sparks}, {Tran},
  {Tsvetanov}, {Volmer}, {White}, \& {Woodruff}}]{Holland+2003_ACS}
{Ford}, H.~C. {et~al.} 2003, in Presented at the Society of Photo-Optical
  Instrumentation Engineers (SPIE) Conference, Vol. 4854, Society of
  Photo-Optical Instrumentation Engineers (SPIE) Conference Series, ed. J.~C.
  {Blades} \& O.~H.~W. {Siegmund}, 81--94

\bibitem[{{Fryxell} {et~al.}(2000){Fryxell}, {Olson}, {Ricker}, {Timmes},
  {Zingale}, {Lamb}, {MacNeice}, {Rosner}, {Truran}, \&
  {Tufo}}]{Fryxell+2000_FLASH}
{Fryxell}, B. {et~al.} 2000, \apjs, 131, 273

\bibitem[{{Goldberg} \& {Bacon}(2005)}]{2005ApJ...619..741G}
{Goldberg}, D.~M., \& {Bacon}, D.~J. 2005, \apj, 619, 741,
  arXiv:astro-ph/0406376

\bibitem[{{Hamana} {et~al.}(2003){Hamana}, {Miyazaki}, {Shimasaku}, {Furusawa},
  {Doi}, {Hamabe}, {Imi}, {Kimura}, {Komiyama}, {Nakata}, {Okada}, {Okamura},
  {Ouchi}, {Sekiguchi}, {Yagi}, \& {Yasuda}}]{2003ApJ...597...98H}
{Hamana}, T. {et~al.} 2003, \apj, 597, 98, arXiv:astro-ph/0210450

\bibitem[{{Hennawi} {et~al.}(2007){Hennawi}, {Dalal}, {Bode}, \&
  {Ostriker}}]{2007ApJ...654..714H}
{Hennawi}, J.~F., {Dalal}, N., {Bode}, P., \& {Ostriker}, J.~P. 2007, \apj,
  654, 714, arXiv:astro-ph/0506171

\bibitem[{{Heymans} {et~al.}(2006){Heymans}, {Van Waerbeke}, {Bacon}, {Berge},
  {Bernstein}, {Bertin}, {Bridle}, {Brown}, {Clowe}, {Dahle}, {Erben}, {Gray},
  {Hetterscheidt}, {Hoekstra}, {Hudelot}, {Jarvis}, {Kuijken}, {Margoniner},
  {Massey}, {Mellier}, {Nakajima}, {Refregier}, {Rhodes}, {Schrabback}, \&
  {Wittman}}]{2006MNRAS.368.1323H}
{Heymans}, C. {et~al.} 2006, \mnras, 368, 1323, arXiv:astro-ph/0506112

\bibitem[{{Hoekstra}(2007)}]{Hoekstra2007}
{Hoekstra}, H. 2007, \mnras, 379, 317, arXiv:0705.0358

\bibitem[{{Hoekstra} {et~al.}(2000){Hoekstra}, {Franx}, \&
  {Kuijken}}]{2000ApJ...532...88H}
{Hoekstra}, H., {Franx}, M., \& {Kuijken}, K. 2000, \apj, 532, 88,
  arXiv:astro-ph/9910487

\bibitem[{{Hoekstra} {et~al.}(1998){Hoekstra}, {Franx}, {Kuijken}, \&
  {Squires}}]{1998ApJ...504..636H}
{Hoekstra}, H., {Franx}, M., {Kuijken}, K., \& {Squires}, G. 1998, \apj, 504,
  636

\bibitem[{{Ilbert} {et~al.}(2009){Ilbert}, {Capak}, {Salvato}, {Aussel},
  {McCracken}, {Sanders}, {Scoville}, {Kartaltepe}, {Arnouts}, {Floc'h},
  {Mobasher}, {Taniguchi}, {Lamareille}, {Leauthaud}, {Sasaki}, {Thompson},
  {Zamojski}, {Zamorani}, {Bardelli}, {Bolzonella}, {Bongiorno}, {Brusa},
  {Caputi}, {Carollo}, {Contini}, {Cook}, {Coppa}, {Cucciati}, {de la Torre},
  {de Ravel}, {Franzetti}, {Garilli}, {Hasinger}, {Iovino}, {Kampczyk},
  {Kneib}, {Knobel}, {Kovac}, {LeBorgne}, {LeBrun}, {F{\`e}vre}, {Lilly},
  {Looper}, {Maier}, {Mainieri}, {Mellier}, {Mignoli}, {Murayama}, {Pell{\`o}},
  {Peng}, {P{\'e}rez-Montero}, {Renzini}, {Ricciardelli}, {Schiminovich},
  {Scodeggio}, {Shioya}, {Silverman}, {Surace}, {Tanaka}, {Tasca}, {Tresse},
  {Vergani}, \& {Zucca}}]{Ilbert+2009_COSMOS}
{Ilbert}, O. {et~al.} 2009, \apj, 690, 1236, 0809.2101

\bibitem[{{Jain} {et~al.}(2000){Jain}, {Seljak}, \&
  {White}}]{2000ApJ...530..547J}
{Jain}, B., {Seljak}, U., \& {White}, S. 2000, \apj, 530, 547,
  arXiv:astro-ph/9901191

\bibitem[{{Jee} {et~al.}(2007){Jee}, {Ford}, {Illingworth}, {White},
  {Broadhurst}, {Coe}, {Meurer}, {van der Wel}, {Ben{\'{\i}}tez}, {Blakeslee},
  {Bouwens}, {Bradley}, {Demarco}, {Homeier}, {Martel}, \&
  {Mei}}]{Jee+2007_CL0024}
{Jee}, M.~J. {et~al.} 2007, \apj, 661, 728, 0705.2171

\bibitem[{{Jing} \& {Suto}(2000)}]{Jing+Suto2000}
{Jing}, Y.~P., \& {Suto}, Y. 2000, \apjl, 529, L69, arXiv:astro-ph/9909478

\bibitem[{{Kaiser}(1995)}]{1995ApJ...439L...1K}
{Kaiser}, N. 1995, \apjl, 439, L1, arXiv:astro-ph/9408092

\bibitem[{{Kaiser} \& {Squires}(1993)}]{1993ApJ...404..441K}
{Kaiser}, N., \& {Squires}, G. 1993, \apj, 404, 441

\bibitem[{{Kaiser} {et~al.}(1995){Kaiser}, {Squires}, \&
  {Broadhurst}}]{1995ApJ...449..460K}
{Kaiser}, N., {Squires}, G., \& {Broadhurst}, T. 1995, \apj, 449, 460,
  arXiv:astro-ph/9411005

\bibitem[{{Kassiola} {et~al.}(1992){Kassiola}, {Kovner}, \&
  {Fort}}]{Kassiola+1992_CL0024}
{Kassiola}, A., {Kovner}, I., \& {Fort}, B. 1992, \apj, 400, 41

\bibitem[{{Kneib} {et~al.}(2003){Kneib}, {Hudelot}, {Ellis}, {Treu}, {Smith},
  {Marshall}, {Czoske}, {Smail}, \& {Natarajan}}]{2003ApJ...598..804K}
{Kneib}, J.-P. {et~al.} 2003, \apj, 598, 804, arXiv:astro-ph/0307299

\bibitem[{{Kodama} {et~al.}(2005){Kodama}, {Tanaka}, {Tamura}, {Yahagi},
  {Nagashima}, {Tanaka}, {Arimoto}, {Futamase}, {Iye}, {Karasawa}, {Kashikawa},
  {Kawasaki}, {Kitayama}, {Matsuhara}, {Nakata}, {Ohashi}, {Ohta}, {Okamoto},
  {Okamura}, {Shimasaku}, {Suto}, {Tamura}, {Umetsu}, \&
  {Yamada}}]{2005PASJ...57..309K}
{Kodama}, T. {et~al.} 2005, \pasj, 57, 309, arXiv:astro-ph/0502444

\bibitem[{{Koo}(1988)}]{Koo1988_CL0024}
{Koo}, D.~C. 1988, {Recent observations of distant matter - Direct clues to
  birth and evolution}, ed. V.~C. {Rubin} \& G.~V. {Coyne}, 513--540

\bibitem[{{Kravtsov} {et~al.}(2006){Kravtsov}, {Vikhlinin}, \&
  {Nagai}}]{Kravtsov+2006_Yx}
{Kravtsov}, A.~V., {Vikhlinin}, A., \& {Nagai}, D. 2006, \apj, 650, 128,
  arXiv:astro-ph/0603205

\bibitem[{{Lapi} \& {Cavaliere}(2009)}]{Lapi+Cavaliere2009}
{Lapi}, A., \& {Cavaliere}, A. 2009, \apjl, 695, L125, 0903.1589

\bibitem[{{Lemze} {et~al.}(2008){Lemze}, {Barkana}, {Broadhurst}, \&
  {Rephaeli}}]{2008MNRAS.386.1092L}
{Lemze}, D., {Barkana}, R., {Broadhurst}, T.~J., \& {Rephaeli}, Y. 2008,
  \mnras, 386, 1092, arXiv:0711.3908

\bibitem[{{Lemze} {et~al.}(2009){Lemze}, {Broadhurst}, {Rephaeli}, {Barkana},
  \& {Umetsu}}]{Lemze+2009}
{Lemze}, D., {Broadhurst}, T., {Rephaeli}, Y., {Barkana}, R., \& {Umetsu}, K.
  2009, \apj, 701, 1336, arXiv:astr-ph/0810.3129

\bibitem[{{Liesenborgs} {et~al.}(2008{\natexlab{a}}){Liesenborgs}, {de Rijcke},
  {Dejonghe}, \& {Bekaert}}]{Liesenborgs+2008_ring}
{Liesenborgs}, J., {de Rijcke}, S., {Dejonghe}, H., \& {Bekaert}, P.
  2008{\natexlab{a}}, \mnras, 386, 307, 0801.4255

\bibitem[{{Liesenborgs} {et~al.}(2008{\natexlab{b}}){Liesenborgs}, {de Rijcke},
  {Dejonghe}, \& {Bekaert}}]{Liesenborgs+2008_CL0024}
------. 2008{\natexlab{b}}, \mnras, 389, 415, 0806.2609

\bibitem[{{Mahdavi} {et~al.}(2008){Mahdavi}, {Hoekstra}, {Babul}, \&
  {Henry}}]{Mahdavi+2008_CL0024}
{Mahdavi}, A., {Hoekstra}, H., {Babul}, A., \& {Henry}, J.~P. 2008, \mnras,
  384, 1567, 0710.4132

\bibitem[{{Massey} {et~al.}(2007){Massey}, {Heymans}, {Berg{\'e}}, {Bernstein},
  {Bridle}, {Clowe}, {Dahle}, {Ellis}, {Erben}, {Hetterscheidt}, {High},
  {Hirata}, {Hoekstra}, {Hudelot}, {Jarvis}, {Johnston}, {Kuijken},
  {Margoniner}, {Mandelbaum}, {Mellier}, {Nakajima}, {Paulin-Henriksson},
  {Peeples}, {Roat}, {Refregier}, {Rhodes}, {Schrabback}, {Schirmer}, {Seljak},
  {Semboloni}, \& {van Waerbeke}}]{2007MNRAS.376...13M}
{Massey}, R. {et~al.} 2007, \mnras, 376, 13, arXiv:astro-ph/0608643

\bibitem[{{Mazzotta} {et~al.}(2004){Mazzotta}, {Rasia}, {Moscardini}, \&
  {Tormen}}]{Mazzotta+2004}
{Mazzotta}, P., {Rasia}, E., {Moscardini}, L., \& {Tormen}, G. 2004, \mnras,
  354, 10, arXiv:astro-ph/0409618

\bibitem[{{Medezinski} {et~al.}(2007){Medezinski}, {Broadhurst}, {Umetsu},
  {Coe}, {Ben{\'{\i}}tez}, {Ford}, {Rephaeli}, {Arimoto}, \&
  {Kong}}]{Medezinski+07}
{Medezinski}, E. {et~al.} 2007, \apj, 663, 717, arXiv:astro-ph/0608499

\bibitem[{{Medezinski} {et~al.}(2009){Medezinski}, {Broadhurst}, {Umetsu},
  {Oguri}, {Rephaeli}, \& {Ben{\'{\i}}tez}}]{Medezinski+2009}
{Medezinski}, E., {Broadhurst}, T., {Umetsu}, K., {Oguri}, M., {Rephaeli}, Y.,
  \& {Ben{\'{\i}}tez}, N. 2009, \apj, in press, arXiv:0906.4791

\bibitem[{{Miyazaki} {et~al.}(2002){Miyazaki}, {Komiyama}, {Sekiguchi},
  {Okamura}, {Doi}, {Furusawa}, {Hamabe}, {Imi}, {Kimura}, {Nakata}, {Okada},
  {Ouchi}, {Shimasaku}, {Yagi}, \& {Yasuda}}]{2002PASJ...54..833M}
{Miyazaki}, S. {et~al.} 2002, \pasj, 54, 833, arXiv:astro-ph/0211006

\bibitem[{{Navarro} {et~al.}(1997){Navarro}, {Frenk}, \&
  {White}}]{1997ApJ...490..493N}
{Navarro}, J.~F., {Frenk}, C.~S., \& {White}, S.~D.~M. 1997, \apj, 490, 493,
  arXiv:astro-ph/9611107

\bibitem[{{Neto} {et~al.}(2007){Neto}, {Gao}, {Bett}, {Cole}, {Navarro},
  {Frenk}, {White}, {Springel}, \& {Jenkins}}]{2007MNRAS.381.1450N}
{Neto}, A.~F. {et~al.} 2007, \mnras, 381, 1450, arXiv:0706.2919

\bibitem[{{Newman} {et~al.}(2009){Newman}, {Treu}, {Ellis}, {Sand}, {Richard},
  {Marshall}, {Capak}, \& {Miyazaki}}]{Newman+2009_A611}
{Newman}, A.~B., {Treu}, T., {Ellis}, R.~S., {Sand}, D.~J., {Richard}, J.,
  {Marshall}, P.~J., {Capak}, P., \& {Miyazaki}, S. 2009, \apj, 706, 1078,
  0909.3527

\bibitem[{{Oguri} \& {Blandford}(2009)}]{Oguri+Blandford2009}
{Oguri}, M., \& {Blandford}, R.~D. 2009, \mnras, 392, 930,
  arXiv:astro-ph/0808.0192

\bibitem[{{Oguri} {et~al.}(2009){Oguri}, {Hennawi}, {Gladders}, {Dahle},
  {Natarajan}, {Dalal}, {Koester}, {Sharon}, \& {Bayliss}}]{Oguri+2009_Subaru}
{Oguri}, M. {et~al.} 2009, \apj, 699, 1038, arXiv:astro-ph/0901.4372

\bibitem[{{Okabe} {et~al.}(2009){Okabe}, {Takada}, {Umetsu}, {Futamase}, \&
  {Smith}}]{2009arXiv0903.1103O}
{Okabe}, N., {Takada}, M., {Umetsu}, K., {Futamase}, T., \& {Smith}, G.~P.
  2009, ArXiv e-prints, arXiv:astro-ph/0903.1103

\bibitem[{{Okabe} \& {Umetsu}(2008)}]{Okabe+Umetsu2008}
{Okabe}, N., \& {Umetsu}, K. 2008, \pasj, 60, 345, arXiv:astro-ph/0702649

\bibitem[{{Okura} {et~al.}(2007){Okura}, {Umetsu}, \& {Futamase}}]{HOLICs1}
{Okura}, Y., {Umetsu}, K., \& {Futamase}, T. 2007, \apj, 660, 995,
  arXiv:astro-ph/0607288

\bibitem[{{Okura} {et~al.}(2008){Okura}, {Umetsu}, \& {Futamase}}]{HOLICs2}
------. 2008, \apj, 680, 1, arXiv:0710.2262

\bibitem[{{Ota} {et~al.}(2004){Ota}, {Pointecouteau}, {Hattori}, \&
  {Mitsuda}}]{2004ApJ...601..120O}
{Ota}, N., {Pointecouteau}, E., {Hattori}, M., \& {Mitsuda}, K. 2004, \apj,
  601, 120, arXiv:astro-ph/0306580

\bibitem[{{Ouchi} {et~al.}(2004){Ouchi}, {Shimasaku}, {Okamura}, {Furusawa},
  {Kashikawa}, {Ota}, {Doi}, {Hamabe}, {Kimura}, {Komiyama}, {Miyazaki},
  {Miyazaki}, {Nakata}, {Sekiguchi}, {Yagi}, \& {Yasuda}}]{Ouchi+2004_SDFRED}
{Ouchi}, M. {et~al.} 2004, \apj, 611, 660, arXiv:astro-ph/0309655

\bibitem[{{Park} {et~al.}(2003){Park}, {Ng}, {Park}, {Liu}, \&
  {Umetsu}}]{Park+2003}
{Park}, C.-G., {Ng}, K.-W., {Park}, C., {Liu}, G.-C., \& {Umetsu}, K. 2003,
  \apj, 589, 67, arXiv:astro-ph/0209491

\bibitem[{{Richard} {et~al.}(2009){Richard}, {Pei}, {Limousin}, {Jullo}, \&
  {Kneib}}]{Richard+2009_A1703}
{Richard}, J., {Pei}, L., {Limousin}, M., {Jullo}, E., \& {Kneib}, J.~P. 2009,
  \aap, 498, 37, 0901.0427

\bibitem[{{Ricker}(2008)}]{Ricker2008}
{Ricker}, P.~M. 2008, \apjs, 176, 293, 0710.4397

\bibitem[{{Ricker} \& {Sarazin}(2001)}]{2001ApJ...561..621R}
{Ricker}, P.~M., \& {Sarazin}, C.~L. 2001, \apj, 561, 621,
  arXiv:astro-ph/0107210

\bibitem[{{Rines} {et~al.}(2000){Rines}, {Geller}, {Diaferio}, {Mohr}, \&
  {Wegner}}]{Rines+2000}
{Rines}, K., {Geller}, M.~J., {Diaferio}, A., {Mohr}, J.~J., \& {Wegner}, G.~A.
  2000, \aj, 120, 2338, arXiv:astro-ph/0007126

\bibitem[{{Schneider} {et~al.}(2000){Schneider}, {King}, \&
  {Erben}}]{2000A&A...353...41S}
{Schneider}, P., {King}, L., \& {Erben}, T. 2000, \aap, 353, 41,
  arXiv:astro-ph/9907143

\bibitem[{{Schneider} \& {Seitz}(1995)}]{1995A&A...294..411S}
{Schneider}, P., \& {Seitz}, C. 1995, \aap, 294, 411, arXiv:astro-ph/9407032

\bibitem[{{Seitz} \& {Schneider}(1997)}]{1997A&A...318..687S}
{Seitz}, C., \& {Schneider}, P. 1997, \aap, 318, 687, arXiv:astro-ph/9601079

\bibitem[{{Shapiro} \& {Iliev}(2000)}]{Shapiro+Iliev2000_CL0024}
{Shapiro}, P.~R., \& {Iliev}, I.~T. 2000, \apjl, 542, L1,
  arXiv:astro-ph/0006353

\bibitem[{{Smail} {et~al.}(1996){Smail}, {Dressler}, {Kneib}, {Ellis}, {Couch},
  {Sharples}, \& {Oemler}}]{Smail+1996_arcs}
{Smail}, I., {Dressler}, A., {Kneib}, J.-P., {Ellis}, R.~S., {Couch}, W.~J.,
  {Sharples}, R.~M., \& {Oemler}, A.~J. 1996, \apj, 469, 508,
  arXiv:astro-ph/9503063

\bibitem[{{Soucail} {et~al.}(2000){Soucail}, {Ota}, {B{\"o}hringer}, {Czoske},
  {Hattori}, \& {Mellier}}]{Soucail+2000_CL0024}
{Soucail}, G., {Ota}, N., {B{\"o}hringer}, H., {Czoske}, O., {Hattori}, M., \&
  {Mellier}, Y. 2000, \aap, 355, 433, arXiv:astro-ph/9911062

\bibitem[{{Takada} \& {Jain}(2003)}]{2003MNRAS.340..580T}
{Takada}, M., \& {Jain}, B. 2003, \mnras, 340, 580, arXiv:astro-ph/0209167

\bibitem[{{Tanaka} {et~al.}(2005){Tanaka}, {Kodama}, {Arimoto}, {Okamura},
  {Umetsu}, {Shimasaku}, {Tanaka}, \& {Yamada}}]{Tanaka+2005_PISCES}
{Tanaka}, M., {Kodama}, T., {Arimoto}, N., {Okamura}, S., {Umetsu}, K.,
  {Shimasaku}, K., {Tanaka}, I., \& {Yamada}, T. 2005, \mnras, 362, 268,
  arXiv:astro-ph/0506713

\bibitem[{{Tyson} {et~al.}(1998){Tyson}, {Kochanski}, \&
  {dell'Antonio}}]{Tyson+1998_CL0024}
{Tyson}, J.~A., {Kochanski}, G.~P., \& {dell'Antonio}, I.~P. 1998, \apjl, 498,
  L107+, arXiv:astro-ph/9801193

\bibitem[{{Umetsu} {et~al.}(2009){Umetsu}, {Birkinshaw}, {Liu}, {Wu},
  {Medezinski}, {Broadhurst}, {Lemze}, {Zitrin}, {Ho}, {Huang}, {Koch}, {Liao},
  {Lin}, {Molnar}, {Nishioka}, {Wang}, {Altamirano}, {Chang}, {Chang}, {Chang},
  {Chen}, {Han}, {Huang}, {Hwang}, {Jiang}, {Kesteven}, {Kubo}, {Li},
  {Martin-Cocher}, {Oshiro}, {Raffin}, {Wei}, \&
  {Wilson}}]{2009ApJ...694.1643U}
{Umetsu}, K. {et~al.} 2009, \apj, 694, 1643, arXiv:astro-ph/0810.0969

\bibitem[{{Umetsu} \& {Broadhurst}(2008)}]{UB2008}
{Umetsu}, K., \& {Broadhurst}, T. 2008, \apj, 684, 177, arXiv:0712.3441

\bibitem[{{Umetsu} {et~al.}(1999){Umetsu}, {Tada}, \&
  {Futamase}}]{1999PThPS.133...53U}
{Umetsu}, K., {Tada}, M., \& {Futamase}, T. 1999, Progress of Theoretical
  Physics Supplement, 133, 53, arXiv:astro-ph/0004400

\bibitem[{{Umetsu} {et~al.}(2007){Umetsu}, {Takada}, \& {Broadhurst}}]{UTB08}
{Umetsu}, K., {Takada}, M., \& {Broadhurst}, T. 2007, Modern Physics Letters A,
  22, 2099, arXiv:astro-ph/0702096

\bibitem[{{Umetsu} {et~al.}(2005){Umetsu}, {Tanaka}, {Kodama}, {Tanaka},
  {Futamase}, {Kashikawa}, \& {Hoshi}}]{2005PASJ...57..877U}
{Umetsu}, K., {Tanaka}, M., {Kodama}, T., {Tanaka}, I., {Futamase}, T.,
  {Kashikawa}, N., \& {Hoshi}, T. 2005, \pasj, 57, 877, arXiv:astro-ph/0506746

\bibitem[{{Vikhlinin} {et~al.}(2009){Vikhlinin}, {Burenin}, {Ebeling},
  {Forman}, {Hornstrup}, {Jones}, {Kravtsov}, {Murray}, {Nagai}, {Quintana}, \&
  {Voevodkin}}]{Vikhlinin+2009_CCC2}
{Vikhlinin}, A. {et~al.} 2009, \apj, 692, 1033, 0805.2207

\bibitem[{{Wright} \& {Brainerd}(2000)}]{2000ApJ...534...34W}
{Wright}, C.~O., \& {Brainerd}, T.~G. 2000, \apj, 534, 34

\bibitem[{{Yagi} {et~al.}(2002){Yagi}, {Kashikawa}, {Sekiguchi}, {Doi},
  {Yasuda}, {Shimasaku}, \& {Okamura}}]{Yagi+2002_SDFRED}
{Yagi}, M., {Kashikawa}, N., {Sekiguchi}, M., {Doi}, M., {Yasuda}, N.,
  {Shimasaku}, K., \& {Okamura}, S. 2002, \aj, 123, 66

\bibitem[{{Zacharias} {et~al.}(2004){Zacharias}, {Monet}, {Levine}, {Urban},
  {Gaume}, \& {Wycoff}}]{NOMAD}
{Zacharias}, N., {Monet}, D.~G., {Levine}, S.~E., {Urban}, S.~E., {Gaume}, R.,
  \& {Wycoff}, G.~L. 2004, in Bulletin of the American Astronomical Society,
  Vol.~36, Bulletin of the American Astronomical Society, 1418--+

\bibitem[{{Zhang} {et~al.}(2005){Zhang}, {B{\"o}hringer}, {Mellier}, {Soucail},
  \& {Forman}}]{Zhang+2005_CL0024}
{Zhang}, Y.-Y., {B{\"o}hringer}, H., {Mellier}, Y., {Soucail}, G., \& {Forman},
  W. 2005, \aap, 429, 85, arXiv:astro-ph/0408545

\bibitem[{{Zhao}(1996)}]{Zhao1996}
{Zhao}, H. 1996, \mnras, 278, 488, arXiv:astro-ph/9509122

\bibitem[{{Zheng} {et~al.}(2009){Zheng}, {Bradley}, {Bouwens}, {Ford},
  {Illingworth}, {Ben{\'{\i}}tez}, {Broadhurst}, {Frye}, {Infante}, {Jee},
  {Motta}, {Shu}, \& {Zitrin}}]{Zheng+2009_CL0024}
{Zheng}, W. {et~al.} 2009, \apj, 697, 1907, 0903.3988

\bibitem[{{Zitrin} \& {Broadhurst}(2009)}]{Zitrin+Broadhurst2009}
{Zitrin}, A., \& {Broadhurst}, T. 2009, \apjl, 703, L132, 0906.5079

\bibitem[{{Zitrin} {et~al.}(2009{\natexlab{a}}){Zitrin}, {Broadhurst},
  {Rephaeli}, \& {Sadeh}}]{Zitrin+2009_0717}
{Zitrin}, A., {Broadhurst}, T., {Rephaeli}, Y., \& {Sadeh}, S.
  2009{\natexlab{a}}, \apjl, 707, L102, arXiv:astro-ph/0907.4232

\bibitem[{{Zitrin} {et~al.}(2009{\natexlab{b}}){Zitrin}, {Broadhurst},
  {Umetsu}, {Coe}, {Ben{\'{\i}}tez}, {Ascaso}, {Bradley}, {Ford}, {Jee},
  {Medezinski}, {Rephaeli}, \& {Zheng}}]{Zitrin+2009_CL0024}
{Zitrin}, A. {et~al.} 2009{\natexlab{b}}, \mnras, 396, 1985, arXiv:0902.3971

\bibitem[{{Zu Hone} {et~al.}(2009{\natexlab{a}}){Zu Hone}, {Lamb}, \&
  {Ricker}}]{Zuhone+2009_CL0024}
{Zu Hone}, J.~A., {Lamb}, D.~Q., \& {Ricker}, P.~M. 2009{\natexlab{a}}, \apj,
  696, 694, 0809.3252

\bibitem[{{Zu Hone} {et~al.}(2009{\natexlab{b}}){Zu Hone}, {Ricker}, {Lamb}, \&
  {Karen Yang}}]{Zuhone+2009_CL0024_X-ray}
{Zu Hone}, J.~A., {Ricker}, P.~M., {Lamb}, D.~Q., \& {Karen Yang}, H.-Y.
  2009{\natexlab{b}}, \apj, 699, 1004, 0808.0930

\bibitem[{{Zwicky}(1959)}]{Zwicky1959}
{Zwicky}, F. 1959, Handbuch der Physik, 53, 390

\end{thebibliography}




\begin{deluxetable}{lc}
\tablecolumns{4}
\tablecaption{
 \label{tab:cluster}
Properties of the Galaxy Cluster Cl0024+1654
} 
\tablewidth{0pt}  
\tablehead{ 
 \multicolumn{1}{c}{Parameter} &
 \multicolumn{1}{c}{Value} 
} 
\startdata  
Optical center position (J2000.0) & \\
\ \ \ \ R.A. ...................................... & 00:26:35.69\\
\ \ \ \ Decl. ..................................... & +17:09:43.12\\
Redshift .................................... & $0.395$\\
X-ray temperature  (keV) ..........& $4.47^{+0.83}_{-0.54}$ (90\% CL)\\
Einstein radius ($\arcsec$) ....................& 30 (at $z=1.675$)
\enddata
\tablecomments{
The optical cluster center is defined as the center of the central
bright elliptical galaxy, or the galaxy 374 in the spectroscopic catalog 
of Reference [1]. Units of right ascension are hours, minutes, and
seconds, and units of declination are degrees, arcminutes, and
arcseconds. The average X-ray temperature $T_X$ is taken from Reference
 [2]. The Einstein radius $\theta_{\rm E}$ for a background source at 
$z=1.675$ is constrained by detailed strong lens modeling by  Reference
 [3].  
}
\tablerefs{ 
 [1] \cite{Czoske+2002};
 [2] \cite{2004ApJ...601..120O};
 [3] \cite{Zitrin+2009_CL0024}.
 }
\end{deluxetable}


\begin{deluxetable}{cccc}
\tablecolumns{4}
\tablecaption{
 \label{tab:subaru}
Subaru Suprime-Cam data
} 
\tablewidth{0pt} 
\tablehead{ 
 \multicolumn{1}{c}{Filter} &
 \multicolumn{1}{c}{Exposure time\tablenotemark{a}} &
 \multicolumn{1}{c}{Seeing\tablenotemark{b}} &
 \multicolumn{1}{c}{$m_{\rm lim}$\tablenotemark{c}}  
\\
 \colhead{} &
 \multicolumn{1}{c}{} &
 \multicolumn{1}{c}{(arcsec)} &
 \multicolumn{1}{c}{(AB mag)} 
} 
\startdata  
 $B$  & $3\times 1200\,{\rm s}$ & 1.27 & 27.1\\
 $R_{\rm c}$  & $11\times 480\,{\rm s}$& 0.80  & 26.8\\ 
 $z'$  &$5\times 300\,{\rm s} + 3\times 60\,{\rm s}$  & 0.82 & 25.3 \\
\enddata
\tablenotetext{a}{Total exposure time in units of s.}
\tablenotetext{b}{Seeing FWHM in the final co-added image.}
\tablenotetext{c}{{}Limiting magnitude for a $3\sigma$ detection within a
 $2\arcsec$ aperture.}
\end{deluxetable}


\begin{deluxetable}{ccrrcccccc}
\tabletypesize{\footnotesize}
\tablecolumns{9} 
\tablecaption{
 \label{tab:color}
Weak lensing galaxy samples
}  
\tablewidth{0pt} 
\tablehead{ 
 \multicolumn{1}{c}{Sample name} &
 \multicolumn{1}{c}{Magnitude limit\tablenotemark{a}} &
 \multicolumn{1}{c}{N} &
 \multicolumn{1}{c}{$\bar{n}_g$\tablenotemark{b}} & 
 \multicolumn{1}{c}{$\overline{\sigma}_g$\tablenotemark{c}} & 
 \multicolumn{1}{c}{$\overline{z}_s$\tablenotemark{d}} &
 \multicolumn{1}{c}{$\overline{z}_{s,\beta}$\tablenotemark{e}} &
 \multicolumn{1}{c}{$\langle \beta\rangle$\tablenotemark{f}}  &
 \multicolumn{1}{c}{$\langle \beta^2\rangle$\tablenotemark{g}} 
\\
 \colhead{} & 
 \multicolumn{1}{c}{(AB mag)} &
 \colhead{} &
 \multicolumn{1}{c}{(arcmin$^{-2}$)} &
 \colhead{} &
 \colhead{} & 
 \colhead{} &
 \colhead{} 
} 
\startdata  
 Red      & $21.0<z'<25.5$ &  8676 & 10.9&  0.447 &
$1.14 \pm 0.09$ &  $1.09 \pm 0.07$ 
 & $0.56\pm 0.02 $ & $0.32$ \\
 Green    & $17.4<z'<25.5$ &  1655 & 2.1 & 0.396 &
$0.46\pm 0.02$ & $0.45\pm 0.02$ & $0.11\pm 0.01$ & $0.03$ \\
 Blue     & $23.0<z'<25.5$ &  5004 & 6.3 & 0.463 &
$ 1.81 \pm 0.17$ & $1.75^{+0.47}_{-0.30}$  & $0.68 \pm 0.04 $ & $0.47$ \\
 Blue+red & ---                        & 13680 & 17.2& 0.453 &
$1.31 \pm 0.09$& $1.29^{+0.16}_{-0.14}$ & $0.61\pm 0.03$ & 0.38 \\
\enddata 
\tablenotetext{a}{Magnitude limits for the galaxy sample.}
\tablenotetext{b}{Mean surface number density of background galaxies in
 the Cl0024+1654 field.} 
\tablenotetext{c}{Mean rms error for the shear estimate per galaxy, 
 $\overline{\sigma}_g\equiv (\overline{\sigma_g^2})^{1/2}$}
\tablenotetext{d}{Mean redshift of the background sample derived with
 the COSMOS photometric catalog.} 
\tablenotetext{e}{Effective source redshift corresponding to the mean
 depth $\langle\beta\rangle$ of the COSMOS background sample.}
\tablenotetext{f}{Distance ratio
 averaged over the COSMOS redshift distribution of the background sample,
 $\langle\beta\rangle=\langle D_{ds}/D_{s}\rangle$.}
\tablenotetext{g}{Distance ratio squared 
averaged over the COSMOS redshift distribution of the background sample,
$\langle\beta^2\rangle=\langle(D_{ds}/D_{s})^2\rangle$.}
\end{deluxetable}


\begin{deluxetable}{ccccccccccc}
\tabletypesize{\footnotesize}
\tablecolumns{10}
\tablecaption{
 \label{tab:nfw}
Summary of the best-fitting NFW/gNFW parameters
} 
\tablewidth{0pt} 
\tablehead{ 
 \multicolumn{1}{c}{Method} &
 \multicolumn{1}{c}{WL\tablenotemark{a}} &
 \multicolumn{1}{c}{SL\tablenotemark{b}} &
 \multicolumn{1}{c}{E.R.\tablenotemark{c}} &
 \multicolumn{1}{c}{$(\theta_{\rm lo}, \theta_{\rm up})$\tablenotemark{d}}&
 \multicolumn{1}{c}{$\alpha$\tablenotemark{e}} &
 \multicolumn{1}{c}{$M_{\rm vir}$\tablenotemark{f}} &
 \multicolumn{1}{c}{$c_{-2}$\tablenotemark{g}} &
 \multicolumn{1}{c}{$\chi^2_{\rm min}/{\rm dof}$} &
 \multicolumn{1}{c}{$\theta_{\rm E}$\tablenotemark{h}}  
\\
 \colhead{} &
 \colhead{} &
 \colhead{} &
 \colhead{} &
 \multicolumn{1}{c}{($\arcmin$)}&
 \colhead{} &
 \multicolumn{1}{c}{$(10^{15}M_\odot/h)$} &
 \colhead{} &
 \colhead{} &
 \multicolumn{1}{c}{(\arcsec)} 
}  
\startdata  
 Strong lensing & --- & 1D $\kappa$ & --- & $(0.17, 0.79)$ & 1 &
 $1.52^{+0.39}_{-0.27}$ &  $7.3^{+1.4}_{-1.3}$ & 0.71/14 & $34^{+13}_{-10}$\\
 Tangential shear & 1D $g_{+}$ & --- & --- & $(0.55,16)$ & 1 &
 $1.14^{+0.22}_{-0.19}$ &  $10.6^{+2.9}_{-2.0}$ & 2.2/8 & $37^{+11}_{-10}$\\
 & 1D $g_{+}$ & --- & Yes & $(0.50,16)$& 1 &
 $1.19^{+0.23}_{-0.20}$ & $8.6^{+1.8}_{-1.4}$ & 4.9/9 & $32^{+10}_{-9}$\\
 $\zeta_{\rm c}$-Statistic & 1D $\kappa$ &  --- & --- & $(0.55,12)$ & 1 &
 $1.08^{+0.21}_{-0.19}$ & $9.7^{+14.6}_{-4.6}$ & $3.6/8$ &
 $33^{+25}_{-20}$\\ 
                           & 1D $\kappa$ &  --- & Yes & $(0.50,12)$& 1 &
 $1.08^{+0.22}_{-0.19}$ & $8.6^{+2.1}_{-1.7}$ & $3.6/9$ &
 $30^{+11}_{-10}$\\ 
                          & 1D $\kappa$ &  1D $\kappa$ & --- & $(0.17,12)$
 & 1 &
 $1.15^{+0.18}_{-0.15}$ & $9.2^{+1.4}_{-1.2}$ & $4.1/20$ &
 $33^{+8}_{-7}$\\ 
                          & 1D $\kappa$ &  1D $\kappa$ & --- & $(0.17,12)$
 & $0.04^{+0.93}_{-0.04}$ &
 $1.06^{+0.19}_{-0.16}$ & $9.8\pm 1.2$ & $3.8/20$ &
 $32^{+9}_{-8}$\\ 
\enddata
\tablenotetext{a}{Type of Subaru weak lensing (WL) data.}
\tablenotetext{b}{Fitting with or without the inner $\kappa$ profile
 derived from the strong lensing (SL) analysis of \cite{Zitrin+2009_CL0024}.}
\tablenotetext{c}{Fitting with or without the inner Einstein-radius
 (E.R.) constraint, $\theta_{\rm E}=30\arcsec$ at 
 $z=1.675$. A $10\%$ error is assumed for $\theta_{\rm E}$.}
\tablenotetext{d}{{}Lower and upper radial limits of 
lensing constraints used for fitting.} 
\tablenotetext{e}{Central cusp slope and its 68.3\%
 confidence interval including the uncertainty in the source redshift
 calibration. For NFW models, $\alpha$ is fixed at $\alpha=1$.}
\tablenotetext{f}{Virial mass and its 68.3\% confidence interval
 including the uncertainty in the source redshift calibration.
} 
\tablenotetext{g}{Halo concentration,
$c_{-2}= c_{\rm vir}/(2-\alpha)$,
and its 68.3\% confidence interval
 including the uncertainty in the source redshift calibration.
}
\tablenotetext{h}{
Einstein radius in units of arcseconds for a background source at
$z=1.675$ as predicted by the best-fit NFW model.
}
\end{deluxetable}


\begin{deluxetable}{cccccc}
\tablecolumns{6} 
\tablecaption{
 \label{tab:2dshear}
Best-fit parameters for the two-component NFW lens model
}  
\tablewidth{0pt} 
\tablehead{ 
 \multicolumn{1}{c}{Halo component} &
 \multicolumn{1}{c}{Profile} &
 \multicolumn{1}{c}{$M_{\rm vir}$} &
 \multicolumn{1}{c}{$c_{\rm vir}$} &
 \multicolumn{1}{c}{$\Delta {\rm R.A.}$} & 
 \multicolumn{1}{c}{$\Delta {\rm Dec.}$} 
\\
 \colhead{} & 
 \colhead{} &
 \multicolumn{1}{c}{($10^{15}M_\odot/h$)} &
 \colhead{} &
 \multicolumn{1}{c}{($\arcmin$)} &
 \multicolumn{1}{c}{($\arcmin$)}  
} 
\startdata  
Central         & NFW  & $1.11\pm 0.18$ & $8.1\pm 1.2$ & 0.0 & 0.0\\
Northwest       & tNFW & $0.128 \pm 0.051$  & $5.0\pm 3.5$ & -2.3 & 2.5 \\
\enddata 
\tablecomments{
Shown are the best-fit parameters and their $1\sigma$ errors for the
two-component NFW lens model (\S \ref{subsec:2dshear}) derived from a
joint fit to the inner $\kappa$ profile from strong lensing
and the outer Subaru two-dimensional distortion data. The quoted errors
include the systematic uncertainty in the redshift distribution of the
background galaxies (Table \ref{tab:color}). 
The central-halo centroid is fixed at the dark-matter center of mass
(\S \ref{subsec:center}). The northwest-halo centroid is fixed at the
peak position of the northwest galaxy clump in the surface number
density distribution of $BR_{\rm c}z'$-selected cluster galaxies.
We have applied a prior to the central halo to represent the constraints
from inner strong-lensing information at $\theta\simlt 48\arcsec$. 
A truncated form of the NFW profile (tNFW) is used for
describing the projected lensing fields of the northwest halo.
The resulting $\chi^2$ value is $\chi^2_{\rm min}\simeq 36761$
with $23306$ degrees of freedom.
}
\end{deluxetable}


\begin{deluxetable}{ccc}
\tablecolumns{3} 
\tablecaption{
 \label{tab:mdelta}
Three-dimensional cluster mass from a deprojection analysis
}  
\tablewidth{0pt} 
\tablehead{ 
 \multicolumn{1}{c}{$\Delta$} &
 \multicolumn{1}{c}{$M_{\Delta}$} &
 \multicolumn{1}{c}{$r_{\Delta}$} 
\\
 \colhead{} & 
 \multicolumn{1}{c}{($10^{15}M_\odot/h$)} &
 \multicolumn{1}{c}{(Mpc$/h$)} 
} 
\startdata  
2500    & $0.42^{+0.08}_{-0.09}$ & $0.46^{+0.03}_{-0.04}$\\
 500    & $0.72^{+0.17}_{-0.20}$ & $0.93^{+0.07}_{-0.10}$\\
 200    & $1.17^{+0.23}_{-0.27}$& $1.49^{+0.09}_{-0.13}$\\
 Virial & $1.21^{+0.18}_{-0.22}$& $1.75^{+0.08}_{-0.11}$
\enddata 
\end{deluxetable}





\clearpage

\begin{figure}[!htb] 
 \begin{center}
    \includegraphics[width=120mm,angle=0]{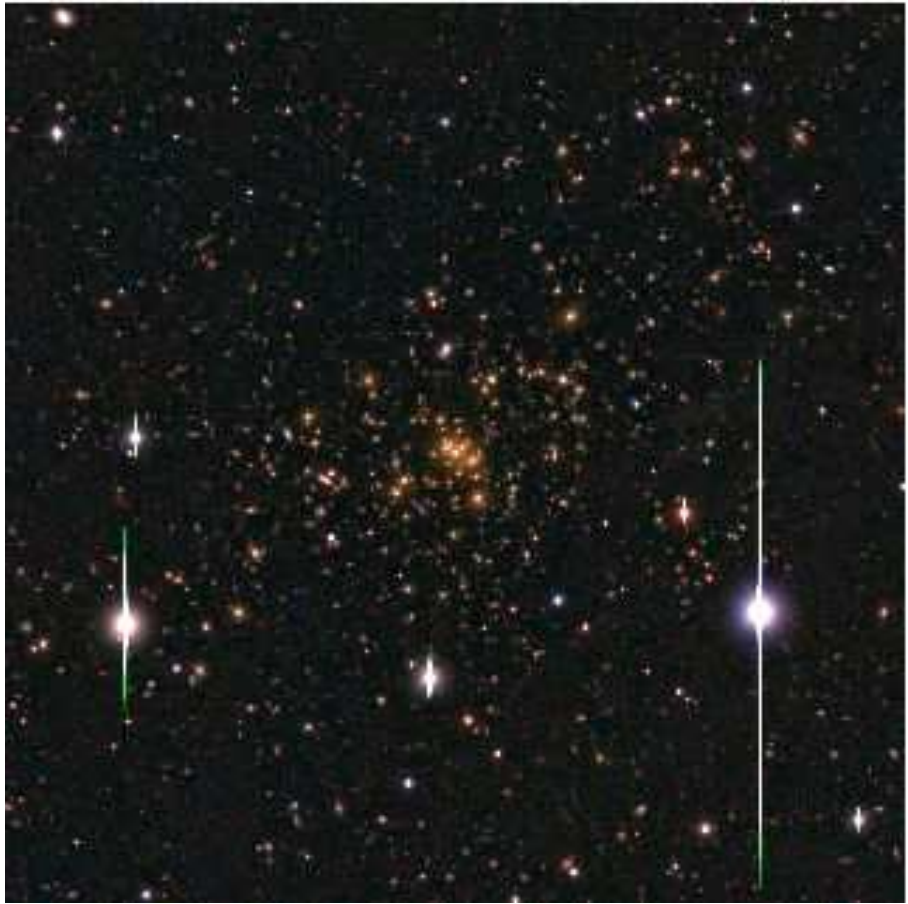}
 \end{center}
\caption{
\label{fig:BRZ}
Subaru $BRz'$ pseudo-color image of the central $8\arcmin\times
 8\arcmin$ region of the galaxy cluster Cl0024+1654 at $z=0.395$.
The side length of the field is $1.8$\,Mpc\,$h^{-1}$ at the cluster
 redshift. 
} 
\end{figure}


\clearpage

\begin{figure}[!htb]
 \begin{center}
    \includegraphics[width=100mm,angle=0]{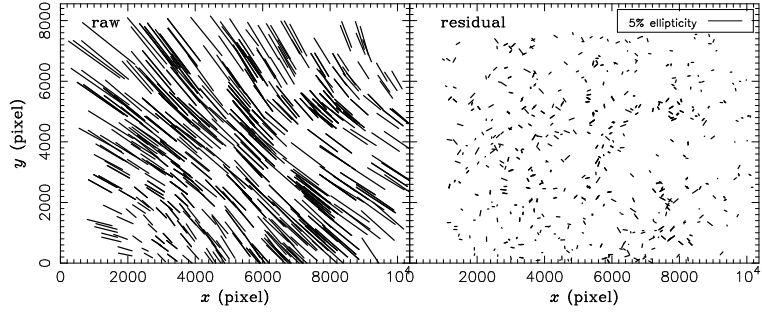}
 \end{center}
\caption{
\label{fig:anisopsf1}
The quadrupole PSF anisotropy field 
as measured from 
stellar ellipticities before and after the PSF anisotropy correction.
The left panel shows the raw ellipticity field of stellar objects,
and the right panel shows the residual ellipticity field after
the PSF anisotropy correction.
The orientation of the sticks indicates the position angle of
the major axis of stellar ellipticity, whereas the length is
 proportional to the modulus of stellar ellipticity. A stick with the
 length of $5\%$ ellipticity is indicated in the top right of the right
 panel. 
} 
\end{figure}


 
\begin{figure}[!htb]
 \begin{center}
    \includegraphics[width=100mm,angle=0]{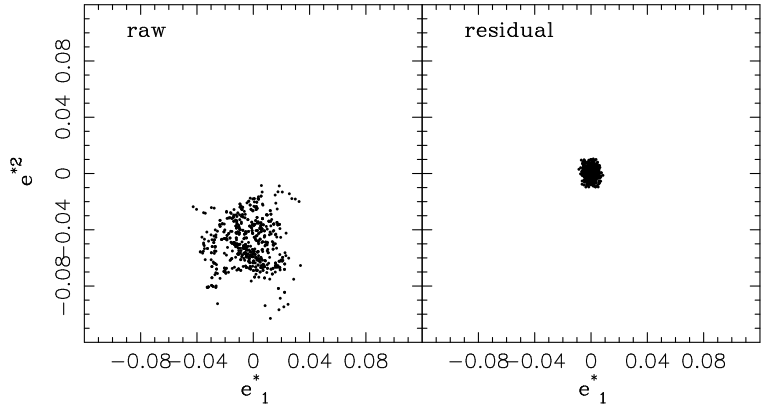}
 \end{center}
\caption{
\label{fig:anisopsf2}
Stellar ellipticity distributions before and after the PSF anisotropy 
correction.
The left panel shows the raw ellipticity components 
$(e_1^*,e_2^*)$ of stellar objects, and the right panel shows
the residual ellipticity components $(\delta e_1^*, \delta e_2^*)$
after the PSF anisotropy correction.
} 
\end{figure}



\begin{figure}[htb]
 \begin{center}
  \includegraphics[width=100mm, angle=0]{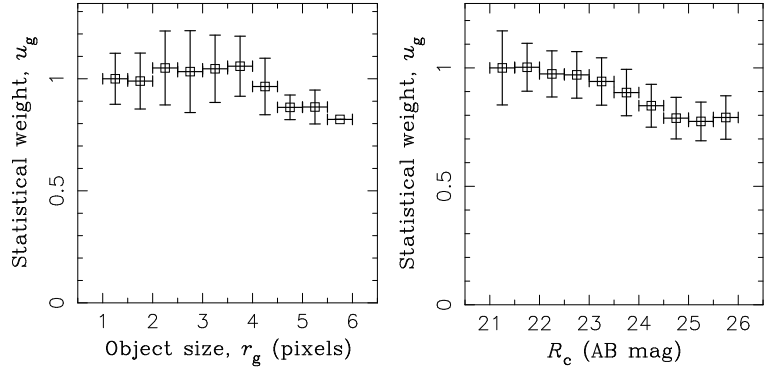}
 \end{center}
\caption{
Mean statistical weight $u_g$ as measured
 from the magnitude-selected galaxy sample,
shown as  a function of the object Gaussian size $r_g$ ({\it left}) and
 of the $R_{\rm   c}$ magnitude ({\it right}). 
In each of the panels, the statistical weight
 $u_g$ is normalized to unity in the first bin.
\label{fig:weight}
}
\end{figure}


\clearpage

\begin{figure}[!htb]
 \begin{center}
    \includegraphics[width=120mm,angle=0]{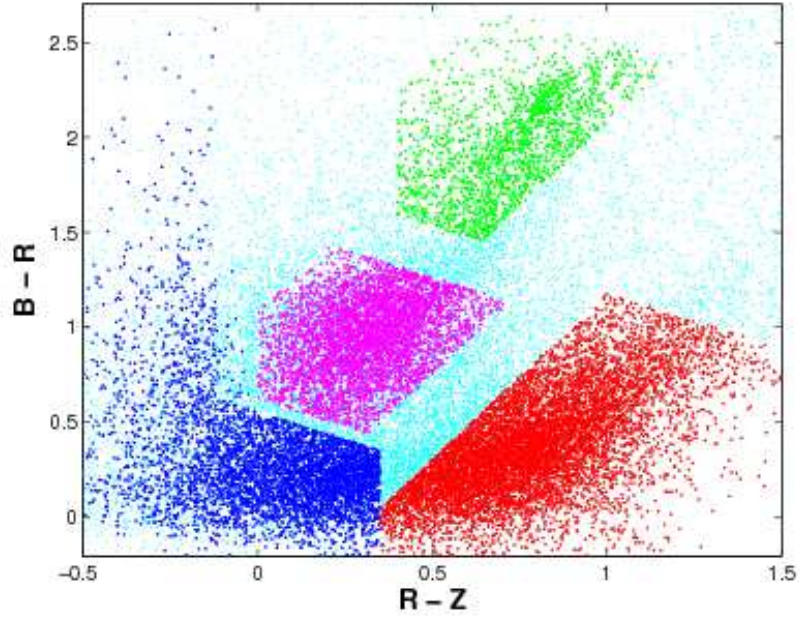}
 \end{center}
\caption{
\label{fig:cc}
Sample selection in the color-color diagram of Cl0024+1654, displaying
the green sample ({\it green points}), 
comprising mostly cluster member galaxies, and the red ({\it red points})
 and blue ({\it blue points}) samples, comprising of background
 galaxies.  
The galaxies that we identify as predominantly foreground lie in between the
 cluster and background galaxies are marked in {\it magenta}. 
} 
\end{figure}



\begin{figure}[!htb]
 \begin{center}
    \includegraphics[width=75mm,angle=0]{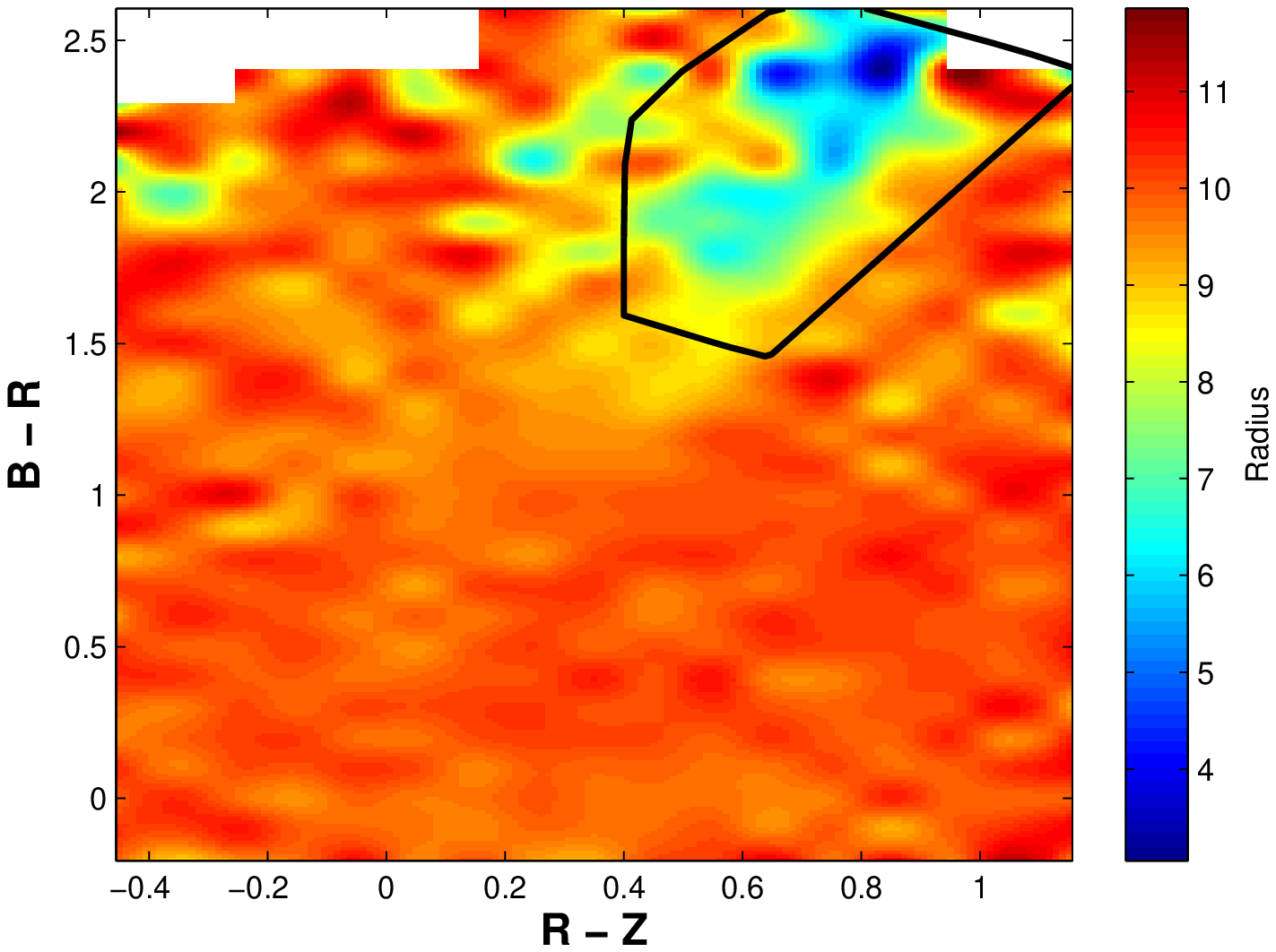}
    \includegraphics[width=75mm,angle=0]{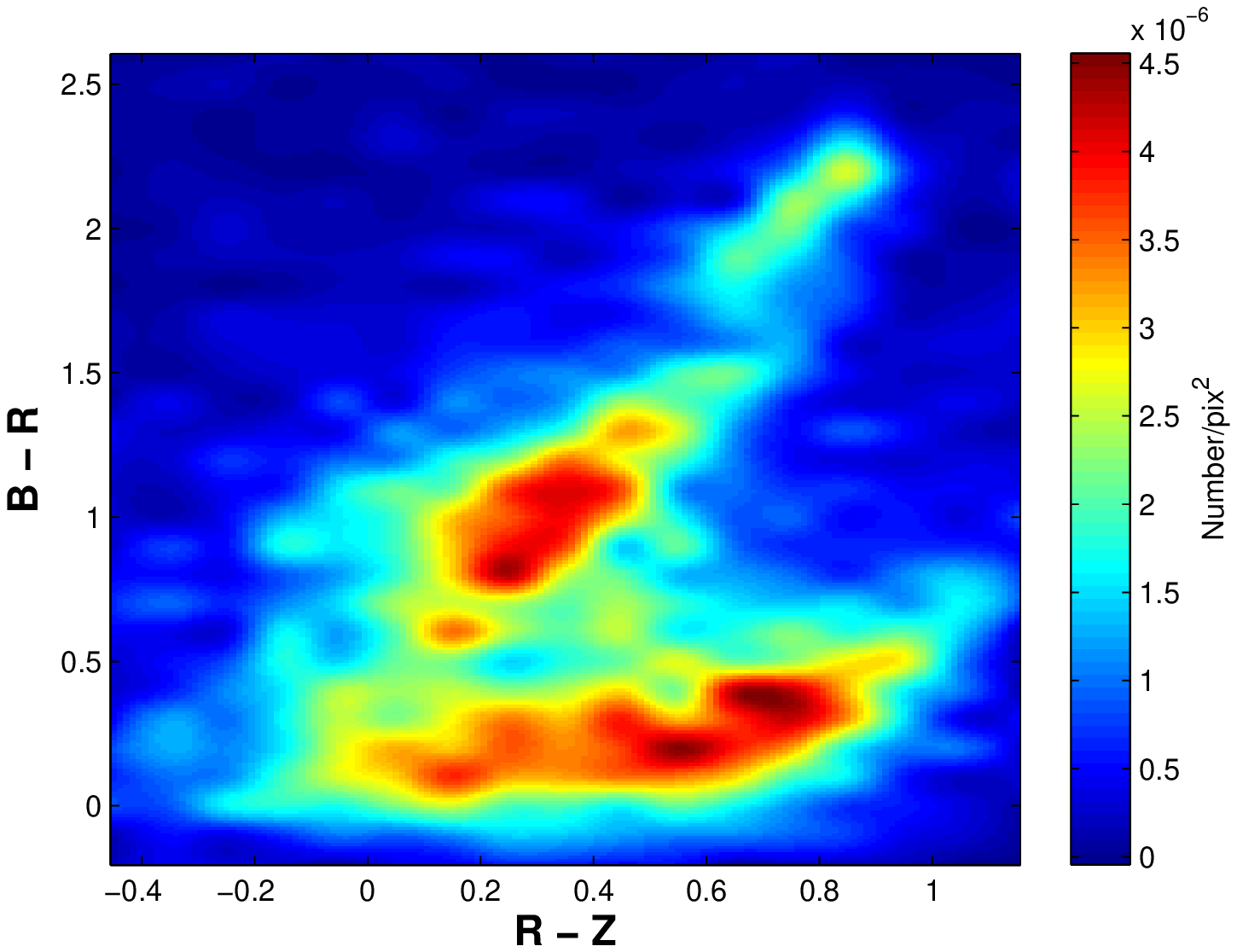}
 \end{center}
\caption{
\label{fig:ccr}
{\it Left}:
Distribution of mean projected distances from the cluster center for all
 galaxies in the Cl0024+1654 field, displayed in color-color (CC) space.
The bluer colors imply smaller mean radii, hence
correspond to the location of the cluster in CC space. The black
box marks the boundaries of the green galaxy sample we select which
conservatively includes all cluster members.
{\it Right}:
Number density of galaxies in CC space.
Several distinct density peaks
are shown to be different galaxy
populations - the reddest peak in the upper-right corner of the
plot depicts the overdensity of cluster galaxies, whose colors are
lying on the red sequence; the middle peak with colors bluer than
the cluster shows the overdensity of foreground galaxies; the 
remaining 
peaks in the bottom part (bluest in $B-R_{\rm c}$) can be demonstrated to
comprise of blue and red (left and right, respectively) background
galaxies.
} 
\end{figure}


\begin{figure}[htb]
 \begin{center}
  \includegraphics[width=100mm, angle=0]{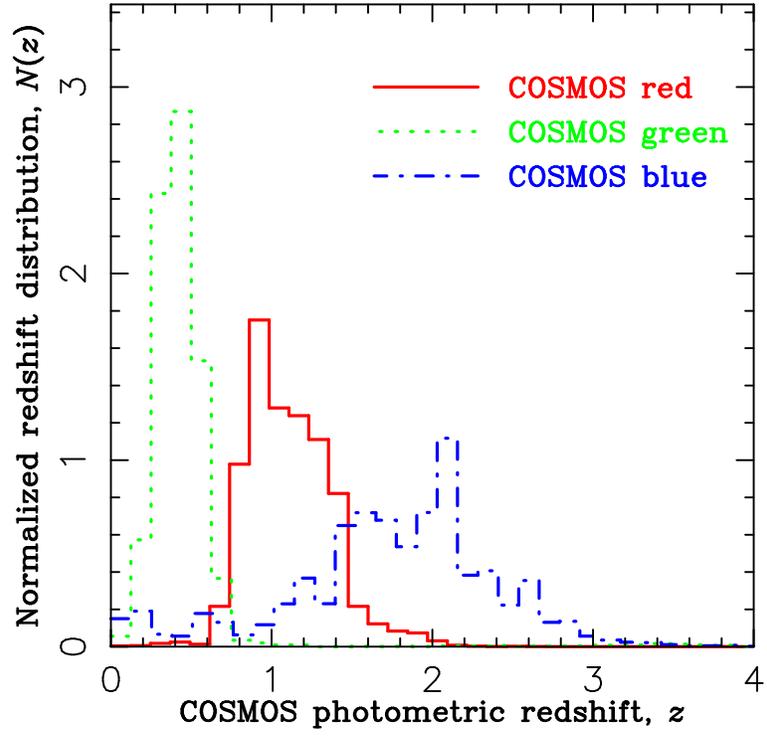}
 \end{center}
\caption{
Photometric redshift distributions $N(z)$ of $BR_{\rm c}z'$ selected
galaxy samples in the 2-deg$^2$ COSMOS field. 
The green ({\it green dotted line}), red ({\it red solid line}), and
 blue ({\it blue dotted-dashed line}) galaxy samples are selected
 according to color-color/magnitude limits used in the weak-lensing
 analysis of Cl0024+1654.
Each distribution is normalized by $\int\!dz\,N(z)=1$.
\label{fig:nz}
}
\end{figure}


\clearpage

\begin{figure}[htb] 
 \begin{center} 
 \includegraphics[width=100mm, angle=0]{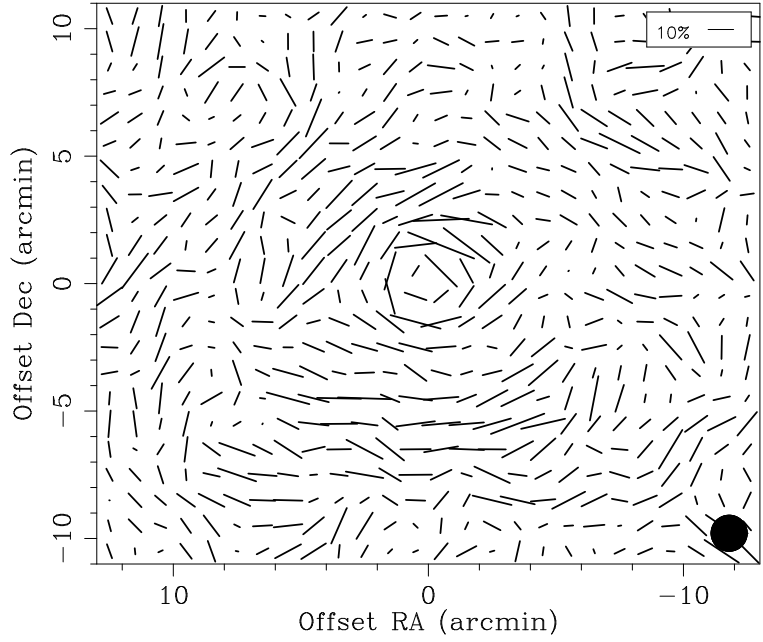}
 \end{center}
\caption{
Gravitational reduced-shear field in Cl0024+1654 obtained from shape
distortions of the blue+red background galaxies, 
smoothed with a Gaussian with
${\rm FWHM}=1.4\arcmin$
for visualization purposes.
A stick with a length of $10\%$ shear is
indicated in the top right corner. 
The filled circle indicates the FWHM of the
Gaussian. The coordinate origin is at the center of the cD galaxy.
\label{fig:gmap}
}
\end{figure}



\begin{figure}[htb]   
 \begin{center}
  \includegraphics[width=140mm, angle=0]{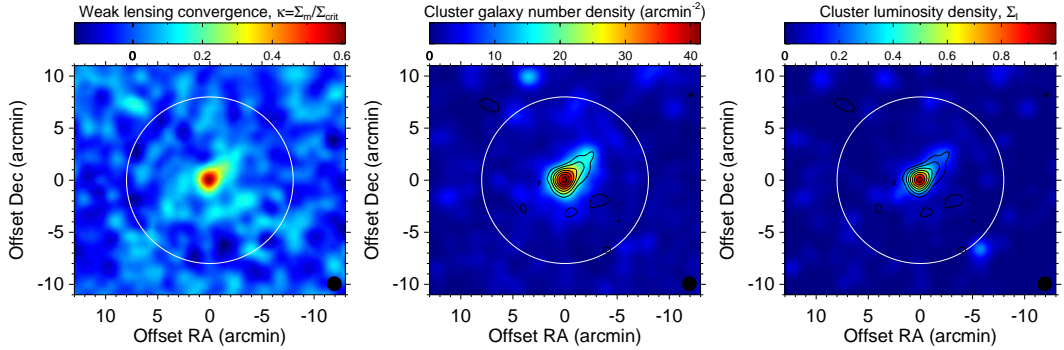}
 \end{center}
\caption{
Comparison of the surface mass density field and the cluster
 galaxy distributions in Cl0024+1654.
{\it Left}: Dimensionless surface mass density field, or the lensing
 convergence  $\kappa(\btheta)=\Sigma_{m}(\btheta)/\Sigma_{\rm crit}$,
 reconstructed from Subaru distortion data.
{\it Middle}: Observed surface number density distribution
 $\Sigma_n(\btheta)$ of $BR_{\rm  c}z'$-selected {\it green} galaxies
 ($17.3<z'<25.5$ AB mag), representing unlensed cluster member galaxies.
{\it Right}: Observed $R_{\rm c}$-band surface luminosity density
 distribution $\Sigma_l(\btheta)$ of the same cluster membership.
The solid circle in each panel indicates the cluster virial
 radius of $\theta_{\rm vir}\simeq 8\arcmin$, or 
 $r_{\rm vir}\simeq 1.8$\,Mpc\,$h^{-1}$ at the cluster redshift of
 $z=0.395$. 
All images are smoothed with a circular Gaussian of FWHM $1.4\arcmin$. 
Also overlaid on the 
$\Sigma_{n}(\btheta)$ and  $\Sigma_l(\btheta)$ maps 
are the $\kappa(\btheta)$ field shown in the left
 panel, given in units of $2\sigma$ reconstruction error from the lowest
 contour level of $3\sigma$.
The field size is $26\arcmin \times 22\arcmin$. 
North is to the top, east to the left.
\label{fig:knl}
}
\end{figure}


\clearpage

\begin{figure}[htb]
 \begin{center}
   \includegraphics[width=70mm, angle=0]{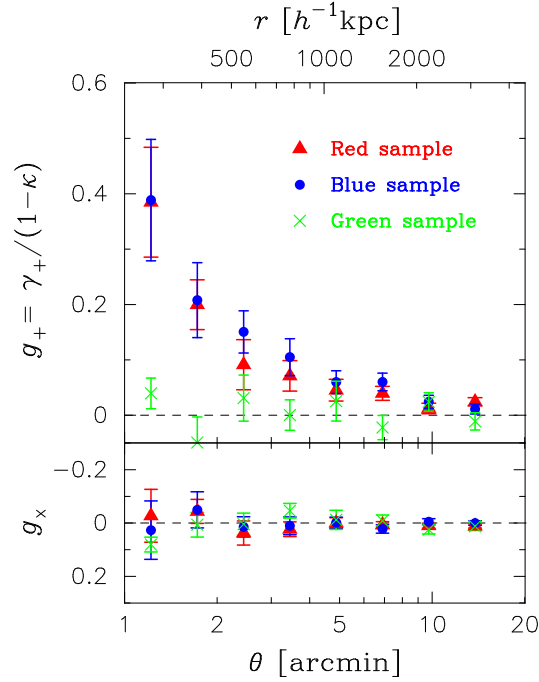}
 \end{center}
\caption{
Azimuthally-averaged radial profiles of the tangential reduced shear 
$g_+$ ({\it upper panel}) and the $45^\circ$ rotated ($\times$)
 component $g_{\times}$ ({\it lower panel})
for our red ({\it triangles}), blue ({\it circles}), 
and green ({\it crosses}), galaxy samples.
The error bars represent $68.3\%$ confidence intervals estimated by 
bootstrap resampling techniques. The red and blue populations show a
 very similar form of the radial distortion profile which declines
 smoothly from the cluster center, remaining positive to the limit of
 our data, $\theta_{\rm max}=16\arcmin$. 
For all of the samples, the $\times$-component is consistent with a null
 signal at all radii, indicating the reliability of our distortion
 analysis. 
\label{fig:rgb}
}
\end{figure}  
 


\begin{figure}[htb]
 \begin{center}
   \includegraphics[width=70mm, angle=0]{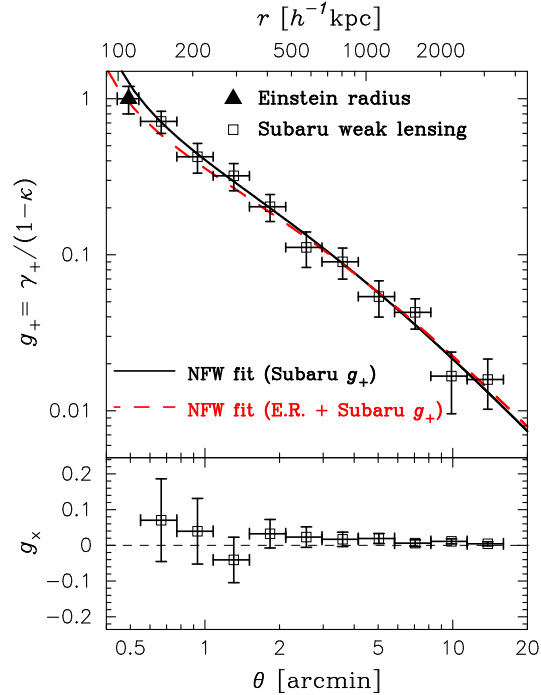}
 \end{center}
\caption{
Azimuthally-averaged radial profile of the tangential reduced shear $g_+$
({\it upper panel}) as measured from Subaru distortion data of our
 composite blue+red background sample. The Einstein radius constraint
 ({\it triangle}) of $\theta_{\rm E}=30\arcsec \pm 3\arcsec$ (at
 $z_s=1.675$), determined from multiply lensed images in HST/ACS/NIC3
 observations \citep{Zitrin+2009_CL0024}, is translated to the
 corresponding depth of the Subaru blue+red background sample (Table
 \ref{tab:color}), using the best-fit NFW model to the Subaru and
 ACS/NIC3 data (see Figure \ref{fig:kprof}), and added to the distortion
 profile ($g_+=1$), marking the point of maximum distortion.
The solid curve shows the best-fit NFW profile for the Subaru $g_{+}$
 measurements. The dashed curve shows the NFW profile from a joint fit
 to the inner Einstein-radius constraint and the outer Subaru $g_+$
 profile. Shown in the bottom panel is the $45^\circ$-rotated
 $g_{\times}$ component. 
\label{fig:gt}
}
\end{figure}


\clearpage

\begin{figure}[htb]
 \begin{center}
   \includegraphics[width=120mm, angle=0]{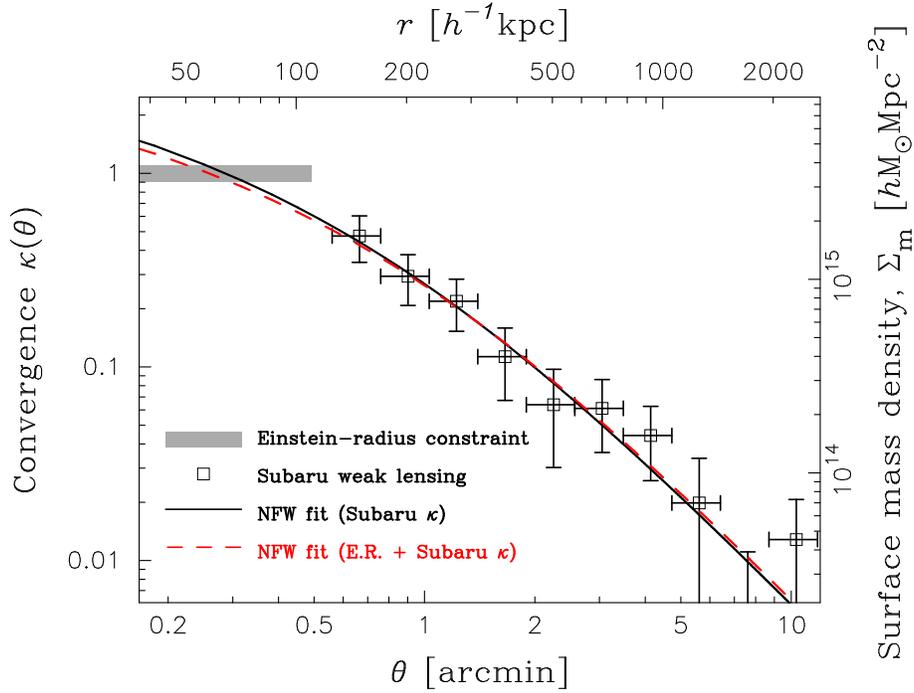}
 \end{center}
\caption{
Radial profile of lensing convergence
 $\kappa(\theta)=\Sigma_{m}(\theta)/\Sigma_{\rm crit}$ as reconstructed
 from Subaru distortion data. The open squares show the results from a
 non-linear extension of aperture mass densitometry based on
 azimuthally-averaged tangential distortion measurements. The error bars
 are correlated.  
The gray-shaded region represents the Einstein-radius constraint on the 
mean interior convergence, $\bar{\kappa}(<\theta_{\rm E})=1$, translated
 into the corresponding depth of the Subaru blue+red background sample. 
The dashed and solid curves are the best-fitting NFW profiles from the
 Subaru $\kappa$ data with and without the inner Einstein radius
 constraint combined, respectively. 
\label{fig:kappa}
}
\end{figure}


  
\begin{figure}[htb]  
 \begin{center}
   \includegraphics[width=120mm, angle=0]{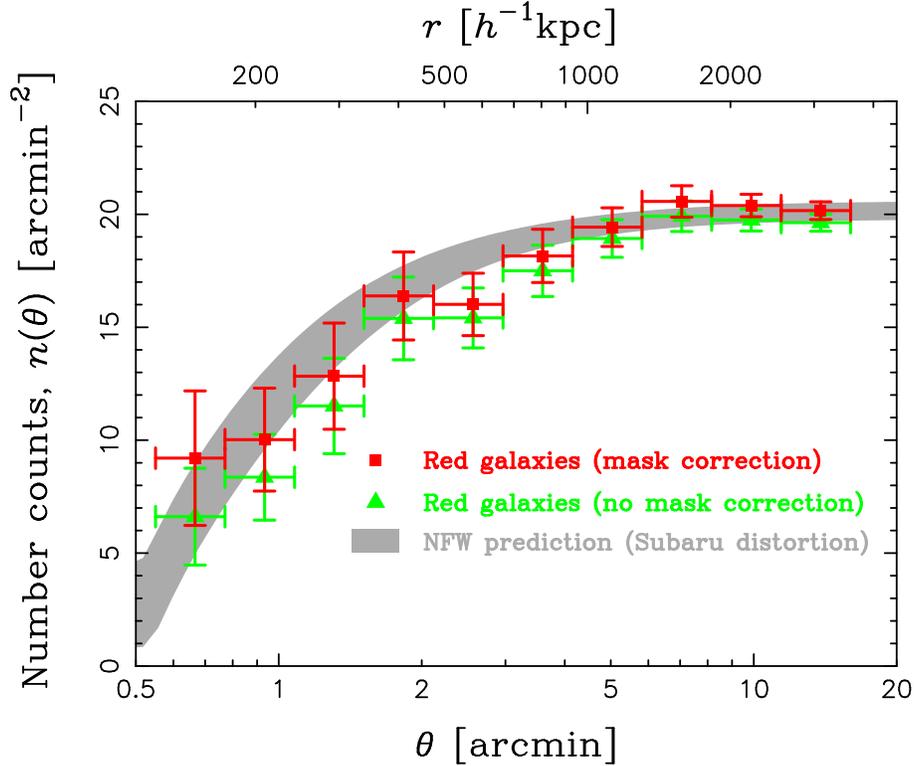}
 \end{center}
\caption{
Number-count profile of $BR_{\rm c}z'$-selected red galaxies ({\it
 squares}) in the background of Cl0024+1654. The triangles show the
 counts without the mask correction due to cluster members and bright
 foreground objects. 
The gray-filled region represents the $68.3\%$ confidence bounds for  
the predicted count depletion curve from an NFW model constrained by our
 Subaru distortion analysis, demonstrating clear consistency between
 these two independent lensing observables.
\label{fig:magbias}
} 
\end{figure}


\clearpage

\begin{figure}[htb]
 \begin{center}
   \includegraphics[width=150mm, angle=0]{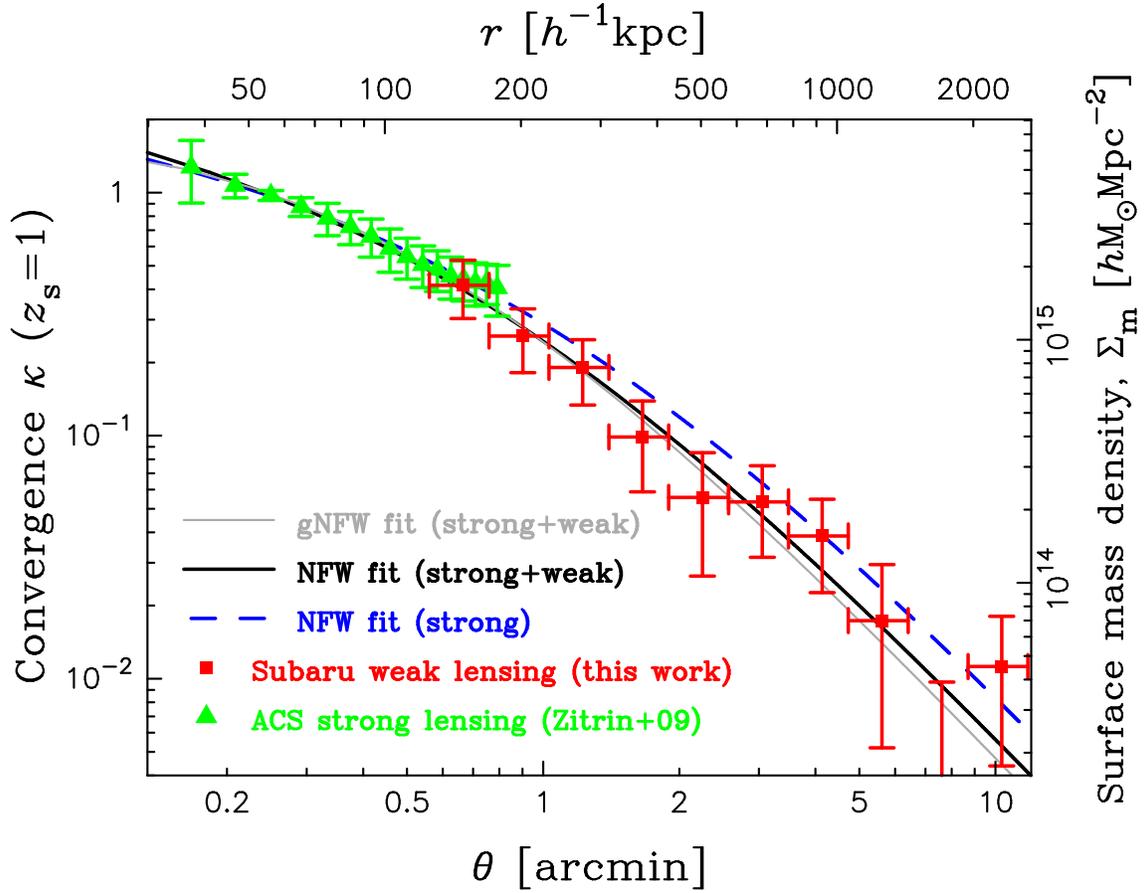}
 \end{center}
\caption{
Radial surface mass density profile of the galaxy cluster
 Cl0024+1654 over a wide range of radius from 40 to 2300\,kpc$\,h^{-1}$
reconstructed from our joint weak and strong lensing
 analysis of Subaru and ACS/NIC3 observations.
All of the radial profiles are scaled to a fiducial redshift of $z_s=1$.
The squares represent our Subaru results (this work) from 
a one-dimensional reconstruction by 
a non-linear extension of aperture mass densitometry
based on azimuthally-averaged tangential distortion measurements.
The error bars are correlated. 
The triangles represent the
inner $\kappa$ profile derived from the strong-lensing analysis 
of \cite{Zitrin+2009_CL0024} based on 33 multiply-lensed images, 
spread fairly evenly over the central region, 
$8\arcsec\simlt \theta \simlt  48\arcsec$.  
The thick ({\it black}) and thin ({\it gray})
solid curves show the best-fitting NFW and gNFW profiles, respectively,
from our  full lensing analysis of ACS/NIC3 and Subaru observations.
The best-fitting NFW profile from the inner 
$\kappa$ profile from the ACS/NIC3 observations
is also shown as a ({\it blue}) dashed curve.
The Subaru weak-lensing 
constraint at the innermost radius 
is fully consistent with the strong-lensing
 information within the measurement uncertainty.
\label{fig:kprof}
}
\end{figure}  


\clearpage

\begin{figure}[htb]
 \begin{center}
   \includegraphics[width=120mm, angle=0]{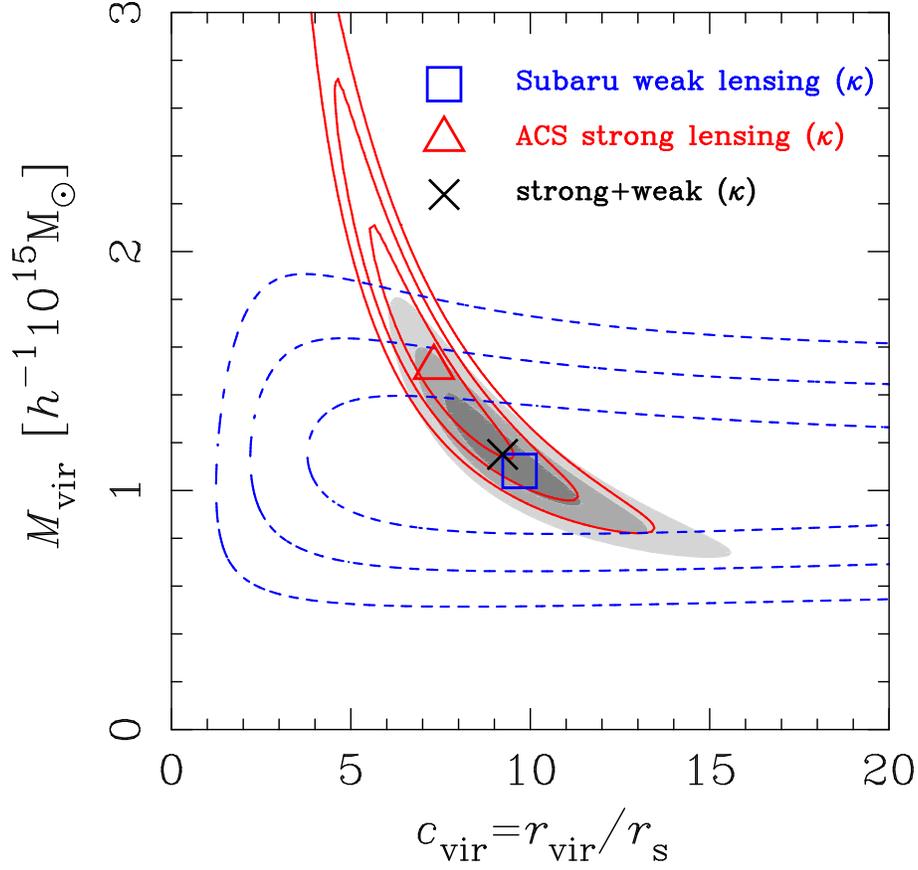}
 \end{center}
\caption{
Joint constraints on the NFW model parameters 
$(c_{\rm vir}, M_{\rm vir})$
for Cl0024+1654 derived from the radial profile of lensing convergence, 
$\kappa(\theta)=\Sigma_{m}(\theta)/\Sigma_{\rm crit}$ (see Figure
 \ref{fig:kprof}). 
The open square shows the best-fit set of the NFW model parameters, and
 the dashed contours show the 68.3\%, 95.4\%, and 99.7\% confidence
 levels in the $c_{\rm vir}$-$M_{\rm vir}$ plane. 
The solid contours show the same confidence levels, but for an NFW fit
 to the inner ($10\arcsec\simlt \theta\simlt 48\arcsec$) $\kappa$
 profile constrained by deep ACS/NIC3 observations of
 \cite{Zitrin+2009_CL0024}.  
The corresponding best-fit set of  $(c_{\rm vir}, M_{\rm  vir})$ is
 shown by the open triangle. Also shown by the filled gray areas are the
 same confidence areas, but obtained for a joint fit to the combined
 ACS/NIC3 and Subaru $\kappa(\theta)$ data. The corresponding best-fit
 set of  $(c_{\rm vir}, M_{\rm  vir})$ is shown by the cross.
\label{fig:CM_kappa}
}
\end{figure}



\begin{figure}[htb]
 \begin{center}
   \includegraphics[width=120mm, angle=0]{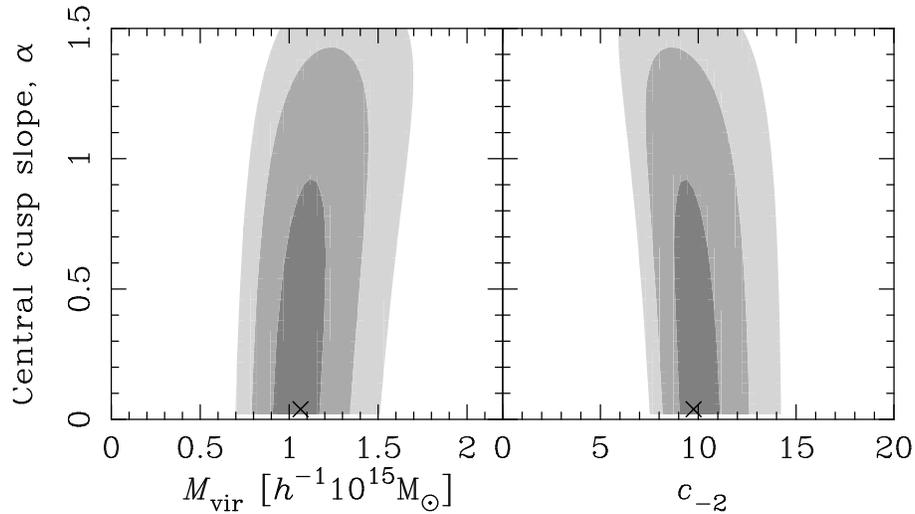}
 \end{center}
\caption{
Constraint on the the gNFW model parameters, namely the central cusp
 slope $\alpha$, the halo virial mass $M_{\rm vir}$, and the halo
 concentration $c_{-2}=c_{\rm vir}/(2-\alpha)$, when all of them are
 allowed to vary, derived from the full lensing profile of CL0024+1654
 shown in Figure \ref{fig:kprof}. 
The left and right panels show the two-dimensional marginalized
 constraints on $(M_{\rm vir},\alpha)$ and $(c_{-2},\alpha)$,
 respectively. In each panel of the figure, the contours show the
 68.3\%, 95.4\%, and 99.7\% confidence levels, and the cross indicates
 the best-fit model parameters. 
\label{fig:CM_gNFW}
}
\end{figure}


\clearpage

\begin{figure}[htb]
 \begin{center}
   \includegraphics[width=150mm, angle=0]{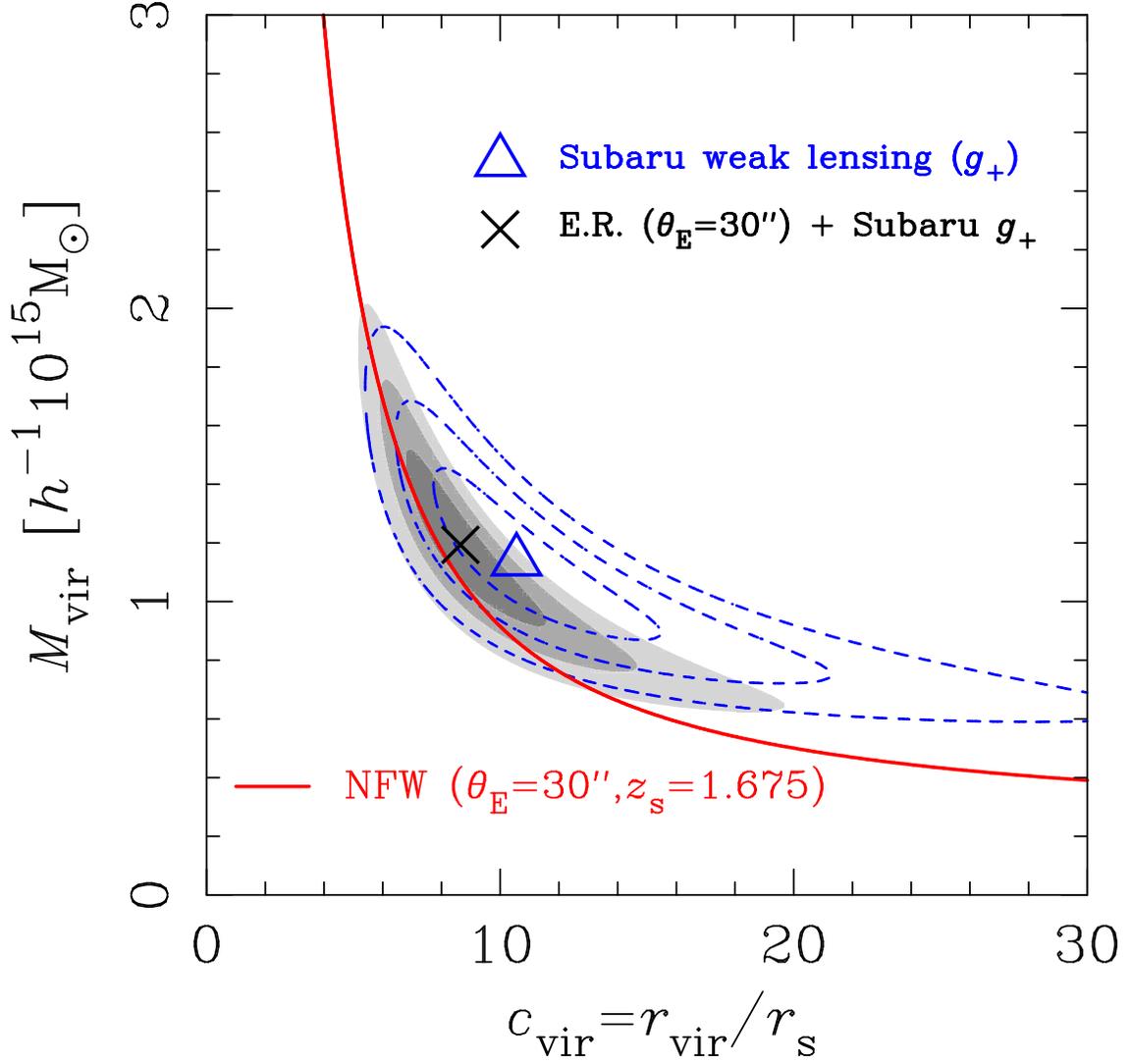}
 \end{center}
\caption{
Joint constraints on the NFW model parameters 
$(c_{\rm vir}, M_{\rm  vir})$ derived from Subaru distortion data. The
 dashed contours show the 68.3\%, 95.4\%, and 99.7\% confidence levels
 in the  $c_{\rm vir}$-$M_{\rm vir}$ plane, estimated from
 $\Delta\chi^2\equiv \chi^2-\chi^2_{\rm min}=2.3, 6.17$, and $11.8$,
 respectively, for the tangential distortion ($g_+$) profile (Figure
 \ref{fig:gt}). 
Also shown by the filled gray areas are the same confidence regions, 
but obtained for a joint fit to the outer Subaru $g_+$ profile and 
the strong-lensing constraint on the location of the Einstein radius,
$\theta_{\rm E}= 30\arcsec\pm 3\arcsec$ at $z_s=1.675$.
The open triangle shows the best-fit set of the NFW model parameters, 
$(c_{\rm vir}, M_{\rm  vir})$, for  the Subaru $g_+$ results. The cross
 shows the best-fit set of $(c_{\rm vir}, M_{\rm  vir})$ for the
 combined Subaru $g_+$ and Einstein-radius  constraints. 
The solid curve shows the NFW $c_{\rm vir}$-$M_{\rm vir}$ relation for
 $\theta_{\rm E}=30\arcsec$ at $z_s=1.675$. 
\label{fig:CM_gt}
}
\end{figure}


\clearpage

\begin{figure}[htb]
 \begin{center}
   \includegraphics[width=110mm, angle=0]{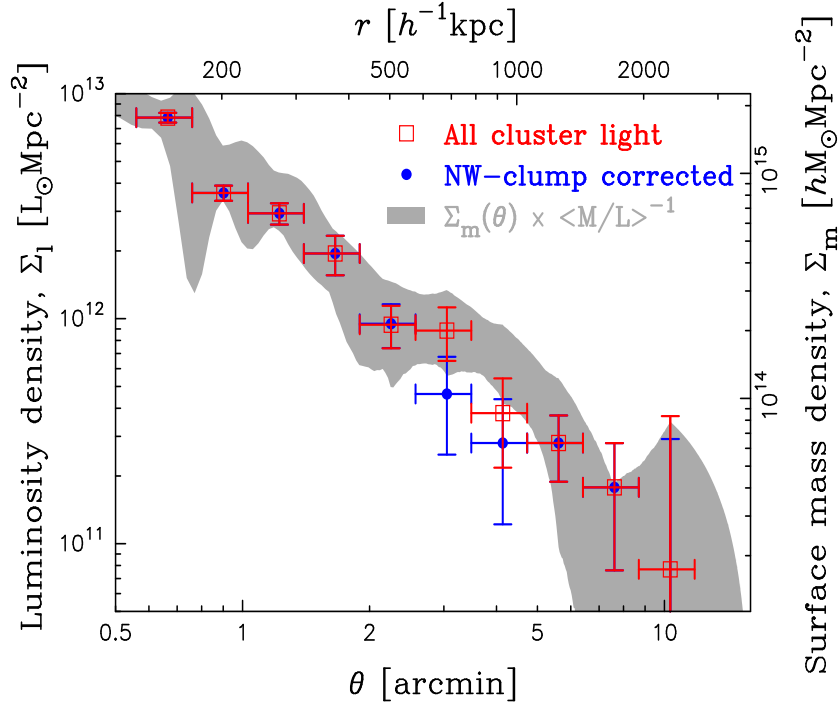}
 \end{center} 
\caption{ 
The $R_{\rm c}$-band surface luminosity density profile
 $\Sigma_l(\theta)$ ({\it squares}) of $BR_{\rm c}z'$-selected cluster
 member galaxies in Cl0024+1654. The gray-filled region represents the
 68.3\% confidence bounds for the surface mass density profile
 $\Sigma_m(\theta)$ reconstructed from our joint weak and strong lensing
 analysis (Figure \ref{fig:kprof}), converted into a luminosity density
 assuming a constant mass-to-light ratio of 
$\langle M/L_R \rangle(< 3\arcmin) \simeq 230  h\,(M/L_R)_\odot$.
Also shown with filled circles is the same cluster luminosity density
 profile, but corrected for the presence of the northwest clump located
 at a projected distance of $\theta \sim 3\arcmin$.
\label{fig:lprof}
} 
\end{figure}



\begin{figure}[htb]
 \begin{center}
   \includegraphics[width=110mm, angle=0]{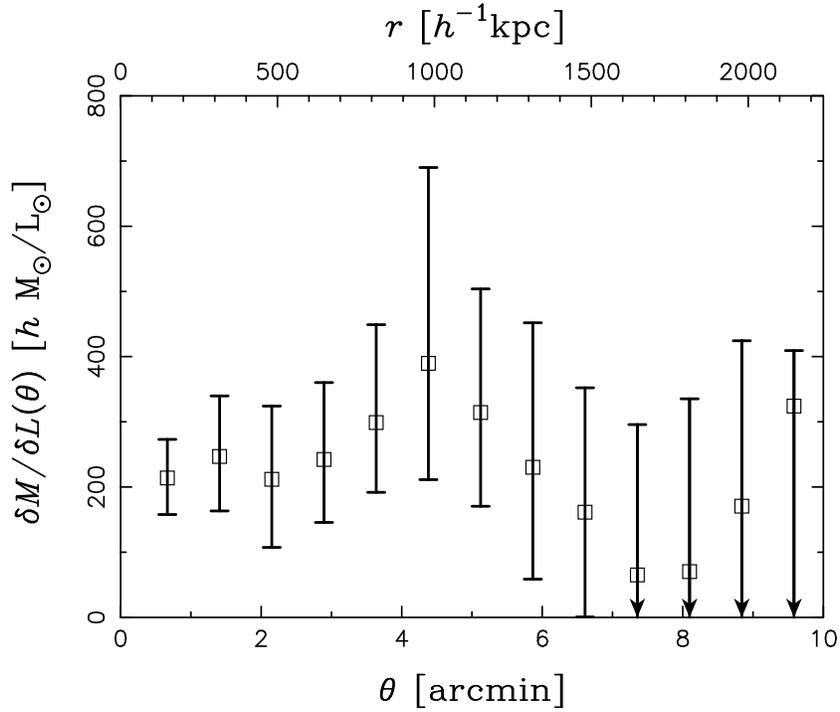}
 \end{center}
\caption{
Model-independent radial profile of the differential mass-to-light ratio  
 $\delta M(\theta)/\delta L_R(\theta)=\Sigma_m(\theta)/\Sigma_l(\theta)$
derived using the surface mass density profile $\Sigma_m(\theta)$ (see
 Figure \ref{fig:kappa}) from the Subaru weak-lensing analysis and the
 $K$-corrected $R_{\rm c}$-band surface luminosity density profile
 $\Sigma_l(\theta)$ measured from $BR_{\rm c}z'$-selected cluster member
 galaxies. 
The error bars represent 68.3\% confidence limits.
\label{fig:ml}
}
\end{figure}


\clearpage

\begin{figure}[htb]
 \begin{center}
   \includegraphics[width=90mm, angle=0]{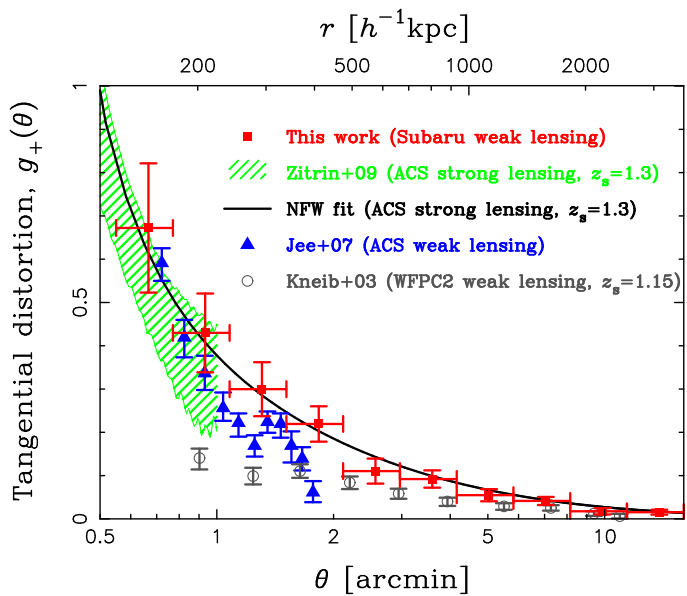}
 \end{center}
\caption{
Comparison of tangential distortion ($g_+$) profiles from different
 lensing studies. No correction is applied to the mean depth of lensing
 observations between different studies. The hatched region shows the
 $68.3\%$ confidence interval for the ACS strong-lensing results based
 on the azimuthally-averaged $\kappa$ profile of
 \cite{Zitrin+2009_CL0024}, scaled to a source redshift of $z_s=1.3$,
 roughly matching the mean depth of the Subaru blue+red background
 sample (see Table \ref{tab:color}). 
Also shown with the solid curve is the best-fit NFW model ($z_s=1.3$)
 from the ACS strong lensing constraints. The squares show our Subaru
 $g_+$ profile as measured from $BR_{\rm c}z'$-selected blue+red
 background galaxies. 
A simple extrapolation of the ACS/NIC3-derived NFW profile ({\it solid
 curve}) fits well with the outer Subaru distortion information ({\it
 squares}) over a wide range of radius, but somewhat overpredicts the
 distortion profile at $3\arcmin \simlt \theta \simlt 5\arcmin$.
The triangles show the $g_+$ profile from the weak lensing analysis of
 \citet[][Figure 13]{Jee+2007_CL0024} based on deep 6-passband HST/ACS
 images, where the data are limited to the positive parity region
 ($\theta\simgt 40\arcsec$) in this comparison.
The open circles show the $g_+$ profile at 
$0.9\arcmin \simlt \theta \simlt 11\arcmin$ from the weak lensing
 analysis of \citet[][Figure 7]{2003ApJ...598..804K} based on a
 sparse-sampled mosaic of 2-band (F450W, F814W) WFPC2 observations.  
 The flattened slope of Kneib et al. $g_+$ profile 
($\overline{z}_s = 1.15\pm 0.3$) at $\theta\simlt 2\arcmin$
 is likely due to contamination of the weak lensing signal by unlensed
 cluster member galaxies. Note that  \cite{2003ApJ...598..804K} 
 made a  correction for this dilution effect in their two-dimensional lens
 modelling. 
\label{fig:gtcomp} 
} 
\end{figure}



\begin{figure}[htb]
 \begin{center}
   \includegraphics[width=90mm, angle=0]{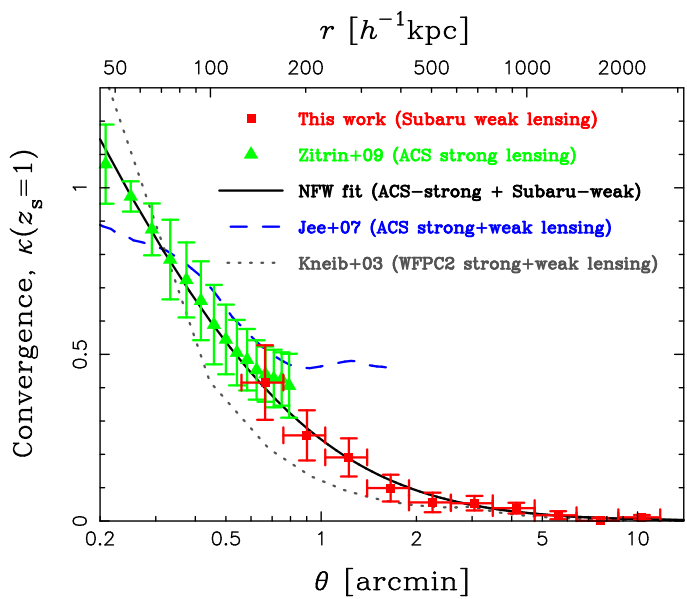}
 \end{center} 
\caption{
Comparison of projected mass density ($\kappa$) profiles of Cl0024+1654
 from different  lensing studies. All of the radial profiles are scaled
 to a fiducial redshift of $z_s=1$. 
The squares represent our model-independent $\kappa(\theta)$ profile
 (this work) reconstructed from Subaru distortion data (Figure
 \ref{fig:kappa}). The error bars are correlated.
The triangles represent the results from the strong-lensing analysis of
 \cite{Zitrin+2009_CL0024} based on deep ACS/NIC3 images. The solid
 curve shows the best-fitting NFW profile from our full lensing analysis
 of ACS/NIC3 and Subaru observations covering a wide range of radii,
 $R\simeq [40,2300]$\,kpc$\,h^{-1}$. 
The dashed curve shows the $\kappa$ profile ($\theta\simlt 1.8\arcmin$) 
 reconstructed from the ACS weak and strong lensing analysis of
 \cite{Jee+2007_CL0024}.  
The dotted curve shows the two-component lens mass model constrained
 from the WFPC2 weak lensing measurements of \cite{2003ApJ...598..804K}
 combined  with the inner strong-lensing constraint on the Einstein
 radius. 
\label{fig:kcomp}
}
\end{figure} 
 

\clearpage

\begin{figure}[htb]
 \begin{center}
   \includegraphics[width=90mm, angle=0]{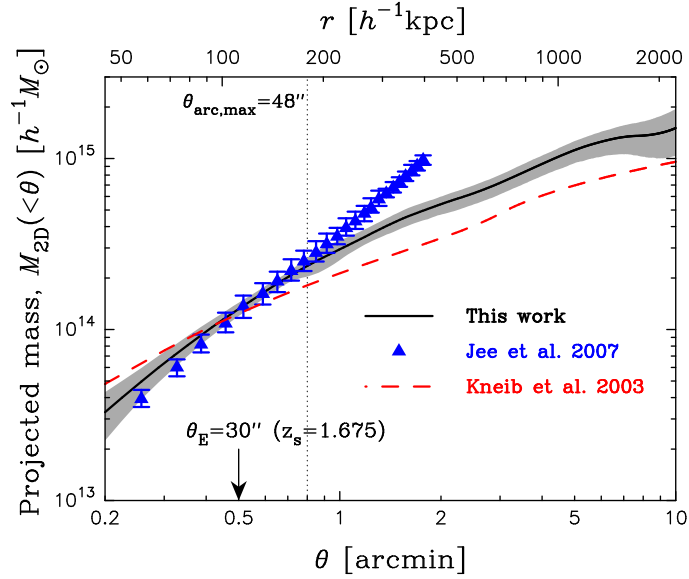}
 \end{center}
\caption{
Comparison of cumulative projected mass profiles $M_{\rm 2D}(<\theta)$
 of the galaxy cluster Cl0024+1654 from different lensing studies.
The solid curve (this work) represents the results from 
our joint weak and strong  lensing analysis of deep Subaru and ACS/NIC3
observations over a wide radial range from $40$\,kpc\,$h^{-1}$ 
to $2300$\,kpc\,$h^{-1}$.
The gray-shaded area shows the $68.3\%$ confidence interval at each
 radius estimated from a Monte-Carlo error analysis taking into account
 the error covariance matrix of our full lensing constraints.
Shown with a dotted vertical line is 
the maximum radius of multiply-lensed images used in the strong lensing
 analysis of \cite{Zitrin+2009_CL0024}.
The triangles represent the results from the ACS weak lensing analysis
of \citet[][Figure 12]{Jee+2007_CL0024}.
The dashed curve shows the best-fit two-component lens model 
of  \cite{2003ApJ...598..804K},
constrained from their WFPC2 weak lensing measurements combined with the
inner Einstein radius constraint, taking into account
the contribution of both the central and northwest clumps identified in
 projection space.
The three mass profiles from different lensing work
are in good agreement at the Einstein radius $\theta_{\rm E}\simeq
 30\arcsec$  
(as indicated
 by the arrow) for the 5-image system ($z_s=1.675$), by which each profile is
 normalized. The projected mass profile of \cite{Jee+2007_CL0024} is in an
 excellent 
 agreement with our joint mass profile out to $\theta\simeq 
 1.1\arcmin$ ($r\simeq 250$\,kpc\,$h^{-1}$ in projection space), but
 increasingly exceeds our profile at $\theta\simgt 1.1\arcmin$ out to
 the limit of their data. 
\label{fig:m2d}
}
\end{figure} 
 


\begin{figure}[htb]
 \begin{center}
   \includegraphics[width=90mm, angle=0]{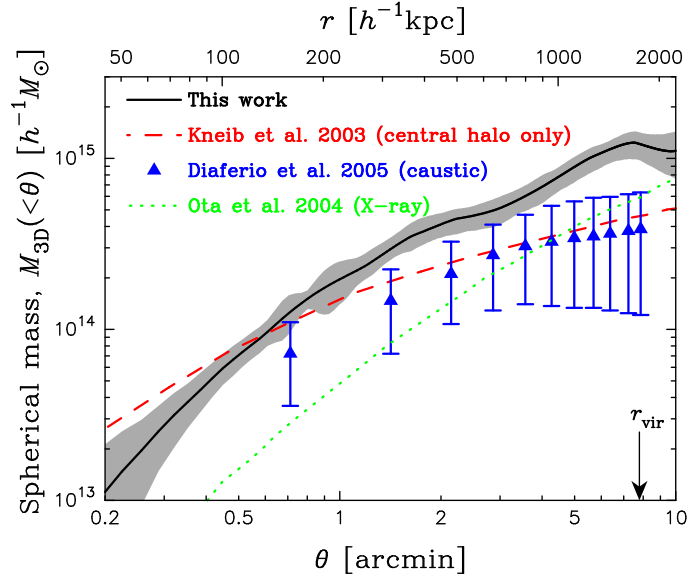}
 \end{center}
\caption{
Comparison of cumulative spherical mass profiles $M_{\rm 3D}(<\theta)$
 of the galaxy cluster Cl0024+1654 from different studies.
The solid curve (this work)
represents the deprojected spherical mass profile 
from our full lensing analysis of Subaru and ACS/NIC3 data
assuming spherical symmetry for the cluster system. 
The gray shaded area shows the $68.3\%$ confidence interval at each
 radius estimated from a Monte-Carlo error analysis taking into account
 the error covariance matrix of our joint Subaru and ACS/NIC3 lensing
constraints. 
Shown with a dashed curve is the spherical NFW model of 
\cite{2003ApJ...598..804K}
for the central component alone (see also Figure \ref{fig:m2d}).  
The triangles represent the results from the dynamical analysis of
\cite{Diaferio+2005}
based on the caustic method.
The dotted curve shows the best-fitting NFW profile from the Chandra
 X-ray analysis of 
\cite{2004ApJ...601..120O}
obtained assuming the hydrostatic
 equilibrium. Note that the maximum limit of the Chandra data of 
\cite{2004ApJ...601..120O}
is about $4\arcmin$ ($900$\,kpc\,$h^{-1}$).
Our joint mass profile is only marginally consistent with the
 dynamical results of 
\cite{Diaferio+2005}
out to $\theta\sim
 3.5\arcmin$, but increasingly exceeds the caustic-based profile at
 large radii. A significant increase in $M_{\rm 3D}(<r)$ is seen in our
 joint mass profile at $3.5\arcmin \simlt \theta \simlt 5\arcmin$, 
indicating an  additional, extended mass component in the outer radius.
\label{fig:m3d}
}
\end{figure}


\clearpage

\begin{figure}[htb]
 \begin{center}
   \includegraphics[width=150mm, angle=0]{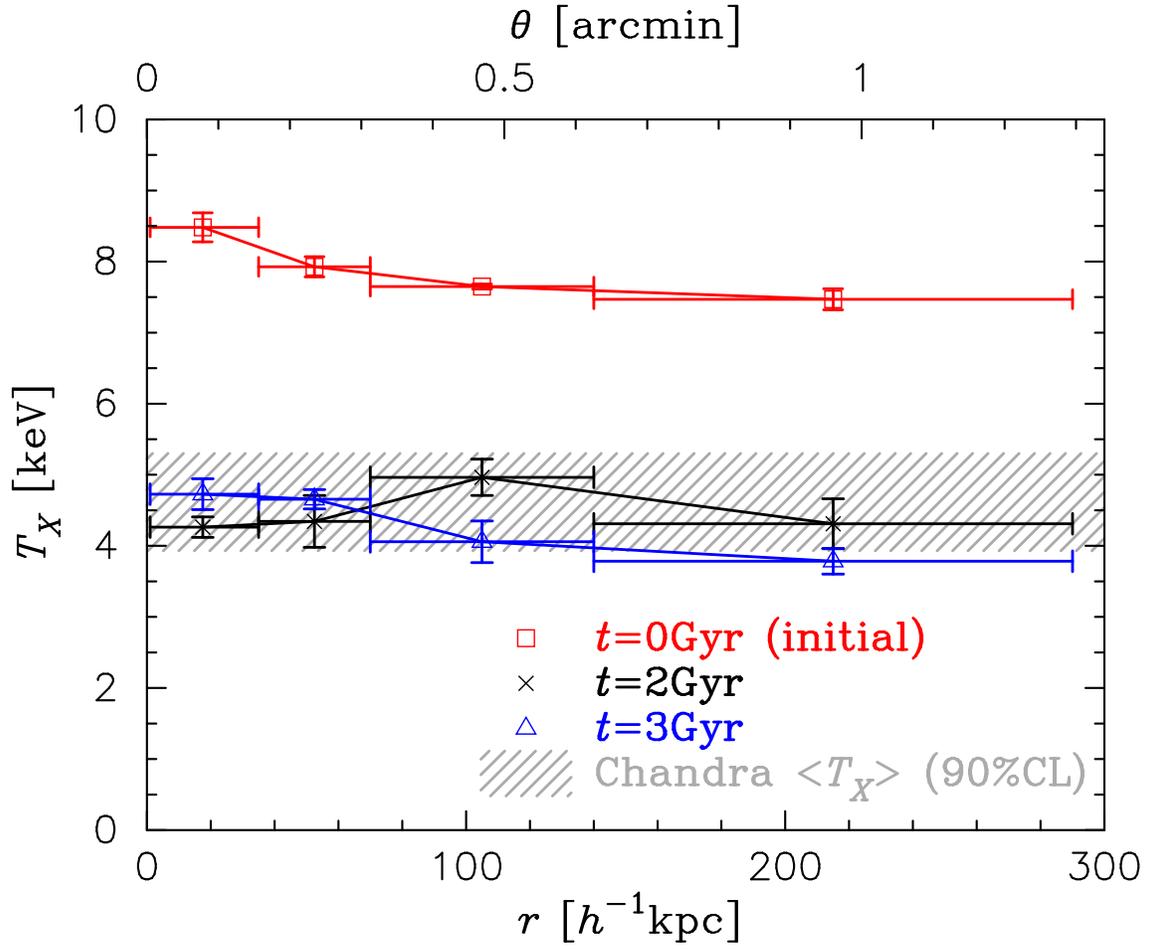}
 \end{center}
\caption{
\label{fig:tx}
Projected spectroscopic-like temperature profiles created from mock
X-ray observations of our simulated head-on 2:1 mass-ratio cluster
 merger, shown at three different times,
$t=0, 2$, and $3$\,Gyr after the beginning of the collision
({\it open squares}, {\it crosses}, and {\it open triangles},
 respectively).
The vertical error bars represent the dispersion due to azimuthal
 averaging. The hatched region shows the 90\% confidence interval for
 the average X-ray temperature,
 $T_X=4.47^{+0.83}_{-0.54}$, from Chandra X-ray observations of
 \cite{2004ApJ...601..120O}. 
}
\end{figure}




\end{document}